\documentclass[%
superscriptaddress,
nofootinbib,
amsmath,amssymb,
prd,
floatfix,
onecolumn,
]{revtex4-2}

\usepackage{graphicx}
\usepackage{float}
\usepackage{subfigure}
\usepackage{color}
\usepackage{dcolumn}
\usepackage{bm}
\usepackage{hyperref}
\usepackage{CJK}
\usepackage{multirow}
\usepackage{ulem} 

\newcommand\tbbint{{-\mkern -16mu\int}}

\newcommand\dbbint{{-\mkern -19mu\int}}

\newcommand\bbint{
{\mathchoice{\dbbint}{\tbbint}{\tbbint}{\tbbint}}
}
\newcommand{\md}{\mathrm{d}}

\newcommand{\be}{\begin{equation}}
\newcommand{\ee}{\end{equation}}
\newcommand{\beq}{\begin{equation}}
\newcommand{\eeq}{\end{equation}}
\newcommand{\bea}{\begin{eqnarray}}
\newcommand{\eea}{\end{eqnarray}}
\newcommand{\ba}{\begin{eqnarray}}
\newcommand{\ea}{\end{eqnarray}}

\begin{document}

\title{Roy equation analyses of $\pi\pi$ scatterings at unphysical pion masses}%



\author{Xiong-Hui Cao}
\affiliation{%
 School of Physics and State Key Laboratory of Nuclear Physics and Technology, Peking University,\\
 Beijing 100871, People's Republic of China
}%
\author{Qu-Zhi Li}
\affiliation{%
 School of Physics and State Key Laboratory of Nuclear Physics and Technology, Peking University,\\
 Beijing 100871, People's Republic of China
}%
\author{Zhi-Hui Guo}
\email[Corresponding author: ]{zhguo@hebtu.edu.cn}
\affiliation{
 Department of Physics and Hebei Key Laboratory of Photophysics Research and Application, Hebei Normal University, Shijiazhuang 050024, People's Republic of China 
}%
\author{Han-Qing Zheng}
\email[Corresponding author: ]{zhenghq@scu.edu.cn}
\affiliation{
 College of Physics, Sichuan University, Chengdu, Sichuan 610065, People's Republic of China
}%

\date{\today}

\begin{abstract}
An extended Roy equation including a bound state pole is used to study $\pi\pi$ scatterings at unphysical large pion masses when $\sigma$ becomes a bound state in one situation and stays as a broad resonance in the other case. The coupled integral equations at large pion masses are solved by taking the lattice driving terms and the Regge amplitudes as inputs. Relying on the solutions of Roy equations that respect unitarity, analyticity and crossing symmetry,  we give predictions to the phase shifts with $IJ=00,11,20$ in the elastic energy region. We then perform analytic continuation into the complex $s$ plane to search for various poles, all of which are inside the validity domain of the Roy equation. This is the first time that lattice data at unphysical large pion masses are analyzed within the rigorous Roy equation method. 
\end{abstract}

\maketitle

\section{Introduction}

Meson-meson scatterings offer a valuable framework to study QCD in the non-perturbative region. 
Roy equation analyses~\cite{Roy:1971tc, Ananthanarayan:2000ht,Buettiker:2003pp,Garcia-Martin:2011iqs} possessing crossing symmetry to meson-meson scatterings that involve rather different types of resonances from channels with different quantum numbers, turn out to be quite useful to put strong constraints on the resonance properties~\cite{Caprini:2005zr,Descotes-Genon:2006sdr,Garcia-Martin:2011nna,Moussallam:2011zg,Pelaez:2020uiw} and the scattering amplitudes~\cite{Colangelo:2001df,Ananthanarayan:2000ht,Garcia-Martin:2011iqs,Caprini:2011ky,Pelaez:2018qny}. 
In addition, the similar Roy-Steiner equation analyses have been introduced into the baryon sector to study the $\pi N$ scattering amplitudes~\cite{Hite:1973pm, Hoferichter:2015dsa, Hoferichter:2015hva, Hoferichter:2015tha} and nucleon resonances~\cite{Cao:2022zhn}. 
For the lightest QCD resonance $\sigma/f_0(500)$, the precise determination of its mass and width is reached upon the use of the rigorous $\pi\pi$ Roy equation~\cite{Caprini:2005zr}, though there has been a long-standing effort aiming at the establishment of its existence in history (for recent reviews, see Refs.~\cite{Pelaez:2015qba, Yao:2020bxx}). 
The convincing results from Roy-like equation analysis are rooted in its rigorous implementation of analyticity and crossing symmetry from the analytic S-matrix theory~\cite{Eden:1966dnq}. 
Crossing symmetry implies delicate relations among the non-resonant force in the $IJ=20$ case, the scalar $\sigma/f_0(500)$ in the $IJ=00$ channel, the vector $\rho(770)$ with $IJ=11$ and other heavier resonance states appearing in $\pi\pi$ scattering. It is demonstrated in Refs.~\cite{Guo:2007ff,Guo:2007hm} that only when resonances in the $s$ and crossed channels are simultaneously included one can obtain consistent results from the matching with chiral perturbation theory ($\chi$PT) in different $IJ$ channels. Such delicate relations among the amplitudes in different channels required by crossing symmetry can be specially useful to constrain the lattice results at unphysical quark masses, which generally bear large uncertainties in the numerical simulations nowadays. This is also one of the key motivations of our study. 

Rapid developments in meson-meson scatterings have been made by lattice QCD simulations, where the scattering phase shifts can be obtained by mapping lattice finite-volume spectra, see a recent review~\cite{Briceno:2017max}. 
Although to tackle unstable hadrons in meson-meson scatterings is challenging in lattice QCD simulations, remarkable progresses have been made not only on the $\rho(770)$~\cite{Dudek:2012xn} but also on the $\sigma/f_0(500)$~\cite{Briceno:2016mjc,Briceno:2017qmb}, where the lattice calculations are typically carried out at unphysical large quark masses. Depending on the channels in question, the amplitudes at large quark masses can be either similar to or drastically different from those at physical masses.   
E.g., the resulting resonance spectra with $m_\pi=391$~MeV~\cite{Dudek:2012xn,Briceno:2016mjc,Briceno:2017qmb} turn out to be rather different from the physical ones: the $\rho(770)$ width becomes around one-order magnitude smaller and the $\sigma$, strikingly, transforms from a broad resonance into a bound state below the two-pion threshold, while the $\pi\pi$ phase shifts with $IJ=20$ at different quark masses share qualitatively similar trends~\cite{Dudek:2010ew,Dudek:2012gj}. These indicate that the fulfillment of crossing symmetry at large lattice masses can be nontrivially different from the situations at physical ones. Such an interesting feature was not addressed in previous works relying on unitarized chiral amplitudes and data-driven $N/D$ method~\cite{Pelaez:2010fj,Albaladejo:2012te,Doring:2016bdr,Danilkin:2020pak}, due to the loss of exact crossing symmetry in those approaches. By contrast, the use of Roy equation that faithfully obey analyticity, unitarity and crossing symmetry, allows us to make a rigorous investigation into this intriguing problem.  

On the other hand, the $\pi\pi$ phase shifts at large quark masses, which although clearly reveal the bound state solution for the $\sigma/f_0(500)$, are still determined with sizable uncertainties in the present lattice simulations~\cite{Briceno:2016mjc,Briceno:2017qmb}. Demanding computing resources will be needed in lattice QCD calculations to reduce the uncertainties. Furthermore, the behavior of $\sigma$ changing from a broad resonance to a typical shallow bound/virtual state has been recognized for a long time~\cite{Hanhart:2008mx} when gradually increasing the pion masses, but the consensus about the exact pole contents is not reached yet~\cite{Pelaez:2010fj,Gao:2022dln,vanBeveren:2022zfx,Gao:2022tlh}, especially in the situation when $\sigma$ turns into a bound state. The coupled integral Roy equations from different $IJ$ channels with crossing symmetry and analyticity can provide useful theoretical constraints to give a more definite conclusion on the various pole contents and to pin down the error bars of the lattice phase shifts, which procedure also gives more reliable phase shifts for future phenomenological studies due to the implementation of crossing symmetry. 
It is noted that to what extent such analyses can constrain the amplitudes at the unphysical large quark mass is still rarely studied in literature. 
Our key task is to carry out the rigorous investigation of such problem within the Roy equation approach. 

The paper is organized as follows. In Sec.~\ref{sec.2} we derive the set of Roy equations that we intend to solve. The procedure is similar to the pioneer work~\cite{Roy:1971tc}, but we add the scalar-isoscalar bound state pole terms used in the dispersive integrals to accommodate the lattice data. Our goal is to show how does (exact) crossing symmetry give a new analytic structure for partial-wave amplitudes. 
After discussing the available lattice inputs and the asymptotic Regge amplitudes for $m_\pi=391~$MeV, the relations between the multiplicity index of the solution within the additional constraints and the unique solution of Roy equations are discussed in detail in Sec.~\ref{sec.3}.
Next, in Sec.~\ref{sec.4} we solve the equations numerically, and some comments on other unitarized methods are also given. The phenomenological discussions on the S-wave scattering lengths, pole information, the pion-mass trajectory for the $\sigma$ pole and the relevant results at $m_\pi=236~$MeV can be found in Sec.~\ref{sec.5}.
The paper ends with summary in Sec.~\ref{sec.6}. 
The demonstration of the existence of a virtual state pole in scalar-isotensor channel and the details to solve Roy equations for $m_\pi=236~$MeV are deferred to App.~\ref{Sec.appij20} and \ref{Sec.app236} respectively.

\section{Extended Roy equations}\label{sec.2}

In order to describe the bound state $\sigma$ in $\pi\pi$ scatterings at large pion masses revealed in Refs.~\cite{Briceno:2016mjc,Briceno:2017qmb}, one needs to modify the coupled dispersive Roy equations by explicitly including in the scattering amplitude the scalar-isoscalar bound state pole terms,  which are absent in the conventional Roy equation for the physical pion case~\cite{Roy:1971tc,Caprini:2005zr}~\footnote{In fact, we have explicitly verified that there would be no sensible solution to Roy equation at $m_\pi\sim 391~$MeV by only including two negative S-wave scattering lengths given in Refs.~\cite{Dudek:2012gj, Briceno:2016mjc} and excluding the bound state $\sigma$ pole term.}. 
The key point is to write a twice subtracted fixed-$t$ dispersion relation with a bound state pole $s_\sigma$ with the quantum number $IJ=00$ for the full amplitude $\vec{T}(s,t,u)$ in the isospin space,
\begin{align}\label{eq.tifull}
    \begin{aligned}
    \vec{T}(s, t, u)= & C_{s t}[\vec{C}(t)+(s-u) \vec{D}(t)]+32\pi g^2_{\sigma\pi\pi}\left(\frac{1}{s_{\sigma}-s}+\frac{1}{s_{\sigma}-u} C_{s u}\right) 
    \begin{pmatrix}
    1\\
    0\\
    0
    \end{pmatrix} \\
    &+\frac{1}{\pi} \int_{4 m_\pi^2}^{\infty} \frac{\md s^{\prime}}{s^{\prime 2}}\left(\frac{s^2}{s^{\prime}-s}+\frac{u^2}{s^{\prime}-u} C_{s u}\right) \operatorname{Im} \vec{T}\left(s^{\prime}, t, u^{\prime}\right)\ .
    \end{aligned}
\end{align}
We will follow the convention of Refs.~\cite{Roy:1971tc,Ananthanarayan:2000ht} for the explicit representation of $\vec{C}(t),\vec{D}(t)$ and the crossing matrices $C_{st},C_{su}$. The bound state scalar $\sigma$ pole accompanied by the $\sigma\pi\pi$ coupling squared $g_{\sigma\pi\pi}^2$ in Eq.~\eqref{eq.tifull}, appears not only in the $s$ channel but also in the crossed $u$ channel for the fixed-$t$ dispersion relation. 
After the partial-wave (PW) projection of the full amplitudes~\eqref{eq.tifull}, one can give the extended Roy equations for the PW amplitudes 
\begin{align}\label{eq.tij}
    \mathrm{Re} t^I_{J}(s)=k_{J}^{I}(s)+\sum_{I^{\prime}=0}^2 \sum_{J^{\prime}=0}^{1} \bbint_{4 m_{\pi}^{2}}^{s_\mathrm{m}} \md s^{\prime} K_{J J^{\prime}}^{I I^{\prime}}\left(s^{\prime}, s\right) {\rm Im} t^{I^{\prime}}_{J^{\prime}}(s^\prime) + d_J^I(s)\ ,
\end{align}
where `$\bbint$' represents the principal value integral, the kernel functions $K^{II'}_{JJ'}(s^\prime,s)$ are the same as those in Ref.~\cite{Ananthanarayan:2000ht}, $s_\mathrm{m}$ stands for the matching point, the driving terms (DTs) $d_J^I(s)$ include the effects of S- and P-waves from higher energy region beyond $s_\mathrm{m}$ and also the higher PWs\footnote{It is arbitrary to choose the value of the matching point $s_\text{m}$ in principle. Above the matching point, the corresponding DTs require the inputs from the experiments, lattice and even Regge models, and the phase shifts below this point can be directly solved numerically using Roy equation (we focus on the low-energy S- and P-waves here). As long as the inputs, such as the various DTs, are provided and suitable numerical methods are taken, we can get the solutions of Roy equations which can be then used to calculate the low-energy S- and P-waves phase shifts below the matching point.}.
The subtraction terms and the $\sigma$ pole terms are collected in $k^I_J(s)$ and they read  
\begin{align}\label{eq.kij}
    \begin{aligned}
    k^0_0(s) &=a^0_0+\frac{s-4 m_\pi^2}{12 m_\pi^2}(2a^0_0-5a^2_0)+\frac{g^2_{\sigma\pi\pi}}{12}\left(\frac{16 m_\pi^2(4 s-s_{\sigma})-4(2 s-s_{\sigma})(s+2 s_{\sigma})}{\left(4 m_\pi^2-s_{\sigma}\right) s_{\sigma}(s_{\sigma}-s)}-\frac{8 L_\sigma}{4 m_\pi^2-s}\right)\ ,\\
    k^1_1(s) &=0\ +\frac{s-4 m_\pi^2}{72 m_\pi^2}(2a^0_0-5a^2_0)+\frac{g^2_{\sigma\pi\pi}}{9}\left(-\frac{\left(4 m_\pi^2-s\right)^2-48 m_\pi^2 s_{\sigma}+12 s^2}{\left(4 m_\pi^2-s\right)\left(4 m_\pi^2-s_{\sigma}\right) s_{\sigma}}+\frac{6\left(s+2 s_{\sigma}-4 m_\pi^2\right) L_\sigma}{\left(4 m_\pi^2-s\right)^2}\right)\ ,\\
    k_0^2(s) &=a^2_0-\frac{s-4 m_\pi^2}{24 m_\pi^2}(2a^0_0-5a^2_0)-\frac{g^2_{\sigma\pi\pi}}{3}\left(\frac{4 m_\pi^2+s-2 s_{\sigma}}{s_{\sigma}(4 m_\pi^2-s_{\sigma})}+\frac{2 L_\sigma}{4 m_\pi^2-s}\right)\ ,
    \end{aligned}
\end{align}
with the logarithm $L_\sigma=\ln \left(\frac{s+s_{\sigma}-4 m_\pi^2}{s_{\sigma}}\right)$. It is easy to verify that $k^I_J(s)$ reduces to the scattering length at $\pi\pi$ threshold due to $\lim\limits_{s\to 4m_\pi^2}(k^0_0, k^1_1, k^2_0)(s)=(a^0_0,0,a^2_0)$. 
It is worth noting that the last terms inside the brackets accompanied by $g^2_{\sigma\pi\pi}$ in Eqs.~\eqref{eq.kij} correspond to the bound state $\sigma$ in the $IJ=00$ channel, which also contributes to the other two channels via crossing. 
We point out that within the various unitarized chiral amplitude approaches~\cite{Hanhart:2008mx,Pelaez:2010fj,Albaladejo:2012te,Doring:2016bdr} and data-driven $N/D$ method~\cite{Danilkin:2020pak} when tuning the pion masses to some specific large values the bound state pole of $\sigma$ can be generated in the $s$ channel, however due to the loss of exact crossing symmetry its effects in the crossed channels are usually neglected. 

Furthermore, the so-called Balachandran-Nuyts-Roskies (BNR) relations~\cite{Balachandran:1968zza, Roskies:1969pe, Roskies:1970uj} derived from crossing symmetry can impose constraints among PW amplitudes with different $IJ$ quantum numbers in the subthreshold energy region between $0$ and $2m_\pi$. Interestingly, as noticed in Ref.~\cite{Gao:2022dln}, the BNR relations could be specially useful for large pion masses when the $\sigma$ becomes a bound state below $\pi\pi$ threshold. 
Only five relations are related to S- and P-waves (see, e.g., Ref.~\cite{Martin:1976}), which are some integral relations of PW amplitudes,
\begin{align}
\begin{aligned}
 & \int_0^{4m_\pi^2}(s-4 m_\pi^2)(3 s-4m_\pi^2)\left[t_0^0(s)+2 t_0^2(s)\right]\md s=0, \\
 & \int_0^{4m_\pi^2}(s-4 m_\pi^2) R_0^0\left[2 t_0^0(s)-5 t_0^2(s)\right]\md s=0, \\
 & \int_0^{4m_\pi^2}(s-4 m_\pi^2) R_1^0\left[2 t_0^0(s)-5 t_0^2(s)\right]\md s-9 \int_0^{4m_\pi^2}(s-4 m_\pi^2)^2 R_0^1 t_1^1(s)~\md s=0, \\
 &  \int_0^{4m_\pi^2}(s-4 m_\pi^2) R_2^0\left[2 t_0^0(s)-5 t_0^2(s)\right]\md s+6 \int_0^{4m_\pi^2}(s-4 m_\pi^2)^2 R_1^1 t_1^1(s)~\md s=0, \\
 & \int_0^{4m_\pi^2}(s-4 m_\pi^2) R_3^0\left[2 t_0^0(s)-5 t_0^2(s)\right]\md s-15 \int_0^{4m_\pi^2}(s-4 m_\pi^2)^2 R_2^1 t_1^1(s)~\md s=0,
\end{aligned}
\end{align}
where $R^j_i$ are polynomials of $s$,
\begin{align}
\begin{aligned}
& R_0^0=1, \quad R_0^1=1, \\
& R_1^0=3 s-4 m_\pi^2, \quad R_1^1=5 s-4 m_\pi^2, \\
& R_2^0=10 s^2-32 s m_\pi^2+16 m_\pi^4, \quad R_2^1=21 s^2-48 s m_\pi^2+16 m_\pi^4, \\
& R_3^0=35 s^3-180 s^2 m_\pi^2+240 s m_\pi^4-64 m_\pi^6\ .
\end{aligned}
\end{align}
The integration region of the BNR relations covers not only the bound state $\sigma$ pole in $t^0_0(s)$ but also part of the left-hand cuts (LHCs) generated by the $\sigma$ in the crossed channel, since the LHCs of PWs are now extended to $(-\infty,4m_\pi^2-s_\sigma]$ instead of $(-\infty,0]$ due to the crossed-channel exchange of $\sigma$. A novel observation is found in our study that the contribution from the $\sigma$ pole term of the $s$ channel in the BNR relation is exactly cancelled by the LHCs generated by the crossed-channel exchanges of $\sigma$. 
This implies that when neglecting the LHCs generated by the bound state $\sigma$ pole in the $\pi\pi$ scattering amplitudes as done in Ref.~\cite{Gao:2022dln} one probably would introduce artificial effects in order to fulfill the BNR relations. 

It is demonstrated here that the rigorous Roy equation analysis enables us to take the full consideration of the bound state $\sigma$ in all channels, as shown in Eqs.~\eqref{eq.kij}. The $\sigma$ pole position $s_{\sigma}$ and its coupling $g_{\sigma\pi\pi}$, together with the scattering lengths $a_0^0$ and $a_0^2$, will be tuned to solve the coupled integral Eqs.~\eqref{eq.tij}.

\section{Inputs to solve Roy equations: lattice data and Regge amplitudes}\label{sec.3}

In this work, our main focus is to determine the phase shifts only in the elastic energy region from the $\pi\pi$ threshold up to the matching point $\sqrt{s_{\mathrm{m}}}=2m_K=1098~$MeV for $m_\pi=391~$MeV, when $\sigma$ becomes a bound state~\cite{Briceno:2016mjc,Briceno:2017qmb}. 
The key inputs of Eqs.~\eqref{eq.tij} are the DTs $d^I_J(s)$, which contain the information of high energy and high PWs.

\subsection{Inputs from lattice calculations}
In practice, the DTs consist of two parts: inputs of 
S-, P- and D-waves from lattice data up to 1.8~GeV, and the higher energy and higher PW contributions.
For the DTs of S-, P-waves in the energy region from $K\bar{K}$ threshold to $1.8~$GeV and D-waves in the energy region from $\pi\pi$ threshold to $1.8~$GeV, we exploit the results from the HadSpec collaboration~\cite{Dudek:2012gj, Dudek:2012xn, Briceno:2017qmb}. Due to the limited lattice resources, HadSpec collaboration does not always provide data up to $1.8~$GeV for all the channels, so it requires us to extrapolate lattice results to $1.8~$GeV. Fortunately, the impact of extrapolation on the final results is minor and almost negligible, due to the high energy suppression $1/s^{\prime 3}$ in the kernel functions $K^{II^\prime}_{JJ^\prime}(s^\prime,s)$ in Eqs.~\eqref{eq.tij}~\cite{Caprini:2005zr}. The various uncertainties from the lattice data themselves and also the extrapolations are then propagated to the final results through bootstrap method.

For the $IJ=00$ channel, the available lattice data are up to around $ 1.5~$GeV~\cite{Briceno:2017qmb}.
An important observation is that the impacts of $m_\pi$ variations gradually decrease with the increase of energy, so physical data can give some insights in the high energy region.
Since physical $\mathrm{Im}t^0_0(s)$ shows a slow downtrend when $\sqrt{s}>1.3~$GeV~\cite{Pelaez:2019eqa}, we adopt a conservative extrapolation to set $\mathrm{Im}t^0_0(s)$ as constants with large uncertainties in the energy region $1.5\sim 1.8~$GeV, and in this way it also accounts for the complicated coupled-channel effects. 
For the $IJ=20$ channel, the available lattice data are up to around $ 1.5~$GeV~\cite{Dudek:2012gj}\footnote{Notice that the isotensor lattice data in Refs.~\cite{Dudek:2012gj} correspond to $m_\pi = 396~$MeV, not
$391~$MeV. However, such mismatch can be nearly ignored, and the reason is twofold. Firstly, we work in the isospin symmetric limit by ignoring the mass difference of the charged pions/kaons and the neutral ones. In practice, the two different 
thresholds for $K^+K^-$ and $K^0\bar{K^0}$ are separated by several MeVs, which are however ignored in the isospin limit. The variation between 391~MeV and 396~MeV is actually at the same level of isospin breaking effects that are neglected in the current study. Second, the $\pi\pi$ phase shifts in the isotensor channel only moderately depend on the pion masses and the small variation of the pion masses is not expected to give noticeable effects. Therefore we claim the effects of the pion mass variation from $m_\pi = 396~$MeV to $m_\pi = 391~$MeV can be ignored in this work. As for matching condition of phase shift $\delta^2_0$ (see the next section), we will simply set the same matching point $\sqrt{s_\text{m}}=2m_K=1098$~MeV as the common one used in the $IJ=00$ and 11 channels.}.
Due to the moderate mass-dependence of the phase shifts in $IJ=20$ channel and the minor inelastic effects below 1.8~GeV at $m_\pi=391$~MeV~\cite{Dudek:2012gj}, we utilize a linear extrapolation of phase shift in the energy region $1.5\sim 1.8~$GeV and assume elastic approximation simultaneously~\cite{Dudek:2012gj}. 
For the $IJ=11$ channel, the available lattice data are only up to $\sqrt{s_0}\simeq 1.1~$GeV~\cite{Dudek:2012xn}. 
By assuming $\delta^1_1(\infty)=\pi$, we use a convenient extrapolation scheme $\delta_1^1(s)=\pi+\left(\delta_1^1\left(s_0\right)-\pi\right) \frac{2}{1+\left(s / s_0\right)^{3 / 2}}$, as proposed in Ref.~\cite{Moussallam:1999aq}. 
Another contribution comes from the D-wave amplitude with the $f_2(1270)$, which turns out to be the most important one among the various higher PW DTs. 
Fortunately, the $\pi\pi$ scattering amplitude in the $IJ=02$ channel is calculated precisely up to $1.8~$GeV~\cite{Briceno:2017qmb}, but for the $IJ=22$ channel, the available lattice data are up to around $ 1.5~$GeV~\cite{Dudek:2012gj}. 
Because the $IJ=22$ channel is a non-resonant case and also shows a slow downward trend, we take the elastic approximation and  extrapolate the phase shifts as a function of energy squared from $1.5$~GeV to $1.8~$GeV.

We verify that the final results are robust with these extrapolations because in the twice subtracted dispersion relation the corresponding contributions from the extrapolated high energy region are suppressed and play a minor role in the final results (see the next section). To be specific, the main conclusions are almost unaffected by these extrapolation methods.

\subsection{Inputs from Regge models}

In addition, higher PWs and the DTs above $1.8~$GeV are estimated by the Regge pole theory~\cite{Martin:1970Elementary, Collins:1977jy}.
Although the physical $\pi\pi$ Regge amplitudes can be constructed by fitting the experimental cross sections as done in Refs.~\cite{Pelaez:2003eh, Garcia-Martin:2011iqs, Caprini:2011ky}, the $\pi\pi$ Regge amplitudes at unphysical large pion masses are poorly known due to lacking of the lattice constraints. 
In this work we will exploit an improved Veneziano-Lovelace-Shapiro model~\cite{Veneziano:1968yb,Lovelace:1968kjy, Shapiro:1969km} to analyze the asymptotic behavior of the scattering amplitude,  see Ref.\cite{Ananthanarayan:2000ht} for more details. 
A Regge trajectory with isospin $I_t$ gives a contribution $\propto s^{\alpha(t)}$ to the $t$ channel isospin amplitude $\operatorname{Im} T^{(I_t)}(s, t)$, which is related to the $s$ channel amplitude $\operatorname{Im} T^{I_s}(s, t)$ via 
\begin{align}
    \operatorname{Im} T^{(I_t)}(s, t)=\sum_{I_s} C_{s t}^{I_t I_s} \operatorname{Im} T^{I_s}(s, t)\ .
\end{align}
The asymptotic behaviors of the $s$ channel isospin amplitudes take the form~\cite{Ananthanarayan:2000ht}
\begin{align}
    \begin{aligned}
    & \operatorname{Im} T^{I_s=0}(s, t)=\frac{1}{3}\beta_P^{\pi\pi}e^{b_P^{\pi\pi}t}\left(\frac{s}{s_1}\right)+\frac{1}{3} \beta_f(t) \left(\frac{s}{s_1}\right)^{\alpha_f(t)}+\beta_\rho(t) \left(\frac{s}{s_1}\right)^{\alpha_\rho(t)}+(t \leftrightarrow u)\ , \\
    & \operatorname{Im} T^{I_s=1}(s, t)=\frac{1}{3}\beta_P^{\pi\pi}e^{b_P^{\pi\pi}t}\left(\frac{s}{s_1}\right)+\frac{1}{3} \beta_f(t) \left(\frac{s}{s_1}\right)^{\alpha_f(t)}+\frac{1}{2}\beta_\rho(t) \left(\frac{s}{s_1}\right)^{\alpha_\rho(t)}-(t \leftrightarrow u)\ , \\
    & \operatorname{Im} T^{I_s=2}(s, t)=\frac{1}{3}\beta_P^{\pi\pi}e^{b_P^{\pi\pi}t}\left(\frac{s}{s_1}\right)+\frac{1}{3} \beta_f(t) \left(\frac{s}{s_1}\right)^{\alpha_f(t)}-\frac{1}{2}\beta_\rho(t) \left(\frac{s}{s_1}\right)^{\alpha_\rho(t)}+(t \leftrightarrow u)\ ,
    \end{aligned}
\end{align}
where the normalization factor is chosen as $s_1=1~\mathrm{GeV}^2$. 
In this model, the $\rho$- and $f$-trajectories are linear and assumed to be degenerate, i.e., $\alpha(t)\equiv\alpha_\rho(t)=\alpha_f(t)=\alpha_0+\alpha_1 t$, and $\alpha_1=\frac{1}{2\left(m_\rho^2-m_\pi^2\right)}=0.87~\mathrm{GeV}^{-2}, \quad \alpha_0=\frac{1}{2}-\alpha_1 m_\pi^2=0.37$, where we have taken $m_\rho= 854.1~$MeV for $m_\pi=391$~MeV~\cite{Dudek:2012xn}.
In addition, the explicit parameterization of the $\rho$- and $f$- residues are $\beta_\rho(t)=\frac{2}{3} \beta_f(t)=\eta\frac{\pi \lambda \alpha_1^{\alpha(t)}}{\Gamma[\alpha(t)]}$~\cite{Ananthanarayan:2000ht} with $\lambda=96 \pi \Gamma_\rho m_\rho^2\left(m_\rho^2-4 m_\pi^2\right)^{-3 / 2}=67.34$, where at $m_\pi=391$~MeV we have taken $\Gamma_\rho= 12.4~$MeV~\cite{Dudek:2012xn}.
As indicated in Ref.~\cite{Ananthanarayan:2000ht}, this model overestimates the magnitude of the Regge residues, thus a significant fraction thereof should be transferred to the Pomeron term.
It is suggested that the value of the strength factor $\eta$ can be set to $0.5\pm0.2$ to estimate the effects from the Pomeron~\cite{Ananthanarayan:2000ht}. 

\begin{figure}[h]
\centering
\includegraphics[width=1\textwidth,angle=-0]{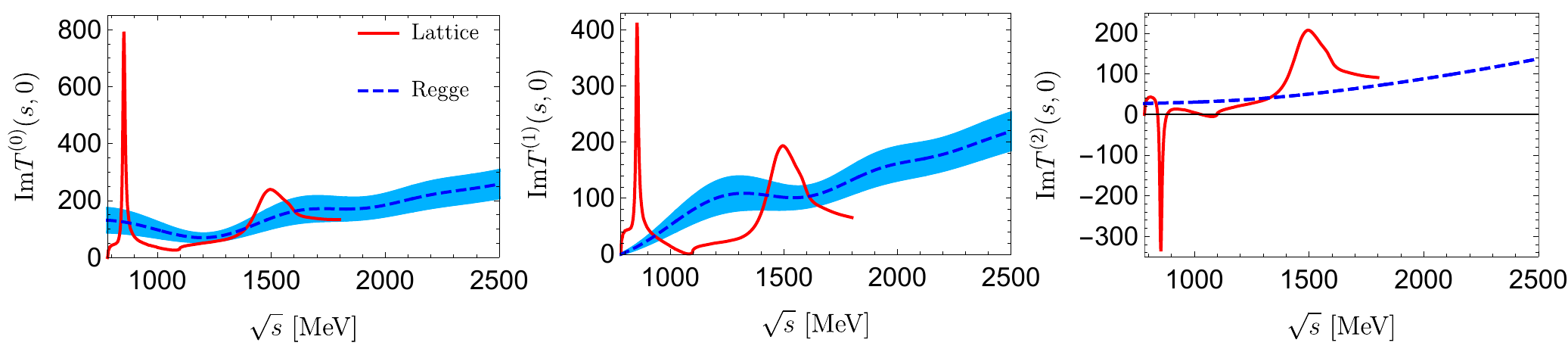} 
\caption{Comparison of $\mathrm{Im}T^{(I_t)}(s,0)$ constructed from lattice
data and the Regge asymptotic amplitudes.} \label{fig.regge}
\end{figure}

Unfortunately, the Pomeron residues $\beta_P^{\pi\pi}$ and $b_P^{\pi\pi}$ are unknown at unphysical large pion masses, and the available lattice data cannot give a direct determination of their values yet. In this work we will rely on the so-called additive-quark rule  of the Pomeron exchange (see e.g.~Sec.3 of \cite{Donnachie:2002en} for details) to estimate the Pomeron residues. The additive-quark rule of Pomeron exchange says that the total cross section of a process $\sigma_{ab}$ (or the imaginary part of the corresponding amplitude) is proportional to the numbers of light valence ( specifically $u, d$) quarks $n_a, n_b$ in the hadrons $a$ and $b$. 
In particular, the residue of Pomeron exchange $\beta_P^{ab}$ satisfies $\beta^{ab}_P(t)\propto n_a n_b$, e.g., $\beta_P^{\pi p}: \beta_P^{p p} \approx 2: 3$. 
It can be also generalized to include $s$ quark. The minor difference is that the coupling between Pomeron and $s$ quark is about $70\%$ of that with $u,d$ quarks.
The additive-quark rule has been verified by various experiments~\cite{Donnachie:2002en}, although its QCD origin has not been fully understood. 
Since the unphysical large pion mass ($\sim 391~$MeV) is not so different from the physical kaon mass ($\sim 496~$MeV), we will take a rough estimation $\beta_P^{\pi\pi}\sim 0.7 \beta_P^{\pi\pi \text{Phy}}=65.8$ and 
$b_P^{\pi\pi}=b_P^{\pi\pi \text{Phy}}$, where $\beta_P^{\pi\pi \text{Phy}}=94$ and $b_P^{\pi\pi \text{Phy}}=2.5~\mathrm{GeV}^{-2}$~\cite{Caprini:2011ky}. 
For illustration, we compare the imaginary part of $T^{(I_t)}(s,0)$ resulting from the lattice data and the Regge asymptotic amplitudes with $\eta=0.5\pm0.2$ in Fig.~\ref{fig.regge}. 

\begin{figure}[h]
   \centering
   \includegraphics[width=1\textwidth,angle=-0]{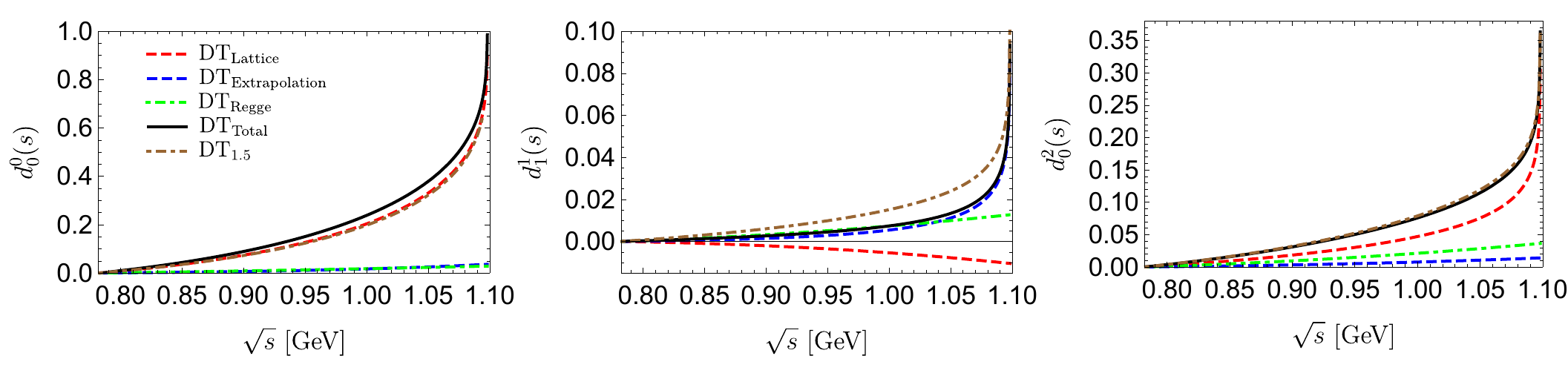} 
   \caption{Decomposition of the DTs in terms of lattice input, extrapolated data and Regge pole theory. The black solid line labeled as DT$_{\rm Total}$ corresponds to the sum of the aforementioned three parts where the Regge contributions are introduced in the energy region above 1.8~GeV. For the brown line labeled as DT$_{\rm 1.5}$, the corresponding Regge contributions are included in the energy region above 1.5~GeV. See the main text for details.} \label{fig.2}
\end{figure}

The DTs are obtained via dispersive integrals above $s_\text{m}$ to infinity. We verify that the contributions by the dispersive integrals above 1.8~GeV from the Regge model are very small in all the three cases with $IJ=00,11,20$ and the DTs are mainly given by S-, P and D-wave contributions below $1.8~$GeV as shown in Fig.~\ref{fig.2}. The sums from the various DTs are shown as black solid lines in Fig.~\ref{fig.2}. The contribution from the asymptotic high energy region and high partial waves estimated by Regge model are around one order of magnitude smaller, which makes our main analyses almost unaffected by the Regge contributions.
In Fig.~\ref{fig.2}, we also compare the contributions from lattice input, the extrapolated data, and Regge asymptotic effects. It is observed that the extrapolated data play a minor role in DTs. The only exception occurs for P-wave, where the main contribution comes from the extrapolated data. However, as depicted later in Fig.~\ref{fig.1}, the DT in the P-wave amplitude Re$t_1^1(s)$, is totally negligible, thus one almost needs not to worry about such an effect from extrapolation. In order to assess the influence of the separation point between the extrapolated data and the Regge contributions, we also try to set the separation point at 1.5~GeV during the calculation and the results are shown in Fig.~\ref{fig.2} together with the curves obtained at 1.8~GeV. The differences between two cases turn out to be very small.

\subsection{Additional constraints and uniqueness of solutions}

As a group of coupled integral equations, the number of independent solutions for Roy equations is dependent on the input phase shifts at the matching point $s_{\mathrm{m}}$, which can be extracted from the HadSpec simulations~\cite{Dudek:2012gj, Dudek:2012xn, Briceno:2017qmb}: $\delta_0^0\left(s_{\mathrm{m}}\right)=(15.5_{-3.5}^{+5.5})^\circ, \delta_1^1\left(s_{\mathrm{m}}\right)=170.1^\circ, \delta_0^2\left(s_{\mathrm{m}}\right)=-(16.3\pm 1.0)^\circ$.
According to the discussion in Refs.~\cite{Gasser:1999hz, Wanders:2000mn,Ananthanarayan:2000ht}, the multiplicity index in this situation is $m=0+1-1=0$, while $m=0$ in the physical case. In the latter case, the subtractions $k^I_J(s)$, i.e. the scattering lengths, in Eqs.~\eqref{eq.kij} are taken as external inputs and the pole terms are absent, which gives the multiplicity index $m=0$ and leads to the unique solution~\cite{Ananthanarayan:2000ht}\footnote{The reality, however, is more complicated. In Ref.~\cite{Ananthanarayan:2000ht}, it was observed that only P-wave amplitude shows a prominent peak around $\sqrt{s_\text{m}}=0.8~$GeV and the solutions of Roy equations in general develop a strong cusp in P-wave. Indeed, such a cusp can be removed by tuning the isotensor scattering length $a^2_0$, while the isoscalar scattering length $a^0_0$ is usually fixed at the value predicted by $\chi$PT. In practice, the cusps in two S-waves are very weak, so that effectively they play negligible roles in constraining parameters. In brief, once $a^0_0$ is fixed at a specific value, then the solutions of Roy equations would become a single parameter family depending on $a_2^0$. In practice, the no-cusp condition on P-wave can constrain $a^2_0$ to reach final central solution (one parameter + one constraint $\Longrightarrow$ the unique solution).}. 

However, in the large pion mass situation, the scattering lengths $a^0_0, a^2_0$, the position and the residue of the bound state pole $s_\sigma, g_{\sigma\pi\pi}$ usually bear comparatively large uncertainties, as discussed in Refs.~\cite{Pelaez:2010fj,Albaladejo:2012te,Briceno:2016mjc,Briceno:2017qmb,Doring:2016bdr,Danilkin:2020pak,Gao:2022dln}, and their precise values are still loosely determined. Since there are four parameters in our case, it is hard to obtain a multi-parameter universal band\footnote{In the physical case, it was proved that the for any reliable value of the scattering length $a_0^0$, the S- and P-wave cusps could be removed by tuning $a^2_0$, resulting in a one-parameter solution family, the so-called universal band~\cite{Ananthanarayan:2000ht}.}.
In practice, it is more reliable to set $s_\sigma, g_{\sigma\pi\pi}, a^0_0, a^2_0$ as free parameters when solving the extended Roy equations, which implies $m=0 \to 4$ and Roy equations will then have a four-parameter solution family. To pin down the unique one in the solution family, four additional independent constrains are required.

We utilize a numerical method based on the constraints of the phase shifts at the matching point $s_\mathrm{m}$~\cite{Hoferichter:2015hva}. 
It requires that the derivatives of the phase shifts at this point either are continuous (no-cusp condition) or have a certain divergence behavior (when an additional strongly coupling channel appears at $s_\mathrm{m}$).
This can provide three constraints on the solutions of the phase shifts in three different channels with $IJ=00,11,20$. 
In practice, the lattice phase shift for P-wave~\cite{Dudek:2012xn} is precise enough to pin down the mass of the $\rho$ resonance directly, because at $m_\pi=391$~MeV $\rho$ becomes a very narrow resonance with the width $\Gamma_\rho\sim 10~$MeV. 
Such condition almost gives a direct constraint on the location $s^\prime$ for $\delta^1_1(s^\prime)=\pi/2$. 
Thus, it is more appropriate to set the position $s^\prime$ where $\delta^1_1(s^\prime)=\pi/2$ as the fourth constraint rather than a theoretical output.
As a result, we are able to fix the two scattering lengths in the $IJ=00$ and $20$ channels, the location and the residue of the $\sigma$ pole by means of the aforementioned four constraints. 
In this way the four parameters $a_0^0, a_2^0, s_\sigma$ and $|g_{\sigma\pi\pi}|$ are not taken as external inputs but correspond to the predictions of this procedure. Our case is analogous to $\pi N$ Roy-Steiner equation study in Ref.~\cite{Hoferichter:2015hva}. We will follow the method in Ref.~\cite{Ananthanarayan:2000ht} to numerically solve Roy equations, and it turns out to be crucial to choose convenient parameterizations for the phase shifts in different channels in order to obtain precise Roy equation solutions. 

\section{Numerical procedures to solve Roy equations}\label{sec.4}

\subsection{Numerical determination of the solutions}

According to Refs.~\cite{Ananthanarayan:2000ht, Buettiker:2003pp, Hoferichter:2015hva}, we pursue the following strategy to solve Roy equations: the phase shifts of each channel in the region $(4m_\pi^2,s_\text{m})$ are conveniently parameterized with a few parameters, which are matched to the input PWs above $s_\text{m}$ in a reasonable way. 
Finally, the process of solving the equations is converted into optimizing these parameters to minimize certain objective functions. One of the crucial steps is to properly parameterize the phase shifts in different channels.

The phase shift $\delta^0_0(s)$ at the $K\bar{K}$ threshold has a strong cusp effect, indicating that the derivative of the phase shift is not continuous and diverges. For a generic Roy solution, the divergence depends on the value of the phase shift at the matching point in the following way~\cite{Gasser:1999hz}:
\begin{align}\label{eq.div}
    \left.\frac{\md}{\md s} \delta_0^0(s)\right|_{s \rightarrow s_\text{m}^{-}} \propto \left(s_\text{m}-s\right)^{\alpha-1}\ , \quad \alpha=\frac{2 \delta_0^0\left(s_\text{m}\right)}{\pi}-2\ .
\end{align} 
Particularly by combining the two-coupled-channel unitarity and the Roy equations, one has~\cite{Moussallam:2011zg}\begin{align}\label{eq.constraint}
    \left.\frac{\md}{\md s} \delta_0^0(s)\right|_{s \rightarrow s_{K}^-}=A\left(s_K-s\right)^{-\frac{1}{2}}\ ,\quad A=\frac{\rho_\pi\left(s_K\right)\left|g_0^0\left(s_K\right)\right|^2}{2 \cos(2 \delta_K) \sqrt{s_K}}\ ,
\end{align}
where $\rho_\pi=\sqrt{1-4m_\pi^2/s}$ and $g^0_0(s)$ is the PW $\pi\pi\to K\bar{K}$ amplitude with $IJ=00$.
In our case, the matching point $s_\text{m}$ coincides with the $K\bar{K}$ threshold $s_K=4m_K^2$. It is expected that the derivative of the phase shift will exhibit a square-root singularity. This divergence is weaker than the generic matching point
divergence \eqref{eq.div} provided the phase shift at threshold is not too
large, i.e. $\delta_0^0\left(s_K\right)<225^{\circ}$, which is indeed fulfilled in the present study~\cite{Briceno:2017qmb}.  Guided by these requirements, a modification of the Schenk parametrization~\cite{Schenk:1991xe} is used for $\delta^0_0(s)$: 
\begin{align}\label{eq.delta0_391}
    \tan \delta_{0}^0(s)=\rho_\pi(s)\left(a_{0}^0+B_{0}^0 q^2+C_{0}^0 q^4+D_{0}^0 q^6\right)\frac{4 m_\pi^2-s_{0}^0}{s-s_{0}^0} \times\frac{\sigma_K\left(s_\pi\right)+\beta}{\sigma_K(s)+\beta}\ ,
\end{align}
where $\sigma_K(s)=\sqrt{s_K / s-1}$ and $\beta=\frac{\sin(4\delta_K)}{4\rho_\pi(s_K)\left|g_0^0(s_K)\right|^2}$. 
In Ref.~\cite{Briceno:2017qmb}, it is analysed that $\left|g_0^0(s_K)\right|^2\sim 1.36$, leading to $\beta\sim 0.23$. 
In practice, we leave it as a constrained parameter, $0.13<\beta<0.33$, in order to get the approximate solution for $s$ close to $s_K$ but not necessarily reproducing the ``exact'' limiting behaviour for $s=s_K$. 
In the $IJ=00$ channel, Eq.~\eqref{eq.constraint} and the matching condition requiring $\delta_0^0(s_\text{m})=\left.\delta_{0}^0(s_\text{m}+0^+)\right|_\text{input}=15.5^\circ$ are two constraints in the optimization process. 
For the $IJ=11$ channel,  a conformal parameterization is adopted~\cite{Garcia-Martin:2011iqs},
\begin{align}\label{eq.delta11}
    \cot \delta_1^1(s)=\frac{\sqrt{s}}{2 q^3}\left(M_R^2-s\right)\left\{\frac{2 m_\pi^3}{M_R^2 \sqrt{s}}+B_0+B_1 w(s)+B_2 w^2(s)\right\}\ ,\quad
    w(s)=\frac{\sqrt{s}-\sqrt{s_0-s}}{\sqrt{s}+\sqrt{s_0-s}}\ .
\end{align}
The matching and no-cusp conditions require $\delta_1^1(s_\text{m})=\left.\delta_1^1(s_\text{m}+0^+)\right|_\text{input}=170.1^\circ$ and $\frac{\md\delta_1^1(s_\text{m})}{\md s}=\left.\frac{\md \delta_1^1(s_\text{m}+0^+)}{\md s}\right|_\text{input}=6.2^\circ~\mathrm{GeV}^{-2}$~\cite{Dudek:2012xn}.
Moreover, the additional constraint $\delta_1^1(s_\rho)=\pi$ corresponds to $\sqrt{s_\rho}=M_R=(854.1\pm1.1)~$MeV~\cite{Dudek:2012xn}. 
The parameterization in the $IJ=20$ channel is similar to Eq.~\eqref{eq.delta0_391},
\begin{align}\label{eq.delta20}
    \tan \delta_0^2(s)=\rho_\pi(s)\left(a_{0}^2+B_{0}^2 q^2+C_{0}^2 q^4+D_{0}^2 q^6\right)\frac{4 m_\pi^2-s_{0}^2}{s-s_{0}^2}\ ,
\end{align}
which is also accompanied by two constraints: $\delta_0^2(s_\text{m})=\left.\delta_0^2(s_\text{m}+0^+)\right|_\text{input}=-16.3^\circ$ and $\frac{\md\delta_0^2(s_\text{m})}{\md s}=\left.\frac{\md \delta_0^2(s_\text{m}+0^+)}{\md s}\right|_\text{input}=-12.2^\circ~\mathrm{GeV}^{-2}$~\cite{Dudek:2012gj}.

\begin{figure}[h]
\centering
\includegraphics[width=0.4\textwidth,angle=-0]{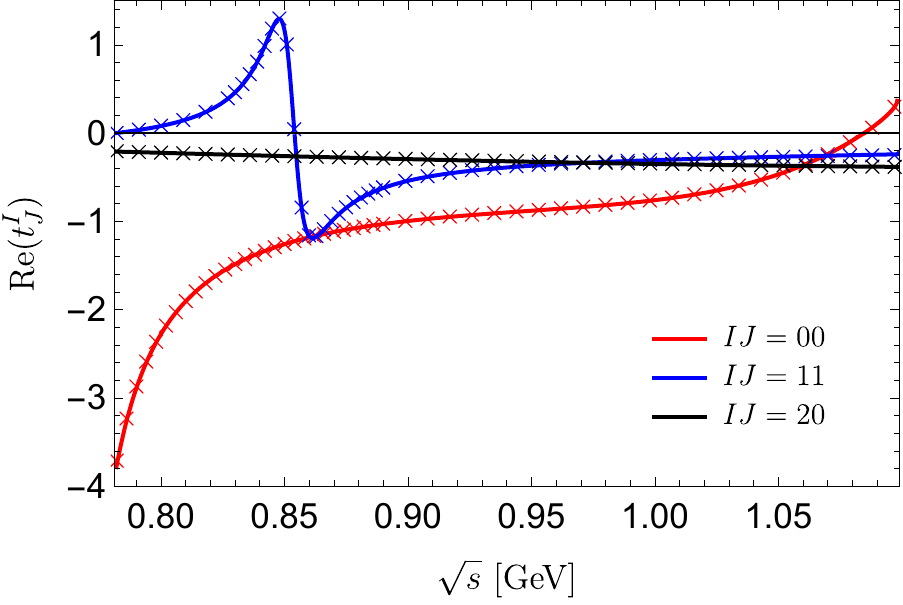} 
\caption{Left-hand sides of the Roy equations (lines) compared to the right-hand sides (points) after minimization for $m_\pi=391~$MeV.} \label{fig.Ret_391}
\end{figure}

\begin{figure}[h]
   \centering
   \includegraphics[width=1\textwidth,angle=-0]{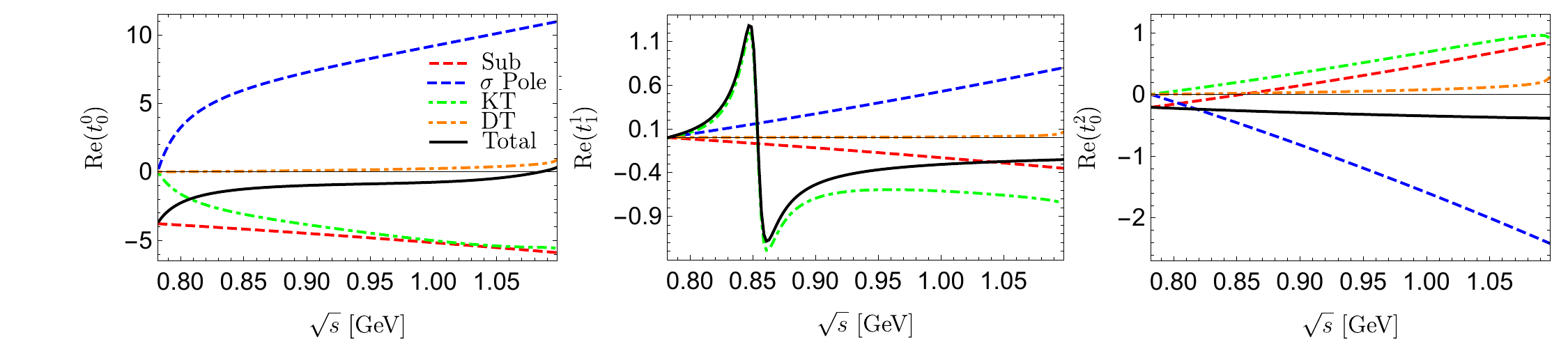} 
   \caption{Decomposition of the right-hand sides of Roy equations~\eqref{eq.tij} into the different contributions. Black solid lines correspond to the sum of all the contributions. Red dashed lines denote the subtraction contribution (``Sub"), whereas the blue dashed lines refer to the $\sigma$ pole terms (``$\sigma$ Pole"). The kernel terms (``KT") are given by the green dot-dashed lines, and finally the driving terms (``DT") are described by the orange dot-dashed lines.} \label{fig.1}
\end{figure}

As discussed above, we need to treat these parameters $\{B_0^0, C_0^0, D_0^0, s^0_0, \beta; B_0, B_1, B_2, s_0; B_0^2, C^2_0, D^2_0, s^2_0\}$ on the same footing as $a^0_0,a^2_0,s_\sigma$ and $g_{\sigma\pi\pi}$. Therefore we are dealing altogether with $5+4+4+4=17$ free variables and $2+3+2=7$ constraints when solving Roy equations. It is natural to re-express the parameters of the phase shifts as a function of the input phase and its derivative at $s_\text{m}$, so that we restrict ourselves to a set of solutions where these conditions are fulfilled automatically. More details about how to match the parameterizations and the lattice input at $s_\text{m}$ can be seen in Sec.5.1 of Ref.~\cite{Hoferichter:2015hva}.
All the parameters are determined from the optimization procedure by minimizing a $\chi^2$-like function,
\begin{align}
    \chi^2=\sum_{I,J} \sum_{i=1}^N \left\{\frac{\operatorname{Re} t_{J}^I\left(s_i\right)-F\left[t_{J}^I\right]\left(s_i\right)}{\xi_J^I}\right\}^2\ ,
\end{align}
where $\xi_J^I$ are the weight factors fixed to $\xi_0^0=\xi^1_1=5\xi^2_0=1$\footnote{Since the amplitude $t^2_0$ is smaller than other amplitudes, setting different weight factors to different channels can accelerate the convergence efficiency in the optimization process. Whatever weight factors are chosen, the final solutions are all the same.}, $\{s_i\}$ denotes a set of energy points between threshold and matching point, and 
$F\left[t_{J}^I\right]$ stands for the right-hand side of the extended Roy equations \eqref{eq.tij}.
We have checked the stability of the solution with respect to the choice of $\xi_J^I$, as well as the number of grid points, which is varied between 20 and 30, and in the end fixed to $N = 25$.
Finally, we obtain $\chi^2\sim 10^{-3}$, indicating that the optimization procedure is converging to a real solution.
The accuracy of the solutions is illustrated in Fig.~\ref{fig.Ret_391}.
Numerical values of the parameters describing the phase shifts of Eqs.~\eqref{eq.delta0_391}~\eqref{eq.delta11} and \eqref{eq.delta20} in the Roy solutions are given in Tab.~\ref{tab.391}.
\begin{table}[h]
    \centering
    \begin{tabular}{ccccccccc}
    \hline\hline
    $a^0_0$\ & \ $B^0_0$\ & \ $C^0_0$\ & \ $D^0_0$\ & \ $s^0_0$\ & \ $\beta$\ & \ $B_0$\ & \ $B_1$\  & \ $B_2$\  \\
    \hline
    $-3.78$ & $4.88 \times 10$ & $-2.04 \times 10^2$ & $2.49 \times 10^2$ & $3.94\times 10$ & $2.61\times 10^{-1}$ & $8.55\times 10^{-1}$ & $6.59\times 10^{-1}$ & $6.81\times 10^{-1}$ \\
    \hline\hline
    $s_0$ & $M_R$ (input) & $a^2_0$ & $B^2_0$ & $C^2_0$ & $D^2_0$ & $s^2_0$ & $s_\sigma$ & $g_{\sigma\pi\pi}$ \\
    \hline
    $1.57$ & $8.54\times 10^{-1}$ & $-2.10\times 10^{-1}$ & $-2.08$ & $5.99\times 10$ & $-2.55\times 10^2$ & $-6.96\times 10$ & $5.76\times 10^{-1}$ & $4.93\times 10^{-1}$ \\
    \hline\hline
    \end{tabular}
    \caption{Parameters for the solutions of the extended Roy equations. All parameters are given in appropriate powers of GeV.} 
    \label{tab.391}
\end{table}
In Fig~\ref{fig.1} we show the effects of different parts in the right-hand side of Roy equations~\eqref{eq.tij}. Notice that in all the channels the $\sigma$ pole terms dominate in the high energy region and are largely canceled by the subtraction and the kernel contributions. The DTs have a minor effect in all the three amplitudes Re$t_J^I(s)$ with $IJ=00,11,20$. 

\begin{figure}[h] 
   \centering
   \includegraphics[width=1\textwidth,angle=-0]{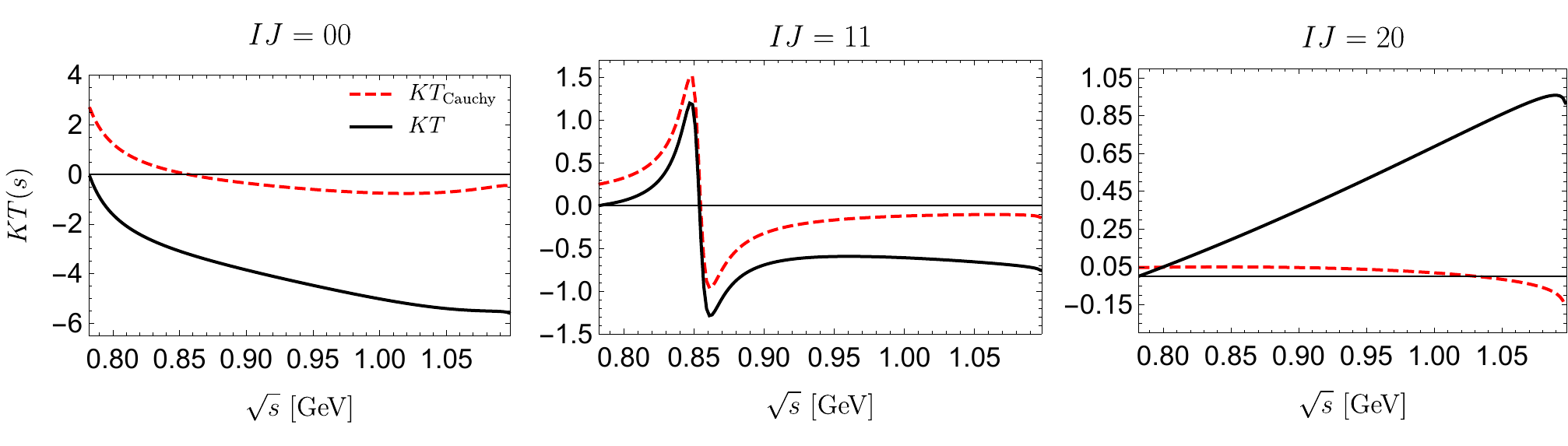} 
   \caption{Comparison between the size of the Cauchy-kernel contribution (``$KT_\text{Cauchy}$'') in \eqref{eq.tij} and the complete kernel contribution (``$KT$'') manifesting crossing symmetry.} \label{fig.3}
\end{figure}

It is an interesting point to compare the size of the Cauchy-kernel contributions in \eqref{eq.tij}. As shown in Fig.~\ref{fig.3}, only P-wave shows ``s-channel dominance'', i.e., the Cauchy-kernel contribution in the P-wave is the main contribution in the kernel terms, $KT(s)$, because the ultra narrow $\rho$ resonance largely dominates the feature of the P-wave. The  reason behind vector meson dominance is the fact that the LHC contributions are kinematically suppressed for the P-wave. In contrast, LHC contributions in S-waves are usually non-negligible. On the other side, by increasing the pion masses, the LHC contributions could still be  relevant and even become more important in special cases. Actually, according to Ref.~\cite{Regge:1958ft}, the LHC effects depend on only the interaction range in potential scattering theory. In the present study, the emergence of the near-threshold bound state $\sigma$ at large pion mass is found to give rather important LHC.  
Most unitarized $\chi$PT amplitudes methods, for instance, the inverse-amplitude method (IAM)~\cite{Truong:1988zp, Dobado:1989qm, Dobado:1996ps}, which is very similar to the P\' ade-approximation method, obtains the resonances via the sum of the s-channel bubble loops and neglects the resonance effects in the crossed channels. To our knowledge, the LHC caused by the bound state $\sigma$ at $m_\pi\sim 391$~MeV has not been addressed by previous studies. Besides, the degree to which the IAM (and other unitarized $\chi$PT methods) can correctly handle (exact) crossing symmetry is a subject under debate~\cite{Boglione:1996uz, Nieves:2001de, Cavalcante:2001yw, Qin:2002hk, Salas-Bernardez:2020hua}. 
At the large $N_C$ limit, it is clearly demonstrated in Refs.~\cite{Guo:2007ff, Guo:2007hm}, that the improper way to include resonances in the crossed channels cannot be correctly matched to  $\chi$PT in the low energy region. According to Ref.~\cite{Qin:2002hk}, unitarized chiral amplitudes usually underestimate the LHC contributions, whose effects are simulated by spurious pole contributions.

\subsection{Error estimations}

The procedure of evaluating theoretical uncertainties consists in performing random variations of the various inputs, which include the D-wave contributions, the floating inputs at the matching point $\delta^I_J(s_\text{m})$, the S- and P-wave lattice phase shifts above the $K\bar{K}$ threshold and the asymptotic Regge contributions.
As one of the key inputs, the lattice result above the inelastic $K\bar{K}$ region still has large uncertainty~\cite{Briceno:2017qmb}, and this prevents us from predicting the lattice phase shifts between $s=4m_\pi^2$ and $s=4m_K^2$ within the Roy equation method as precise as the physical situations. Nevertheless, we can still give predictions to the phase shifts in Fig.~\ref{fig.phase} after solving the Roy equations that respect crossing symmetry. 
For the complete error estimations of the Roy-type equations in the physical case, see the discussions in Refs.~\cite{Ananthanarayan:2000ht, Buettiker:2003pp, Hoferichter:2015hva}.

In the present study, we analyze the uncertainties contributed by the variations of matching phase shifts $\delta_0^0(s_\text{m}), \delta_0^2(s_\text{m})$, the lattice DTs below $1.8~$GeV and the asymptotic Regge amplitudes. Besides, the rough estimation of the ``cusp'' parameter $\beta=0.23\pm 0.1$ in Eq.~\eqref{eq.delta0_391} can also give rise to some non-negligible uncertainties.
According to  Ref.~\cite{Briceno:2017qmb}, we roughly set $\delta^0_0(s_\text{m})=(15.5^{+5.5 }_{-3.5})^\circ$.
For $\delta^2_0(s_\text{m})$~\cite{Dudek:2012gj}, we perform ``global" fits based on a K-matrix parameterization in the energy region $782<\sqrt{s}<1550~$MeV and ``local" fits in which one considers separately a small energy region surrounding the matching point $1000<\sqrt{s}<1200$~MeV (see Ref.~\cite{Buettiker:2003pp} for more details).
In the small energy region, an approximation to $\delta^2_0(s)$ as a function of quadratic polynomial of energy $\sqrt{s}$ is enough. 
We consider the differences of $\delta^2_0(s_\text{m})$ obtained from the two different fits as an additional source of uncertainty in our study.
In summary, the difference of the phase shifts between these two fits at the matching point is about $1^\circ$, thus we set $\delta^2_0(s_\text{m})=-(16.3\pm1.0)^\circ$.
It is explicitly verified that variations of the inputs in the energy region $\sqrt{s}>1.8$~GeV have negligibly small influences. So we will mainly analyze the inputs in the energy region $1.1<\sqrt{s}<1.8~$GeV, especially for the result from the $IJ=00$ channel, which turns out to dominate the uncertainties among the DTs above the $K\bar{K}$ threshold. 
We utilize various extrapolations of $\delta^0_0(s)$ in the energy region $1.44<\sqrt{s}<1.8~$GeV to test the robustness of the solutions. Based on these variations of inputs, the uncertainties of the phase shifts in the $IJ=00,11,20$ channels and the pole positions are obtained using the bootstrap approach.

\section{Phenomenological discussions at large pion masses}\label{sec.5}

\subsection{Results for phase shifts and the S-wave scattering lengths}

\begin{figure}[h]
\centering
\includegraphics[width=1\textwidth,angle=0]{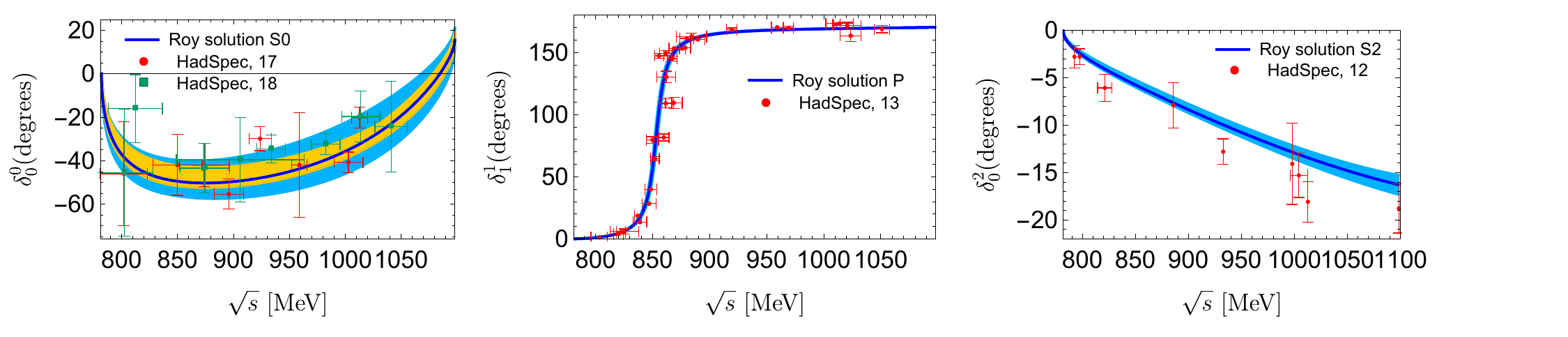}
\caption{$\pi\pi$ phase shifts at $m_\pi=391$~MeV from Roy equation solutions: S0, P and S2 stand for the results of the $IJ=00,11,20$ channels, respectively. For the sources of the shaded error bands, see the main text for details. The lattice data are taken from Refs.~\cite{Dudek:2012gj, Dudek:2012xn, Briceno:2016mjc, Briceno:2017qmb}.} \label{fig.phase}
\end{figure}

Relying on the aforementioned solutions of extended Roy equations, we are ready to reveal the corresponding phenomenological consequences at large pion masses. The $\pi\pi$ phase shifts that respect crossing symmetry at $m_\pi=391$~MeV are provided in Fig.~\ref{fig.phase}, where blue shaded uncertainty areas are obtained by including all the error sources from DTs $d^I_J(s)$, such as the D-wave contributions, the floating inputs at the matching point $\delta^I_J(s_\mathrm{m})$, S- and P-wave lattice inputs in the energy region above the $K\bar{K}$ threshold and the asymptotic Regge amplitudes. 
The resulting uncertainty for $IJ=00$ is obviously larger than those in $IJ=20$ and $11$ channels. 
It is verified that in our study the uncertainties of phase shifts with $IJ=00$ are dominated by input phase shift $\delta^0_0(s)$ at the matching point $\sqrt{s_{\mathrm{m}}}=1098$~MeV and the parameter $\beta$ in Eq.~\eqref{eq.delta0_391}. This can be clearly seen by artificially assigning smaller errors to these two constraints, e.g., $\delta^0_0(s_\text{m})=(15.5^{+3.5}_{-1.5})^\circ$ and $\beta=0.23\pm 0.05$, the error band of phase shift $\delta^0_0(s)$ will then considerably shrink to the yellow region as shown in Fig.~\ref{fig.phase}. Our results of phase shifts clearly give a useful constraint for future lattice QCD simulations and phenomenological studies. 

The corresponding parameters that give the solutions in Fig.~\ref{fig.phase} are  
\begin{align}\label{eq.sol}
    a_0^0 &= -(3.8^{+1.1}_{-1.2})\ , \quad a_0^2=-(0.21^{+0.02}_{-0.03}) \ , \nonumber\\
    \sqrt{s_\sigma} & = 759^{+7}_{-16}~\mathrm{MeV}\ , \quad 
    \left|g_{\sigma\pi\pi} \right|= 493^{+27}_{-46}~\mathrm{MeV}\ .
\end{align}
Our determination for the $\sigma$ mass agrees with the $N/D$ determination of $758(5)$~MeV from Ref.~\cite{Danilkin:2020pak}, and are also roughly compatible with other results in Refs.~\cite{Pelaez:2010fj,Briceno:2017qmb,Doring:2016bdr,Gao:2022dln} after taking into account the uncertainties. 
We find that the value of the scalar-isoscalar scattering length $a^0_0$ has a significant correlation with the $\sigma$ mass in numerical optimization, probably because the $\sigma$ is too close to the threshold. 
It directly leads to the presence of a ``platform'' near the numerical solution~\eqref{eq.sol}, which signals the existence of flat directions in the four-dimensional-parameter space to which the Roy equation constraints are only weakly sensitive~\footnote{Such behavior in the parameter space has been thoroughly investigated in the Roy-Steiner equation analyses of $\pi N$ scattering~\cite{Hoferichter:2015hva}.}. 
The presence of this ``platform'' gives a possible explanation about the spread values for $a^0_0$ and $s_\sigma$ from different approaches~\cite{Danilkin:2020pak,Pelaez:2010fj,Briceno:2017qmb,Doring:2016bdr,Gao:2022dln}. 

\subsection{Pole contents in the complex plane}

\begin{figure}[h]
\centering
\includegraphics[width=0.5\textwidth,angle=-0]{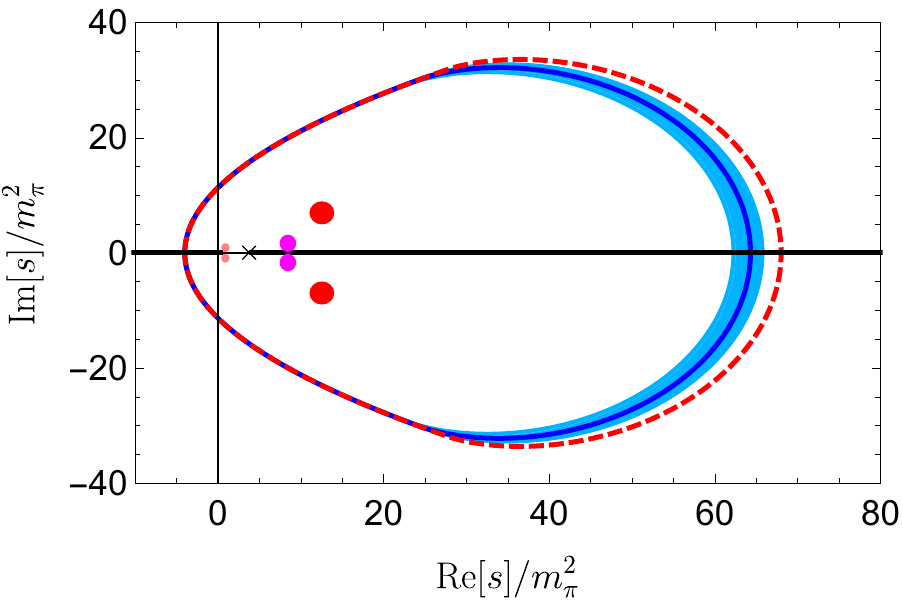} 
\caption{Validity domain of extended Roy equation for $m_\pi=391$~MeV. 
The dashed red boundary represents the validity domain by dropping the effects of the bound state $\sigma$, and the blue boundary corresponds to the complete validity domain within uncertainty from the location of the $\sigma$.
The poles in the validity domain in the second RS are from left to right, as shown in Eq.~\eqref{eq.f0pole}.} \label{fig.roydomain}
\end{figure}

\begin{figure}[h]
\centering
\includegraphics[width=0.5\textwidth,angle=-0]{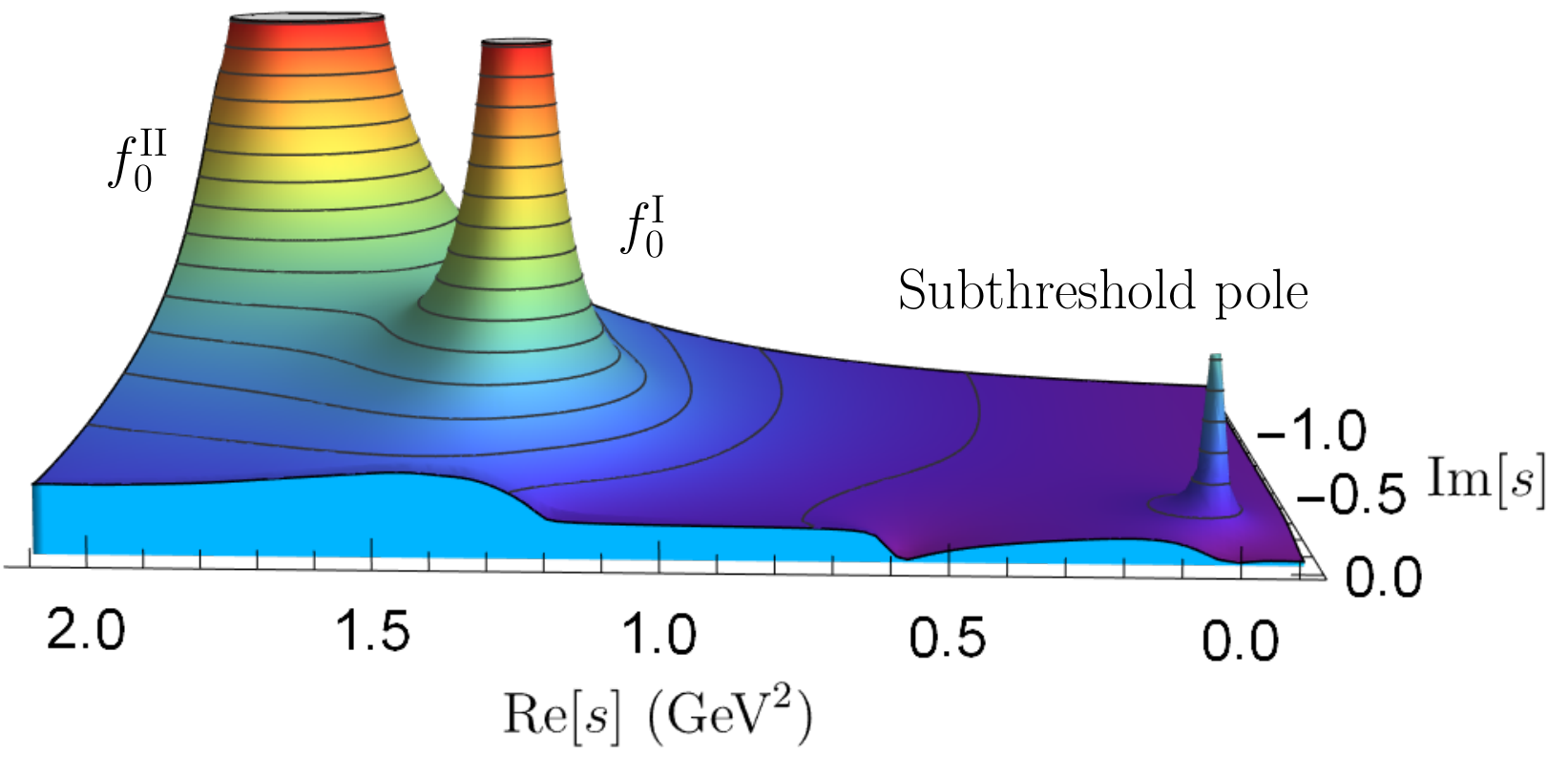} 
\caption{$\left|S^{0~\text{II}}_0(s)\right|$ obtained from the extended Roy equation, analytically continued to the lower-half
complex $s$ plane.} \label{fig.contour}
\end{figure}

Next we perform the analytic continuation into the complex $s$ plane to look for poles in the second Riemann sheet (RS). 
In the PW amplitude with $IJ=00$, apart from the bound state pole for $\sigma$ in the physical RS, we further find several other poles in the second RS, whose positions are 
\begin{align}\label{eq.f0pole}
    \sqrt{s_{\rm sub}} = (269^{+40}_{-25})-i(211^{+26}_{-23})~{\rm MeV},  \quad 
    \sqrt{s_{f_{0}^{\mathrm{I}}}} = (1142^{+53}_{-46})-i(112_{-45}^{+59})~{\rm MeV}, \quad \sqrt{s_{f_0^{\mathrm{II}}}}=(1434^{+167}_{-223})-i(371^{+97}_{-49}) ~{\rm MeV}\ .  
\end{align}
The coupled-channel analysis by explicitly including $\pi\pi, K\bar{K}$ and $\eta\eta$ in Ref.~\cite{Briceno:2017qmb} reveals a pole in the second RS at $(1166\pm45) -\frac{i}{2} (181\pm 68)$~MeV (advocated as the $f_0(980)$ resonance in the previous reference), which is consistent with the $f_0^{\mathrm{I}}$ pole in Eq.~\eqref{eq.f0pole}. While, the broad pole $f^{\text{II}}_0$ in our determination~\eqref{eq.f0pole} could correspond to a second RS shadow pole of the long-debated $f_0(1370)$ resonance~\cite{ParticleDataGroup:2022pth, Pelaez:2022qby}. Notice that the position of the $f_0^{\rm II}$ pole is already above the $\eta\eta$ threshold, therefore it is possible that this pole position could be visibly shifted when including the inelastic $K\bar{K}$ and $\eta\eta$ amplitudes. Although to explicitly include the latter heavier states as dynamical channels is clearly beyond the scope of this study that exploits the Roy equation method in the elastic case, we try to estimate the high energy influence on the heavy $f_0^{\rm I}$ and $f_0^{\rm II}$ poles by varying the DTs. Notice that both $f_0^\text{I}$ and $f_0^\text{II}$ are wider than in the physical case, one possible explanation may be that, unlike the $\rho$ meson (which may be  understood as an $SU(2)$ isospin gauge boson~\cite{Bando:1984ej} -- hence its relation to $m_\pi$ could be  simple and trivial), $f_0^{\text{I}}$ may be more appropriately described as a $K \bar K$ molecule~\cite{Locher:1997gr, Baru:2003qq, Su:2007au}. Hence its mass and decay phase space also depend on $m_\pi$, and there is no simple expectation on the $m_\pi$ dependence of its width. It is also verified that all the poles in Eq.~\eqref{eq.f0pole} fall in the validity domain of the Roy equation, see Fig.~\ref{fig.roydomain}~\footnote{The validity domain relies both on the Lehmann-Martin ellipse and the double spectral function of $\pi\pi$ scatterings. When there is a bound state $s_\sigma$ ($<4m_\pi^2$), the right extremity $r(s^\prime)$ of $\pi\pi$ Lehmann-Martin ellipse (i.e. the double spectral function) changes from $\min\{16s^\prime m_\pi^2/(s^\prime-4m_\pi^2), 4s^\prime m_\pi^2/(s^\prime-16m_\pi^2)\}$~\cite{Caprini:2005zr} to $\min\{4s_\sigma\left(1-s_\sigma/(s^\prime-4m_\pi^2)\right), 4\left(m_\pi^2-s_\sigma^2/(s^\prime-4 s_\sigma)\right)\}$.}.
An intermediate task is to discern how the different poles can affect the amplitudes on the real axis. For this purpose, we give in Fig.~\ref{fig.contour} the contour plot for the S-matrix with $IJ=00$ in the second RS, i.e. $S^{0~\text{II}}_0(s)=1/S^0_0(s)=1/(1-2\sqrt{4m_\pi^2/s-1}~t^0_0(s))$. The prominent pole structures corresponding to $f_0^{\rm I}$ and $f_0^{\rm II}$ can be clearly seen in this figure. 
The remaining question is how to understand the subthreshold complex pole close to $s=0$, i.e. the broad pole $\sqrt{s_{\rm sub}} = (269^{+40}_{-25})-i(211^{+26}_{-23})~{\rm MeV}$ in Eq.~\eqref{eq.f0pole}. It should be reiterated that this subthreshold complex pole is inside the validity domain of the extended Roy equations, as shown in Fig.~\ref{fig.roydomain}.  

Recently a near threshold virtual state pole, apart from the bound state pole of $\sigma$, was introduced in Ref.~\cite{Gao:2022dln} within the PKU parameterization of S-matrix formalism\footnote{Actually, the general discussions about the existence of a virtual state (resulting from the two conjugate $\sigma$ poles in the physical case) were previously given in Ref.~\cite{Hanhart:2008mx} based on $\chi$PT and IAM.}, in order to simultaneously describe the recent lattice phase shifts and fulfill crossing symmetries imposed by the BNR relations at $m_\pi=391$~MeV. This virtual state pole is later challenged by the authors of Ref.~\cite{vanBeveren:2022zfx}, who claim that the virtual state pole does not exist when including the dynamics in the energy region above the inelastic $K\bar{K}$ or even $\eta\eta$ channels. Our study provides a more complete picture about the pole contents for $\pi\pi$ scatterings at $m_\pi=391$~MeV. Two broad resonance poles above $K\bar{K}$ threshold, namely $f_0^{\rm I}$ and $f_0^{\rm II}$, are found in our amplitudes. Below the $\pi\pi$ threshold, compared with the virtual state pole on the real axis as introduced in Ref.~\cite{Gao:2022dln}, our study reveals a pair of broad complex poles in complex plane in the $IJ=00$ amplitude. One reason behind this discrepancy could be that in Ref.~\cite{Gao:2022dln} the LHC contributed by the bound state $\sigma$ pole is omitted, which can play important roles in the fulfillment of the BNR relations, since the integral region of the BNR relations covers part of the $\sigma$-induced LHC.  
\begin{figure}[h]
\centering
\includegraphics[width=0.5\textwidth,angle=-0]{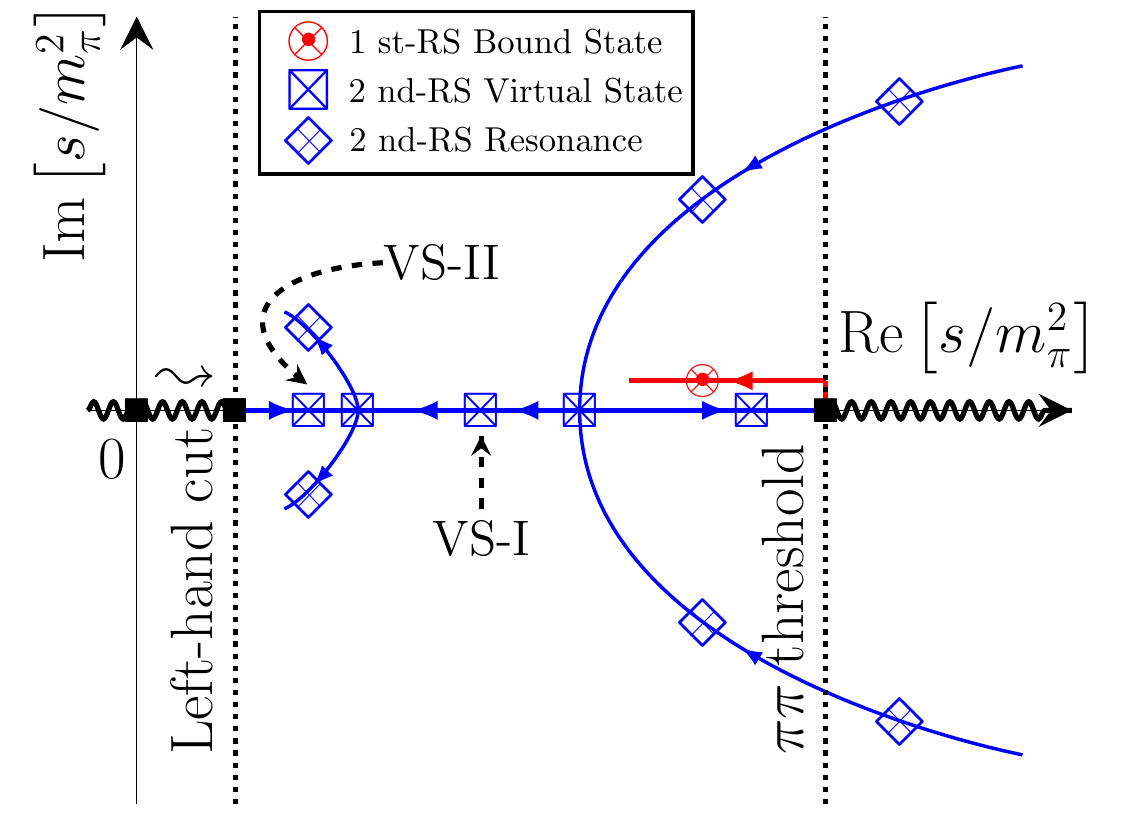} 
\caption{The qualitative trajectory of the $\sigma$ pole on the second RS of the $s$ plane with varying $m_\pi$. See the main text for the meaning of the labels `VS-I,II'.} \label{fig.trajectory}
\end{figure}

We are not able to trace the continuous $\sigma$ pole trajectory with different values of $m_\pi$, because there are not enough lattice inputs. Therefore we focus on two special cases: $m_\pi=391~$MeV and $m_\pi=236~$MeV (more details see below), the former corresponds to a bound state $\sigma$ and the latter corresponds to a broad resonance. 
The pole contents in Eq.~\eqref{eq.f0pole} could already imply a more involved pion-mass trajectory for the $\sigma$ pole as demonstrated in Fig.~\ref{fig.trajectory}, rather than the simple ones illustrated in Refs.~\cite{Hanhart:2008mx,Gao:2022dln}.  

When gradually increasing the pion masses from its physical value, the pair of broad physical $\sigma$ resonance poles will move toward the real axis from the complex plane and meet on the real axis below the threshold $s_\text{th}=4m_\pi^2$ becoming a pair of virtual state poles at a specific value of $m_\pi$ (see, e.g. Refs.~\cite{Hanhart:2008mx,Pelaez:2010fj, Albaladejo:2012te}).
By further increasing the pion masses, one of the virtual state pole (denoted as VS-I) will move left along the real axis, and the other one moves right across the threshold to the first RS and becomes a bound state pole. At the same time, the bound state pole will cause a new LHC singularity via crossing as shown in Eqs.~\eqref{eq.kij}, and the corresponding branch point at $s_{\text{th}}-s_\sigma$ extends to the real axis above $s=0$. 
From Eqs.~\eqref{eq.tij} and ~\eqref{eq.kij}, it can be proved that the S-matrix $S^0_0(s)$ will change from positive infinity to negative infinity\footnote{Note that in Eqs.~\eqref{eq.kij}, $\lim\limits_{s\to 4m_\pi^2-s_\sigma+0^+}k^0_0(s)\to -\infty$.} when approaching the LHC caused by the bound state $\sigma$. 
The sharp change of $S^0_0(s)$ from $+\infty$ to $-\infty$ in the vicinity of $s_\text{th}-s_\sigma$ implies that it must cross the real axis once in the range $s_\text{th}-s_\sigma <s < s_\text{th}$. The interception point corresponds to a zero for $S_0^0(s)$ in the first RS, and it also denotes a virtual state pole for the S-matrix in the second RS\footnote{Analyticity and unitarity tell us that $S^\text{II}(s)=1/S(s)$, namely the first sheet zero exactly corresponds to a second sheet pole.}. In another words, it indicates that an additional virtual state pole (denoted as VS-II) is generated from the $\sigma$-induced LHC, which completely comes from the analysis of crossing symmetry. 
Finally, it is natural to conjecture that the two virtual state poles, i.e. VS-I (evolved from the physical $\sigma$ resonance) and VS-II (generated from the new LHC), will collide at a specific value of $m_\pi$, evolve into complex poles by further increasing $m_\pi$ and they finally give rise to the pair of subthreshold complex poles we find here at $m_\pi=391~$MeV. Therefore, we consider that the pair of subthreshold complex poles corresponds to ``companion pole'' of the bound state pole ``$\sigma$'', since one of its origins comes from the conjectured virtual pole caused by the LHCs of the bound state $\sigma$. It is worth emphasizing again that the trajectory in Fig.~\ref{fig.trajectory} should be considered as a semi-conjecture -- only the poles at two specific values of $m_\pi$, i.e., $m_\pi=391~$MeV and $m_\pi=236~$MeV are derived from Roy equation analyses  of lattice results. It is the existence of the broad pole on the complex plane below the $\pi\pi$ threshold that drives us to conclude that one additional virtue pole VS-II should be generated from the LHC caused by the bound state $\sigma$ at $m_\pi=391$~MeV. Clearly, our study provides a new insight into the pole trajectories of $\sigma$ as a function of the pion mass.

\begin{figure}[h]
\centering
\includegraphics[width=0.45\textwidth,angle=-0]{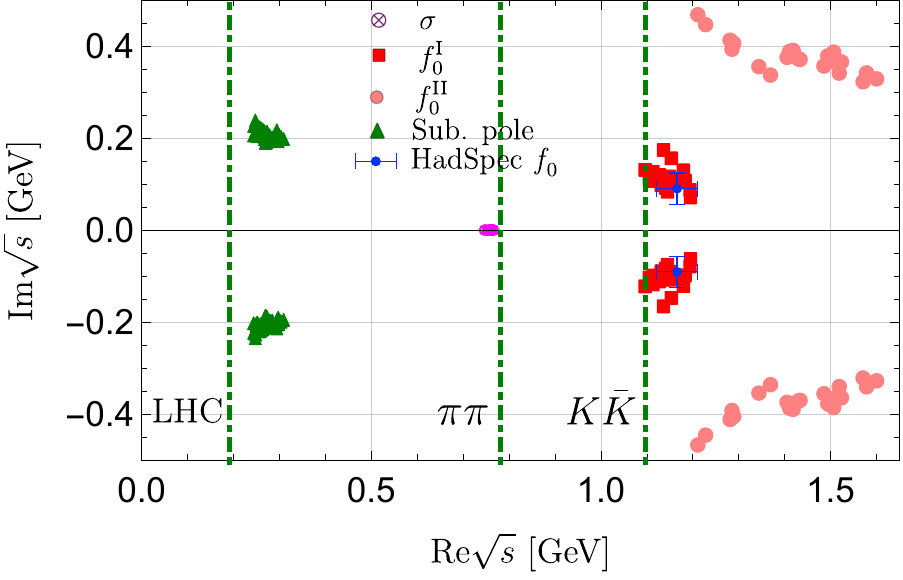} 
\caption{Poles of the amplitudes with $IJ=00$ from the extended Roy equation studies. The green dot-dashed lines denote the positions of the left-hand cut (LHC) contributed by the $\sigma$ and the thresholds of $\pi\pi$ and $K\bar{K}$.} \label{fig.00pole}
\end{figure}

In Fig.~\ref{fig.00pole}, we plot the pole locations in the complex $\sqrt{s}$ plane for the $IJ=00$ amplitudes at $m_\pi=391~$MeV. For comparison, the result of $f_0$ poles reported by the HadSpec collaboration~\cite{Briceno:2017qmb} are also shown.  
Our determination of the $f_0^\text{I}$ is consistent with its values within uncertainties.
Interestingly, we find that by dropping the contributions from the Regge amplitudes and the inputs in the $1.44<\sqrt{s}<1.8~$GeV from the $IJ=00$ channel, all the poles are barely affected except the $f^\text{II}_0$ one. This explicitly demonstrates that the Roy equation solutions in the low energy region are insensitive to the inputs in high energy region. Meanwhile, it also indicates that the heavy $f^\text{II}_0$ pole could be noticeably affected by the dynamics in the high energy region, and the coupled-channel effects can be important to precisely pin down the properties of this resonance. Therefore, the result of the $f^\text{II}_0$ pole in \eqref{eq.f0pole} from Roy equation analysis should be considered as qualitative only.

The resonance pole position in the $IJ=11$ amplitude reads $\sqrt{s_\rho} =(853.3^{+1.1}_{-1.1})-i(6.7_{-0.7}^{+0.2})~$MeV, which is in agreement with the result of Ref.~\cite{Dudek:2012xn}. 
For the non-resonant channel with $IJ=20$, we also find a virtual state pole in the second RS at $\sqrt{s_{v,IJ=20}} =435_{-12}^{+4}$~MeV, which value is somewhat larger than that in Ref.~\cite{Gao:2022dln} but compatible with the prediction of next-to-next-to-leading order (NNLO) $\chi$PT within the uncertainties that will be addressed later in the next subsection. Actually such a virtual state pole also exists in the case of physical mass, it is a prediction by combining the current algebra result, relativistic kinematics and the S-matrix theory~\cite{Ang:2001bd, Dai:2019zao}. We provide an illustrative explanation about the simple fact that there should be a virtual state pole in the $IJ=20$ amplitude in App.~\ref{Sec.appij20}. 
We also give the coupling $|g_{\pi \pi}|$ of all poles mentioned above extracted from the residues of the amplitudes $t^I_J(s)$ at the pole $g_{\pi \pi}^2=\lim _{s \rightarrow s_0}\left(s_0-s\right) t^I_J(s)$ in Tab.~\ref{tab.1}.
\begin{table}[h]
   \centering
   \begin{tabular}{ccc}
      \hline \hline Poles & Roy equation & K-matrix \\
      \hline $\sigma$ & $493^{+27}_{-46}$ & $521\pm 23$ \\
      $f_0^\text{I}$ & $783^{+171}_{-129}$ & $710\pm 140$ \\
      $f_0^\text{II}$ & $1189^{+322}_{-260}$ & $-$\\
      Sub. pole (I=0) & $112^{+23}_{-30}$ & $-$\\
      $\rho$ & $162^{+2}_{-2}$ & $162\pm 4$ \\
      VS. pole ($I=2$) & $165^{+3}_{-7}$ & $-$ \\
      \hline \hline
   \end{tabular}
   \caption{Comparison of the residues $|g_{\pi\pi}|$ (all in MeV) of various poles from Roy equation analyses and the K-matrix approaches in lattice studies at $m_\pi=391~$MeV~\cite{Wilson:2015dqa,Briceno:2017qmb}. }\label{tab.1}
\end{table}

\subsection{Remarks about the results at $m_\pi=236$~MeV from Roy equation analyses}

In addition, the HadSpec collaboration has also performed the simulation at $m_\pi=236$~MeV~\cite{Wilson:2015dqa,Briceno:2016mjc}. 
The key difference between the two sets of simulations at $m_\pi=236$~MeV and $391$~MeV is that the phase shifts with $IJ=00$ at $m_\pi=236$~MeV reconcile with the broad resonance description for $\sigma$, in contrast with the bound state behavior at $m_\pi=391$~MeV. 
In principle, it would be straightforward to take a Roy equation analysis for the lattice data at $m_\pi=236$~MeV. 
However, in practice, due to the lack of the lattice inputs of $IJ=20$ phase shifts and the DTs (especially the amplitudes above the $K\bar{K}$ threshold in the $IJ=00$ case), our predictions at $m_\pi=236$~MeV are considered to be less substantial comparing with the Roy equation analyses at $m_\pi=391$~MeV.

\begin{figure}[h]
\centering
\includegraphics[width=0.9\textwidth,angle=0]{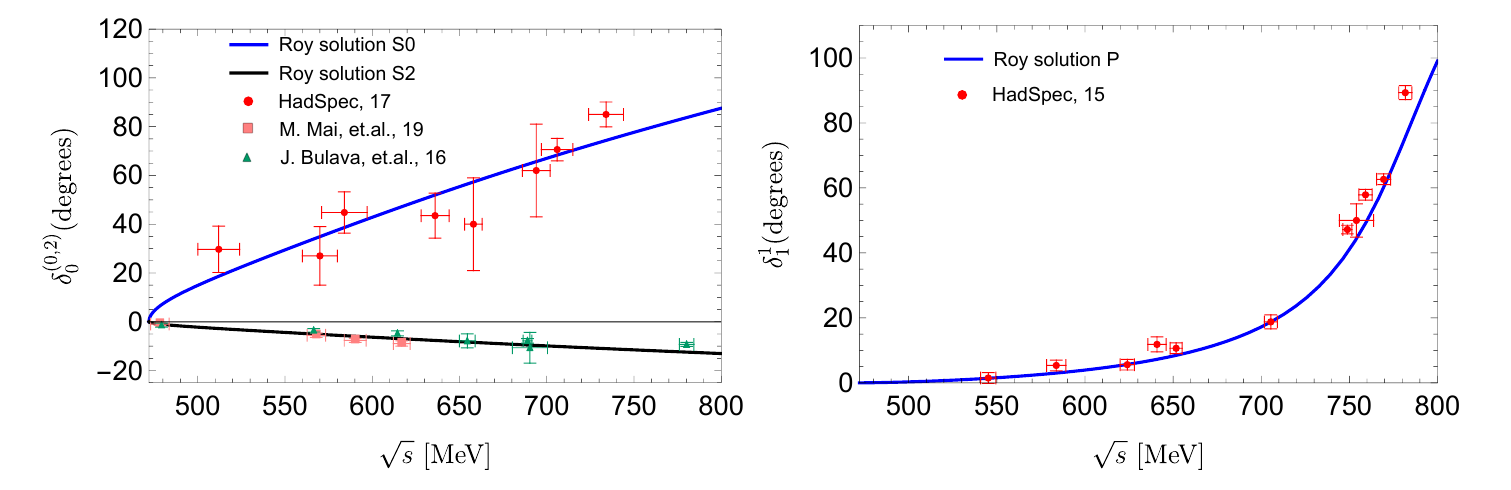}
\caption{Roy equation solutions S0, S2, P for the phase shifts of the $IJ=00,20,11$ channels,  respectively. The data come from Refs.~\cite{Wilson:2015dqa,Briceno:2016mjc,Bulava:2016mks, Mai:2019pqr}.} \label{fig.phase_236}
\end{figure}

The phase shifts with $IJ=00,11,20$ predicted by the Roy equations at $m_\pi=236$~MeV
are shown in Fig.~\ref{fig.phase_236}. 
As discussed previously, since the crucial inputs to solve the Roy equations in the case of $m_\pi=236$~MeV are still not available, we consider the calculation in this case a preliminary attempt. Therefore in this work we only give the central solutions of Roy equations at $m_\pi=236$~MeV, without providing the error analyses. The numerical procedures and the relevant ingredients to solve Roy equations for the case of $m_\pi=236$~MeV are more or less similar to the discussions in the previous section for $m_\pi=391$~MeV, although there are some subtleties regarding the inputs above the matching point. The details about the numerical discussions to solve Roy equations at $m_\pi=236$~MeV are relegated to the App.~\ref{Sec.app236}. We focus on the phenomenological outputs from the Roy equation solutions here. 

The $\pi\pi$ phase shifts below the matching point $\sqrt{s_\text{m}}=800$~MeV at $m_\pi=236$~MeV based on the Roy equation solutions that respect crossing symmetry and unitarity are shown in Fig.~\ref{fig.phase_236}. The lattice determinations of the phase shifts of all the three channels seem compatible with our Roy equation results. We also extrapolate the amplitudes from Roy equation analyses into complex $s$ plane to search the various poles at $m_\pi=236$~MeV. The pole positions and their residues read: $\sqrt{s_{\sigma}} =543-i 250~\mathrm{MeV}, |g_{\pi\pi}|=624~\mathrm{MeV}; \sqrt{s_\rho}=785-i43~\mathrm{MeV}, |g_{\pi\pi}|=289~\mathrm{MeV}; \sqrt{s_{v,IJ=20}} =117~\mathrm{MeV}, |g_{\pi\pi}|=49~\mathrm{MeV}$. 
The present determination of the $\rho$ position is consistent with Refs.~\cite{Wilson:2015dqa,Gao:2022dln}. Notice that the $\sigma$ pole positions at $m_\pi=236$~MeV from various approaches still span a broad range~\cite{Albaladejo:2012te,Briceno:2016mjc, Doring:2016bdr, Danilkin:2020pak, Gao:2022dln, Danilkin:2022cnj}. Again a virtual state pole is also found in the amplitude with $IJ=20$ at $m_\pi=236$~MeV. 

Adler zeros in $\pi\pi$ scattering amplitudes are characteristic predictions of chiral symmetry and can be considered as important constraints to various model setups. At NNLO of the two-flavor $\chi$PT,  the analytic PW amplitudes of the $\pi\pi$ scatterings are available in Ref.~\cite{Niehus:2020gmf}. By taking the values of the low energy constants $F=85.96(42)~\mathrm{MeV}, l_1^r=-4.03(63) \times 10^{-3}, l_2^r=1.87(21) \times 10^{-3}$, $l_3^r=0.8(3.8) \times 10^{-3}$ and $l_4^r=6.2(1.3) \times 10^{-3}, r^r_1=-0.6 \times 10^{-4}, r^r_2=1.3 \times 10^{-4}, r^r_3=-1.7 \times 10^{-4}, r^r_4=-1.0 \times 10^{-4}, r^r_5=1.1 \times 10^{-4}, r^r_6=0.3 \times 10^{-4}, r^r_F=0.0 \times 10^{-3}$ from Refs.~\cite{Bijnens:1995yn,Bijnens:1997vq,Bijnens:1998fm,Bijnens:1999hw,Bijnens:2014lea} and $\mu=0.77$~GeV, we can straightforwardly calculate the Adler zeros in $IJ=00,20$ channels and the S-matrix zero below the $\pi\pi$ threshold in the first RS (corresponding to the virtual state pole in the second RS) in $IJ=20$ channel. The results are given in Tab.~\ref{tab.comparison}, where the error bars are conservatively estimated, since different low energy constants are assumed to be uncorrelated. For the error bars from Roy equations, they are obtained by taking the same inputs as previously discussed. It is worth noting that the Adler zero, $s_{A,IJ=00}$, moves to complex plane in Roy equation analysis at $m_\pi=391~$MeV due to the appearance of the LHCs generated by the $\sigma$ in the crossed channel, which also hints that the situation in the $IJ=00$ channel at $m_\pi=391~$MeV is certainly beyond the range of applicability of $\chi$PT due to the new LHCs generated by the bound state pole.
\begin{table}[h]
    \centering
    \begin{tabular}{ccccc}
    \hline\hline
    & \multicolumn{2}{c}{$m_\pi=236~$MeV} & \multicolumn{2}{c}{$m_\pi=391~$MeV} \\
    \hline
    &\ Roy equation\ & \ $\chi \text{PT}_\text{NNLO}$\ & \ Roy equation\ & \ $\chi \text{PT}_\text{NNLO}$\ \\
    \hline
    $\sqrt{s_{A,IJ=00}}$ & $162$ & $140^{+46}_{-29}$ & $(206^{+29}_{-18})\pm i (218^{+3}_{-18})$ & $225^{+131}_{-115}$ \\
    \hline
    $\sqrt{s_{A,IJ=20}}$ & $326$ & $334^{+13}_{-16}$ & $601^{+8}_{-17}$ & $546^{+41}_{-73}$ \\
    \hline
    $\sqrt{s_{v,IJ=20}}$ & $117$ & $167^{+8}_{-9}$ & $435_{-12}^{+4}$ & $410^{+30}_{-41}$ \\
    \hline\hline
    \end{tabular}
    \caption{The resulting positions of Adler zeros in $IJ=00,20$ channels and the virtual state pole in $IJ=20$ channel from the Roy equation and NNLO $\chi$PT at $m_\pi=236,391~$MeV. All numbers are given in units of MeV.}
    \label{tab.comparison}
\end{table}

The future lattice simulations in the energy region above $K\bar{K}$ threshold for the $IJ=00$ case at $m_\pi=236~$MeV are expected to be the key ingredient to improve the accuracy of predictions for phase shifts and pole contents in the Roy equation analyses.

\section{Summary}\label{sec.6}

In this work we derive an extended Roy equation by including a bound state pole and apply this formalism to $\pi\pi$ scatterings at unphysical large pion mass when the $\sigma$ becomes a bound state. 
By taking the lattice phase shifts above the $K\bar{K}$ threshold in the $IJ=00,11,20$ channels, the Regge amplitudes and the D-wave contributions as the inputs of the driving terms in Roy equation, we obtain the phase shifts in the elastic region by solving the coupled integral equations at $m_\pi=391$~MeV. We then extrapolate the amplitudes into the complex $s$ plane to search for the poles. The pole positions of $\sigma$ and $f_0(980)$ from our studies are similar to those of HadSpec collaboration. In addition, we also find two additional types of poles for the $IJ=00$ channel in the second Riemann sheet: a pair of subthreshold complex poles near $s=0$ and a broad resonance pole $f_0^{\text{II}}$, where the former may correspond to a ``companion pole'' of the bound state pole ``$\sigma$'' and the latter could correspond to a second Riemann sheet (shadow) pole of the $f_0(1370)$ at large pion mass case.  
We have shown that the constraints from crossing symmetry play a crucial role in $\pi\pi$ scatterings at large pion masses, especially when there exists a bound state pole. Our predictions to the phase shifts at large pion masses are now consistent with the requirement of crossing symmetry, therefore they can be considered as a set of reference values for future phenomenological studies. Similar Roy equation analyses are also carried out for the situation at $m_\pi=236$~MeV, in which we give predictions to phase shifts, resonance poles and Adler zeros. 

Anticipated improvements in the precision of lattice QCD calculations will definitely increase the needs to rigorously extract the resonance information. In this work, we have demonstrated that the  sophisticated dispersive Roy equation can provide a powerful and rigorous tool to analyze lattice data. To our knowledge, this is also the first time that lattice data at unphysical large pion masses are analyzed by the model-independent Roy equation method, which strictly respects crossing symmetry. It is interesting to perform similar Roy equation analyses to more complicated $\pi K$ and even $\pi N$ scatterings at unphysical quark masses, which can be helpful to understand  chiral symmetry of low-energy QCD.
\\
\\
\textbf{Note added:} While this manuscript was under referee's review,  a preprint~\cite{Rodas:2023twk} appeared, in which the authors also explicitly analysed $\pi\pi$ lattice data ($m_\pi\sim 283$ and $239~$MeV) in the context of Roy equation constraints, which are consistent with the result we find here, although their numerical approach is different from ours.

\begin{acknowledgments}
The authors thank Zhi-Yong Zhou and Zhi-Guang Xiao for enlightening discussions. We are also grateful to the anonymous referee for helpful remarks which led us to add some additional material to the earlier versions of the manuscript.
This work is supported by the Natural Science Foundation of China (NSFC) under contracts No.~11975028, No.~11975090, No.~12150013, and the Science Foundation of Hebei Normal University with contract No.~L2023B09. 
\end{acknowledgments}

\begin{widetext}
\section*{Appendix}

\subsection{Demonstration of the existence of a virtual state pole in $IJ=20$ channel}\label{Sec.appij20}

We do find a virtual state pole in the $S$-matrix on the second Riemann sheet for the $\pi\pi$ isotensor channel. This  phenomenon was firstly discussed in \cite{PhysRev.123.692} (rediscovered in $\pi\pi$ scatterings~\cite{Zhou:2004ms, Dai:2019zao}, in $\pi N$ scatterings~\cite{Li:2021oou, Cao:2022zhn}). 

Taking for example the $\pi\pi$ scattering amplitude $t^2_0(s)$ to illustrate, the partial wave S-matrix $S^2_0(s)=1-2\sqrt{4m_\pi^2/s-1}~t^2_0(s)$ is a real function in the range between the threshold $4m_\pi^2$ and the branch point $s_L$ of the left-hand cut, where $s_L=0$ for $m_\pi=139,236~$MeV and $s_L=4m_\pi^2-s_\sigma$ for $m_\pi=391~$MeV. Since there are no bound states, $S^2_0(s)$ is bounded in the range between $s_L$ and $4m_\pi^2$. Furthermore, since there is no anomalous threshold, $S^2_0(s)=1$ at $4m_\pi^2$. According to Eq.~\eqref{eq.kij}, it can be proved that $S^2_0(s)$ approaches negative infinity when $s$ gets close to the branch point of the left-hand cut. Therefore $S^2_0(s)$ must have at least one zero on the first Riemann sheet as shown in Fig.~\ref{fig.4}. Analyticity and unitarity tell us that $S^\text{II}(s)=1/S(s)$, namely the first sheet zero exactly corresponds to a second sheet pole.  In this way, we demonstrate that the zero of $S_0^2(s)$ in the range ($s_L,4m_\pi^2$] corresponds to a virtual state pole of $S^\text{II}(s)$ on the second sheet.
\begin{figure}[H]
   \centering
    \includegraphics[width=0.35\textwidth,angle=-0]{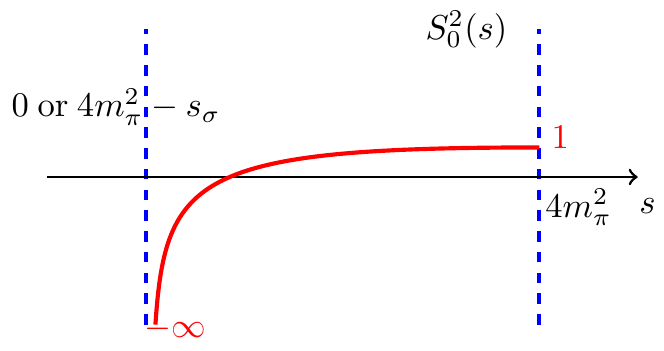} 
   \caption{The PW S-matrix $S^2_0(s)$ below the $\pi\pi$ threshold. Note that $S^2_0(s)$ is real below the $\pi\pi$ threshold. The intersection point between $S^2_0(s)$ and $s$ axis, corresponds to the virtual state position.}\label{fig.4}
\end{figure}
The above conclusion can be confirmed from the prediction of $\chi$PT at $s=0^+$, which leads to $t^2_0(0^+)>0$. In the energy region of $(s_L,4m_\pi^2]$, using $S^2_0(s)=1-2\sqrt{4m_\pi^2/s-1}~t^2_0(s)$ we can obtain $S^2_0(s)\overset{s\to 0^+}{\sim} -t^2_0(0^+)s^{-1/2} \overset{s\to 0^+}{\to} -\infty$. While at $s=4m_\pi^2$, $S^2_0(s)=1-2\sqrt{4m_\pi^2/s-1}~t^2_0(s)$ approaches to 1. This leads to the conclusion that there must be at least one zero for the S-matrix $S_0^2(s)$ in the physical Riemann sheet in the range $(s_L,4m_\pi^2]$, which in turn means that there is a virtual pole in the second RS.  

We have verified that the virtual state pole $s_0$ appearing in $S^\text{II}(s_0)$, i.e. when $S(s_0)=0$, is quite stable by taking the different kinds of $t_0^2(s)$ in the S-matrix $S^2_0(s)=1-2\sqrt{4m_\pi^2/s-1}~t^2_0(s)$, namely by using the leading-order current algebra result, the perturbative $O(p^4)$ and $O(p^6)$ $\chi$PT amplitudes, the unitarized IAM expression and the Roy equation solution for $t_0^2(s)$. All the discussions given above strongly supports and explains our numerical findings of such virtual state pole in the Roy equation analyses. Therefore this enables us to conclude that the virtual state pole in the $IJ=20$ channel is a pure prediction by combining $\chi$PT, relativistic kinematics and S-matrix theory.

\subsection{Procedures and inputs to numerically solve Roy equations at $m_\pi=236~$MeV}\label{Sec.app236}

\subsubsection{Inputs of the driving terms} 

For the lattice simulations at $m_\pi=236$~MeV, the phase shifts $\delta^0_0$ above $800~$MeV are still not available. Since they are crucial inputs when solving the Roy equation, it could be difficult to get robust predictions to the low-energy phase shifts in the case of $m_\pi=236$~MeV. 
Fortunately, according to the results in Ref.~\cite{Briceno:2016mjc}, the phase shifts $\delta^0_0$ at $m_\pi=236$~MeV are only slightly larger than the physical ones, and the kaon mass $m_K=501~$MeV is also close to its physical value $496~$MeV in this case. Therefore, we will take a very rough estimation by simply using the physical phase shifts and inelasticities above the $K\bar{K}$ threshold up to $1.4~$GeV from Ref.~\cite{Pelaez:2019eqa} and smoothly extrapolate the phase shifts between the matching point $\sqrt{s_{\mathrm{m}}}=800~$MeV and the $K\bar{K}$ threshold.
In addition, there are no lattice phase shifts with $IJ=20$ at $m_\pi=236~$MeV from the HadSpec collaboration. By taking into account of the moderate pion mass-dependence of the phase shifts in the $IJ=20$ channel~\cite{Dudek:2010ew,Nebreda:2011di,Dudek:2012gj}, we will estimate such phase shifts from Refs.~\cite{Bulava:2016mks, Mai:2019pqr} which perform the lattice simulations at similar pion masses with $m_\pi=224,230~$MeV. 
For the D-wave contributions to the DTs, we show that $d^I_J$ (especially $d^0_0$) is dominated by the contribution from the resonance $f_2(1270)$, whose mass and width can be estimated by chiral extrapolation of the resonance $\chi$PT~\cite{Chen:2023ybr}: $m_{f_2}\simeq  1330~\mathrm{MeV}, \Gamma_{f_2\to\pi\pi}\simeq 150~$MeV at $m_\pi=236$~MeV. 
Using the narrow width approximation, expanding the relevant kernel in the inverse powers of $s^\prime=m_{f_2}^2$ and
retaining only the leading term at the order of $1/s^{\prime 3}$, we obtain~\cite{Caprini:2005zr}
\begin{align}
    &d_{0,\text{D}}^0(s) \simeq \frac{5\left(s-4 m_\pi^2\right)\left(11 s+4 m_\pi^2\right) \Gamma_{f_2 \rightarrow \pi \pi}}{9 m_{f_2}^4 \sqrt{m_{f_2}^2-4 m_\pi^2}},\nonumber\\ 
    &d_{1,\text{D}}^1(s) \simeq -\frac{5\left(s-4 m_\pi^2\right)s \Gamma_{f_2 \rightarrow \pi \pi}}{9 m_{f_2}^4 \sqrt{m_{f_2}^2-4 m_\pi^2}},\nonumber\\ 
    &d_{0,\text{D}}^2(s) \simeq \frac{10\left(s-4 m_\pi^2\right)(s+2m_\pi^2) \Gamma_{f_2 \rightarrow \pi \pi}}{9 m_{f_2}^4 \sqrt{m_{f_2}^2-4 m_\pi^2}}\ .
\end{align} 
Since our current work in the $m_\pi=236~$MeV case is a preliminary attempt, we will simply neglect the Regge contributions above $1.4~$GeV, whose effects are believed to be much less relevant than the previous assumptions about the inputs of phase shifts above the $K\bar{K}$ threshold. 

\subsubsection{Details of the optimization strategy}

For the lattice simulations at $m_\pi=236$~MeV, the phase shifts at the matching point $\sqrt{s_\text{m}}=800~$MeV are~\cite{Wilson:2015dqa,Briceno:2016mjc,Bulava:2016mks, Mai:2019pqr}: $\delta_0^0\left(s_{\mathrm{m}}\right)=87.5^{\circ}, \delta_1^1\left(s_{\mathrm{m}}\right)=99.1^{\circ}, \delta_0^2\left(s_{\mathrm{m}}\right)=-13.0^{\circ}$, which lead to the multiplicity index $m=0+1-1=0$. 
Due to the similarity between this situation and the physical pion mass case, we adopt an analogous optimization strategy following Ref.~\cite{Ananthanarayan:2000ht}. However, since the scattering length $a_0^0$ at $m_\pi=236$~MeV is still poorly known, we will take it as a free parameter. For the scattering length $a_0^2$ at $m_\pi=236$~MeV, it can be accurately determined  from the NLO $\chi$PT~\cite{Albaladejo:2012te}. 
Another constraint in the $IJ=11$ channel, i.e. $\frac{\md\delta_1^1(s_\text{m})}{\md s}=\left.\frac{\md \delta_1^1(s_\text{m}+0^+)}{\md s}\right|_\text{input}=12.9~{\rm rad}\cdot\mathrm{GeV}^{-2}$~\cite{Wilson:2015dqa}, will be included as well. 
During the optimization process, we adopt a Schenk-like parametrization for $\delta_0^0(s)$,
\begin{align}
    \tan \delta_0^0(s)=\rho_\pi(s)\left(a_{0}^0+B_{0}^0 q^2+C_{0}^0 q^4+D_{0}^0 q^6\right)\frac{4 m_\pi^2-s_{0}^0}{s-s_{0}^0}\ ,
\end{align}
and the parametrizations for $\delta_1^1(s)$ and $\delta_0^2(s)$ are same as Eqs.~\eqref{eq.delta11} and \eqref{eq.delta20}. 
The accuracy of the solutions is illustrated in Fig.~\ref{fig.Ret_236}.
\begin{figure}[h]
\centering
\includegraphics[width=0.42\textwidth,angle=-0]{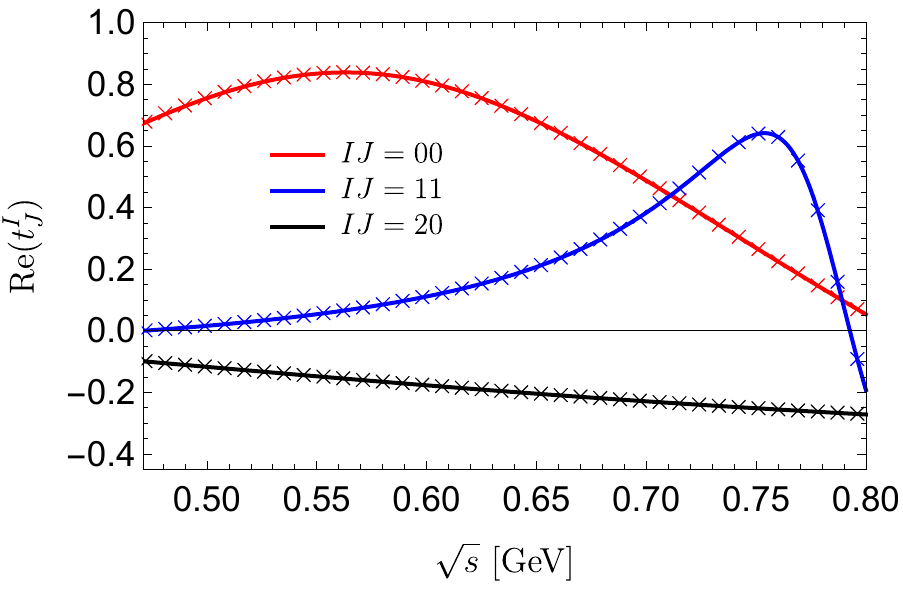} 
\caption{Left-hand sides of the Roy equations (lines) compared to the right-hand sides (points) after minimization at $m_\pi=236~$MeV.} \label{fig.Ret_236}
\end{figure}

Numerical values of the parameters that give the Roy solutions in Fig.~\ref{fig.Ret_236} are  collected in Tab.~\ref{tab.236}.
\begin{table}[h]
    \centering
    \begin{tabular}{cccccccc}
    \hline\hline
    $a^0_0$\ & \ $B^0_0$\ & \ $C^0_0$\ & \ $D^0_0$\ & \ $s^0_0$\ & \ $B_0$\ & \ $B_1$\ & \ $B_2$ \\
    \hline
    $6.75\times 10^{-1}$ & $1.27 \times 10$ & $-7.18 \times 10$ & $1.45 \times 10^2$ & $6.61\times 10^{-1}$ & $1.10$ & $6.01\times 10^{-2}$ & $-1.65\times 10^{-1}$ \\
    \hline\hline
    $s_0$ & $M_R$ & $a^2_0$ (input) & $B^2_0$ & $C^2_0$ & $D^2_0$ & $s^2_0$ \\
    \hline
    $1.22$ & $7.93\times 10^{-1}$ & $-1.00\times 10^{-1}$ & $-2.67$ & $1.11\times 10$ & $-2.60\times 10$ & $-1.74\times 10^2$\\
    \hline\hline
    \end{tabular}
    \caption{Parameters for the solutions of the extended Roy equations at $m_\pi=236$~MeV. All parameters are given in appropriate powers of GeV.}
    \label{tab.236}
\end{table}

\end{widetext}


\bibliography{apssamp}

\providecommand{\noopsort}[1]{}\providecommand{\singleletter}[1]{#1}%
\begin{thebibliography}{85}%
\makeatletter
\providecommand \@ifxundefined [1]{%
 \@ifx{#1\undefined}
}%
\providecommand \@ifnum [1]{%
 \ifnum #1\expandafter \@firstoftwo
 \else \expandafter \@secondoftwo
 \fi
}%
\providecommand \@ifx [1]{%
 \ifx #1\expandafter \@firstoftwo
 \else \expandafter \@secondoftwo
 \fi
}%
\providecommand \natexlab [1]{#1}%
\providecommand \enquote  [1]{``#1''}%
\providecommand \bibnamefont  [1]{#1}%
\providecommand \bibfnamefont [1]{#1}%
\providecommand \citenamefont [1]{#1}%
\providecommand \href@noop [0]{\@secondoftwo}%
\providecommand \href [0]{\begingroup \@sanitize@url \@href}%
\providecommand \@href[1]{\@@startlink{#1}\@@href}%
\providecommand \@@href[1]{\endgroup#1\@@endlink}%
\providecommand \@sanitize@url [0]{\catcode `\\12\catcode `\$12\catcode
  `\&12\catcode `\#12\catcode `\^12\catcode `\_12\catcode `\%12\relax}%
\providecommand \@@startlink[1]{}%
\providecommand \@@endlink[0]{}%
\providecommand \url  [0]{\begingroup\@sanitize@url \@url }%
\providecommand \@url [1]{\endgroup\@href {#1}{\urlprefix }}%
\providecommand \urlprefix  [0]{URL }%
\providecommand \Eprint [0]{\href }%
\providecommand \doibase [0]{https://doi.org/}%
\providecommand \selectlanguage [0]{\@gobble}%
\providecommand \bibinfo  [0]{\@secondoftwo}%
\providecommand \bibfield  [0]{\@secondoftwo}%
\providecommand \translation [1]{[#1]}%
\providecommand \BibitemOpen [0]{}%
\providecommand \bibitemStop [0]{}%
\providecommand \bibitemNoStop [0]{.\EOS\space}%
\providecommand \EOS [0]{\spacefactor3000\relax}%
\providecommand \BibitemShut  [1]{\csname bibitem#1\endcsname}%
\let\auto@bib@innerbib\@empty
\bibitem [{\citenamefont {Roy}(1971)}]{Roy:1971tc}%
  \BibitemOpen
  \bibfield  {author} {\bibinfo {author} {\bibfnamefont {S.~M.}\ \bibnamefont
  {Roy}},\ }\bibfield  {title} {\bibinfo {title} {{Exact integral equation for
  pion pion scattering involving only physical region partial waves}},\ }\href
  {https://doi.org/10.1016/0370-2693(71)90724-6} {\bibfield  {journal}
  {\bibinfo  {journal} {Phys. Lett. B}\ }\textbf {\bibinfo {volume} {36}},\
  \bibinfo {pages} {353} (\bibinfo {year} {1971})}\BibitemShut {NoStop}%
\bibitem [{\citenamefont {Ananthanarayan}\ \emph {et~al.}(2001)\citenamefont
  {Ananthanarayan}, \citenamefont {Colangelo}, \citenamefont {Gasser},\ and\
  \citenamefont {Leutwyler}}]{Ananthanarayan:2000ht}%
  \BibitemOpen
  \bibfield  {author} {\bibinfo {author} {\bibfnamefont {B.}~\bibnamefont
  {Ananthanarayan}}, \bibinfo {author} {\bibfnamefont {G.}~\bibnamefont
  {Colangelo}}, \bibinfo {author} {\bibfnamefont {J.}~\bibnamefont {Gasser}},\
  and\ \bibinfo {author} {\bibfnamefont {H.}~\bibnamefont {Leutwyler}},\
  }\bibfield  {title} {\bibinfo {title} {{Roy equation analysis of pi pi
  scattering}},\ }\href {https://doi.org/10.1016/S0370-1573(01)00009-6}
  {\bibfield  {journal} {\bibinfo  {journal} {Phys. Rept.}\ }\textbf {\bibinfo
  {volume} {353}},\ \bibinfo {pages} {207} (\bibinfo {year}
  {2001})}\BibitemShut {NoStop}%
\bibitem [{\citenamefont {Buettiker}\ \emph {et~al.}(2004)\citenamefont
  {Buettiker}, \citenamefont {Descotes-Genon},\ and\ \citenamefont
  {Moussallam}}]{Buettiker:2003pp}%
  \BibitemOpen
  \bibfield  {author} {\bibinfo {author} {\bibfnamefont {P.}~\bibnamefont
  {Buettiker}}, \bibinfo {author} {\bibfnamefont {S.}~\bibnamefont
  {Descotes-Genon}},\ and\ \bibinfo {author} {\bibfnamefont {B.}~\bibnamefont
  {Moussallam}},\ }\bibfield  {title} {\bibinfo {title} {{A new analysis of pi
  K scattering from Roy and Steiner type equations}},\ }\href
  {https://doi.org/10.1140/epjc/s2004-01591-1} {\bibfield  {journal} {\bibinfo
  {journal} {Eur. Phys. J. C}\ }\textbf {\bibinfo {volume} {33}},\ \bibinfo
  {pages} {409} (\bibinfo {year} {2004})}\BibitemShut {NoStop}%
\bibitem [{\citenamefont {Garcia-Martin}\ \emph
  {et~al.}(2011{\natexlab{a}})\citenamefont {Garcia-Martin}, \citenamefont
  {Kaminski}, \citenamefont {Pelaez}, \citenamefont {Ruiz~de Elvira},\ and\
  \citenamefont {Yndurain}}]{Garcia-Martin:2011iqs}%
  \BibitemOpen
  \bibfield  {author} {\bibinfo {author} {\bibfnamefont {R.}~\bibnamefont
  {Garcia-Martin}}, \bibinfo {author} {\bibfnamefont {R.}~\bibnamefont
  {Kaminski}}, \bibinfo {author} {\bibfnamefont {J.~R.}\ \bibnamefont
  {Pelaez}}, \bibinfo {author} {\bibfnamefont {J.}~\bibnamefont {Ruiz~de
  Elvira}},\ and\ \bibinfo {author} {\bibfnamefont {F.~J.}\ \bibnamefont
  {Yndurain}},\ }\bibfield  {title} {\bibinfo {title} {{The Pion-pion
  scattering amplitude. IV: Improved analysis with once subtracted Roy-like
  equations up to 1100 MeV}},\ }\href
  {https://doi.org/10.1103/PhysRevD.83.074004} {\bibfield  {journal} {\bibinfo
  {journal} {Phys. Rev. D}\ }\textbf {\bibinfo {volume} {83}},\ \bibinfo
  {pages} {074004} (\bibinfo {year} {2011}{\natexlab{a}})}\BibitemShut
  {NoStop}%
\bibitem [{\citenamefont {Caprini}\ \emph {et~al.}(2006)\citenamefont
  {Caprini}, \citenamefont {Colangelo},\ and\ \citenamefont
  {Leutwyler}}]{Caprini:2005zr}%
  \BibitemOpen
  \bibfield  {author} {\bibinfo {author} {\bibfnamefont {I.}~\bibnamefont
  {Caprini}}, \bibinfo {author} {\bibfnamefont {G.}~\bibnamefont {Colangelo}},\
  and\ \bibinfo {author} {\bibfnamefont {H.}~\bibnamefont {Leutwyler}},\
  }\bibfield  {title} {\bibinfo {title} {{Mass and width of the lowest
  resonance in QCD}},\ }\href {https://doi.org/10.1103/PhysRevLett.96.132001}
  {\bibfield  {journal} {\bibinfo  {journal} {Phys. Rev. Lett.}\ }\textbf
  {\bibinfo {volume} {96}},\ \bibinfo {pages} {132001} (\bibinfo {year}
  {2006})}\BibitemShut {NoStop}%
\bibitem [{\citenamefont {Descotes-Genon}\ and\ \citenamefont
  {Moussallam}(2006)}]{Descotes-Genon:2006sdr}%
  \BibitemOpen
  \bibfield  {author} {\bibinfo {author} {\bibfnamefont {S.}~\bibnamefont
  {Descotes-Genon}}\ and\ \bibinfo {author} {\bibfnamefont {B.}~\bibnamefont
  {Moussallam}},\ }\bibfield  {title} {\bibinfo {title} {{The K*0 (800) scalar
  resonance from Roy-Steiner representations of pi K scattering}},\ }\href
  {https://doi.org/10.1140/epjc/s10052-006-0036-2} {\bibfield  {journal}
  {\bibinfo  {journal} {Eur. Phys. J. C}\ }\textbf {\bibinfo {volume} {48}},\
  \bibinfo {pages} {553} (\bibinfo {year} {2006})}\BibitemShut {NoStop}%
\bibitem [{\citenamefont {Garcia-Martin}\ \emph
  {et~al.}(2011{\natexlab{b}})\citenamefont {Garcia-Martin}, \citenamefont
  {Kaminski}, \citenamefont {Pelaez},\ and\ \citenamefont {Ruiz~de
  Elvira}}]{Garcia-Martin:2011nna}%
  \BibitemOpen
  \bibfield  {author} {\bibinfo {author} {\bibfnamefont {R.}~\bibnamefont
  {Garcia-Martin}}, \bibinfo {author} {\bibfnamefont {R.}~\bibnamefont
  {Kaminski}}, \bibinfo {author} {\bibfnamefont {J.~R.}\ \bibnamefont
  {Pelaez}},\ and\ \bibinfo {author} {\bibfnamefont {J.}~\bibnamefont {Ruiz~de
  Elvira}},\ }\bibfield  {title} {\bibinfo {title} {{Precise determination of
  the f0(600) and f0(980) pole parameters from a dispersive data analysis}},\
  }\href {https://doi.org/10.1103/PhysRevLett.107.072001} {\bibfield  {journal}
  {\bibinfo  {journal} {Phys. Rev. Lett.}\ }\textbf {\bibinfo {volume} {107}},\
  \bibinfo {pages} {072001} (\bibinfo {year} {2011}{\natexlab{b}})},\ \Eprint
  {https://arxiv.org/abs/1107.1635} {arXiv:1107.1635 [hep-ph]} \BibitemShut
  {NoStop}%
\bibitem [{\citenamefont {Moussallam}(2011)}]{Moussallam:2011zg}%
  \BibitemOpen
  \bibfield  {author} {\bibinfo {author} {\bibfnamefont {B.}~\bibnamefont
  {Moussallam}},\ }\bibfield  {title} {\bibinfo {title} {{Couplings of light
  I=0 scalar mesons to simple operators in the complex plane}},\ }\href
  {https://doi.org/10.1140/epjc/s10052-011-1814-z} {\bibfield  {journal}
  {\bibinfo  {journal} {Eur. Phys. J. C}\ }\textbf {\bibinfo {volume} {71}},\
  \bibinfo {pages} {1814} (\bibinfo {year} {2011})},\ \Eprint
  {https://arxiv.org/abs/1110.6074} {arXiv:1110.6074 [hep-ph]} \BibitemShut
  {NoStop}%
\bibitem [{\citenamefont {Pel\'aez}\ and\ \citenamefont
  {Rodas}(2020)}]{Pelaez:2020uiw}%
  \BibitemOpen
  \bibfield  {author} {\bibinfo {author} {\bibfnamefont {J.~R.}\ \bibnamefont
  {Pel\'aez}}\ and\ \bibinfo {author} {\bibfnamefont {A.}~\bibnamefont
  {Rodas}},\ }\bibfield  {title} {\bibinfo {title} {{Determination of the
  lightest strange resonance $K_0^*(700)$ or $\kappa$, from a dispersive data
  analysis}},\ }\href {https://doi.org/10.1103/PhysRevLett.124.172001}
  {\bibfield  {journal} {\bibinfo  {journal} {Phys. Rev. Lett.}\ }\textbf
  {\bibinfo {volume} {124}},\ \bibinfo {pages} {172001} (\bibinfo {year}
  {2020})},\ \Eprint {https://arxiv.org/abs/2001.08153} {arXiv:2001.08153
  [hep-ph]} \BibitemShut {NoStop}%
\bibitem [{\citenamefont {Colangelo}\ \emph {et~al.}(2001)\citenamefont
  {Colangelo}, \citenamefont {Gasser},\ and\ \citenamefont
  {Leutwyler}}]{Colangelo:2001df}%
  \BibitemOpen
  \bibfield  {author} {\bibinfo {author} {\bibfnamefont {G.}~\bibnamefont
  {Colangelo}}, \bibinfo {author} {\bibfnamefont {J.}~\bibnamefont {Gasser}},\
  and\ \bibinfo {author} {\bibfnamefont {H.}~\bibnamefont {Leutwyler}},\
  }\bibfield  {title} {\bibinfo {title} {{$\pi \pi$ scattering}},\ }\href
  {https://doi.org/10.1016/S0550-3213(01)00147-X} {\bibfield  {journal}
  {\bibinfo  {journal} {Nucl. Phys. B}\ }\textbf {\bibinfo {volume} {603}},\
  \bibinfo {pages} {125} (\bibinfo {year} {2001})},\ \Eprint
  {https://arxiv.org/abs/hep-ph/0103088} {arXiv:hep-ph/0103088} \BibitemShut
  {NoStop}%
\bibitem [{\citenamefont {Caprini}\ \emph {et~al.}(2012)\citenamefont
  {Caprini}, \citenamefont {Colangelo},\ and\ \citenamefont
  {Leutwyler}}]{Caprini:2011ky}%
  \BibitemOpen
  \bibfield  {author} {\bibinfo {author} {\bibfnamefont {I.}~\bibnamefont
  {Caprini}}, \bibinfo {author} {\bibfnamefont {G.}~\bibnamefont {Colangelo}},\
  and\ \bibinfo {author} {\bibfnamefont {H.}~\bibnamefont {Leutwyler}},\
  }\bibfield  {title} {\bibinfo {title} {{Regge analysis of the pi pi
  scattering amplitude}},\ }\href
  {https://doi.org/10.1140/epjc/s10052-012-1860-1} {\bibfield  {journal}
  {\bibinfo  {journal} {Eur. Phys. J. C}\ }\textbf {\bibinfo {volume} {72}},\
  \bibinfo {pages} {1860} (\bibinfo {year} {2012})}\BibitemShut {NoStop}%
\bibitem [{\citenamefont {Pelaez}\ and\ \citenamefont
  {Rodas}(2018)}]{Pelaez:2018qny}%
  \BibitemOpen
  \bibfield  {author} {\bibinfo {author} {\bibfnamefont {J.~R.}\ \bibnamefont
  {Pelaez}}\ and\ \bibinfo {author} {\bibfnamefont {A.}~\bibnamefont {Rodas}},\
  }\bibfield  {title} {\bibinfo {title} {{$\pi \pi \rightarrow K {\bar{K}}$
  scattering up to 1.47 GeV with hyperbolic dispersion relations}},\ }\href
  {https://doi.org/10.1140/epjc/s10052-018-6296-9} {\bibfield  {journal}
  {\bibinfo  {journal} {Eur. Phys. J. C}\ }\textbf {\bibinfo {volume} {78}},\
  \bibinfo {pages} {897} (\bibinfo {year} {2018})}\BibitemShut {NoStop}%
\bibitem [{\citenamefont {Hite}\ and\ \citenamefont
  {Steiner}(1973)}]{Hite:1973pm}%
  \BibitemOpen
  \bibfield  {author} {\bibinfo {author} {\bibfnamefont {G.~E.}\ \bibnamefont
  {Hite}}\ and\ \bibinfo {author} {\bibfnamefont {F.}~\bibnamefont {Steiner}},\
  }\bibfield  {title} {\bibinfo {title} {{New dispersion relations and their
  application to partial-wave amplitudes}},\ }\href
  {https://doi.org/10.1007/BF02722827} {\bibfield  {journal} {\bibinfo
  {journal} {Nuovo Cim. A}\ }\textbf {\bibinfo {volume} {18}},\ \bibinfo
  {pages} {237} (\bibinfo {year} {1973})}\BibitemShut {NoStop}%
\bibitem [{\citenamefont {Hoferichter}\ \emph
  {et~al.}(2015{\natexlab{a}})\citenamefont {Hoferichter}, \citenamefont
  {Ruiz~de Elvira}, \citenamefont {Kubis},\ and\ \citenamefont
  {Mei\ss{}ner}}]{Hoferichter:2015dsa}%
  \BibitemOpen
  \bibfield  {author} {\bibinfo {author} {\bibfnamefont {M.}~\bibnamefont
  {Hoferichter}}, \bibinfo {author} {\bibfnamefont {J.}~\bibnamefont {Ruiz~de
  Elvira}}, \bibinfo {author} {\bibfnamefont {B.}~\bibnamefont {Kubis}},\ and\
  \bibinfo {author} {\bibfnamefont {U.-G.}\ \bibnamefont {Mei\ss{}ner}},\
  }\bibfield  {title} {\bibinfo {title} {{High-Precision Determination of the
  Pion-Nucleon \ensuremath{\sigma} Term from Roy-Steiner Equations}},\ }\href
  {https://doi.org/10.1103/PhysRevLett.115.092301} {\bibfield  {journal}
  {\bibinfo  {journal} {Phys. Rev. Lett.}\ }\textbf {\bibinfo {volume} {115}},\
  \bibinfo {pages} {092301} (\bibinfo {year} {2015}{\natexlab{a}})}\BibitemShut
  {NoStop}%
\bibitem [{\citenamefont {Hoferichter}\ \emph {et~al.}(2016)\citenamefont
  {Hoferichter}, \citenamefont {Ruiz~de Elvira}, \citenamefont {Kubis},\ and\
  \citenamefont {Mei\ss{}ner}}]{Hoferichter:2015hva}%
  \BibitemOpen
  \bibfield  {author} {\bibinfo {author} {\bibfnamefont {M.}~\bibnamefont
  {Hoferichter}}, \bibinfo {author} {\bibfnamefont {J.}~\bibnamefont {Ruiz~de
  Elvira}}, \bibinfo {author} {\bibfnamefont {B.}~\bibnamefont {Kubis}},\ and\
  \bibinfo {author} {\bibfnamefont {U.-G.}\ \bibnamefont {Mei\ss{}ner}},\
  }\bibfield  {title} {\bibinfo {title} {{Roy\textendash{}Steiner-equation
  analysis of pion\textendash{}nucleon scattering}},\ }\href
  {https://doi.org/10.1016/j.physrep.2016.02.002} {\bibfield  {journal}
  {\bibinfo  {journal} {Phys. Rept.}\ }\textbf {\bibinfo {volume} {625}},\
  \bibinfo {pages} {1} (\bibinfo {year} {2016})}\BibitemShut {NoStop}%
\bibitem [{\citenamefont {Hoferichter}\ \emph
  {et~al.}(2015{\natexlab{b}})\citenamefont {Hoferichter}, \citenamefont
  {Ruiz~de Elvira}, \citenamefont {Kubis},\ and\ \citenamefont
  {Mei\ss{}ner}}]{Hoferichter:2015tha}%
  \BibitemOpen
  \bibfield  {author} {\bibinfo {author} {\bibfnamefont {M.}~\bibnamefont
  {Hoferichter}}, \bibinfo {author} {\bibfnamefont {J.}~\bibnamefont {Ruiz~de
  Elvira}}, \bibinfo {author} {\bibfnamefont {B.}~\bibnamefont {Kubis}},\ and\
  \bibinfo {author} {\bibfnamefont {U.-G.}\ \bibnamefont {Mei\ss{}ner}},\
  }\bibfield  {title} {\bibinfo {title} {{Matching pion-nucleon Roy-Steiner
  equations to chiral perturbation theory}},\ }\href
  {https://doi.org/10.1103/PhysRevLett.115.192301} {\bibfield  {journal}
  {\bibinfo  {journal} {Phys. Rev. Lett.}\ }\textbf {\bibinfo {volume} {115}},\
  \bibinfo {pages} {192301} (\bibinfo {year} {2015}{\natexlab{b}})}\BibitemShut
  {NoStop}%
\bibitem [{\citenamefont {Cao}\ \emph {et~al.}(2022)\citenamefont {Cao},
  \citenamefont {Li},\ and\ \citenamefont {Zheng}}]{Cao:2022zhn}%
  \BibitemOpen
  \bibfield  {author} {\bibinfo {author} {\bibfnamefont {X.-H.}\ \bibnamefont
  {Cao}}, \bibinfo {author} {\bibfnamefont {Q.-Z.}\ \bibnamefont {Li}},\ and\
  \bibinfo {author} {\bibfnamefont {H.-Q.}\ \bibnamefont {Zheng}},\ }\bibfield
  {title} {\bibinfo {title} {{A possible subthreshold pole in S$_{11}$ channel
  from \ensuremath{\pi}N Roy-Steiner equation analyses}},\ }\href
  {https://doi.org/10.1007/JHEP12(2022)073} {\bibfield  {journal} {\bibinfo
  {journal} {JHEP}\ }\textbf {\bibinfo {volume} {12}},\ \bibinfo {pages}
  {073}},\ \Eprint {https://arxiv.org/abs/2207.09743} {arXiv:2207.09743
  [hep-ph]} \BibitemShut {NoStop}%
\bibitem [{\citenamefont {Pelaez}(2016)}]{Pelaez:2015qba}%
  \BibitemOpen
  \bibfield  {author} {\bibinfo {author} {\bibfnamefont {J.~R.}\ \bibnamefont
  {Pelaez}},\ }\bibfield  {title} {\bibinfo {title} {{From controversy to
  precision on the sigma meson: a review on the status of the non-ordinary
  $f_0(500)$ resonance}},\ }\href
  {https://doi.org/10.1016/j.physrep.2016.09.001} {\bibfield  {journal}
  {\bibinfo  {journal} {Phys. Rept.}\ }\textbf {\bibinfo {volume} {658}},\
  \bibinfo {pages} {1} (\bibinfo {year} {2016})},\ \Eprint
  {https://arxiv.org/abs/1510.00653} {arXiv:1510.00653 [hep-ph]} \BibitemShut
  {NoStop}%
\bibitem [{\citenamefont {Yao}\ \emph {et~al.}(2021)\citenamefont {Yao},
  \citenamefont {Dai}, \citenamefont {Zheng},\ and\ \citenamefont
  {Zhou}}]{Yao:2020bxx}%
  \BibitemOpen
  \bibfield  {author} {\bibinfo {author} {\bibfnamefont {D.-L.}\ \bibnamefont
  {Yao}}, \bibinfo {author} {\bibfnamefont {L.-Y.}\ \bibnamefont {Dai}},
  \bibinfo {author} {\bibfnamefont {H.-Q.}\ \bibnamefont {Zheng}},\ and\
  \bibinfo {author} {\bibfnamefont {Z.-Y.}\ \bibnamefont {Zhou}},\ }\bibfield
  {title} {\bibinfo {title} {{A review on partial-wave dynamics with chiral
  effective field theory and dispersion relation}},\ }\href
  {https://doi.org/10.1088/1361-6633/abfa6f} {\bibfield  {journal} {\bibinfo
  {journal} {Rept. Prog. Phys.}\ }\textbf {\bibinfo {volume} {84}},\ \bibinfo
  {pages} {076201} (\bibinfo {year} {2021})}\BibitemShut {NoStop}%
\bibitem [{\citenamefont {Eden}\ \emph {et~al.}(1966)\citenamefont {Eden},
  \citenamefont {Landshoff}, \citenamefont {Olive},\ and\ \citenamefont
  {Polkinghorne}}]{Eden:1966dnq}%
  \BibitemOpen
  \bibfield  {author} {\bibinfo {author} {\bibfnamefont {R.~J.}\ \bibnamefont
  {Eden}}, \bibinfo {author} {\bibfnamefont {P.~V.}\ \bibnamefont {Landshoff}},
  \bibinfo {author} {\bibfnamefont {D.~I.}\ \bibnamefont {Olive}},\ and\
  \bibinfo {author} {\bibfnamefont {J.~C.}\ \bibnamefont {Polkinghorne}},\
  }\href@noop {} {\emph {\bibinfo {title} {{The analytic S-matrix}}}}\
  (\bibinfo  {publisher} {Cambridge Univ. Press},\ \bibinfo {address}
  {Cambridge},\ \bibinfo {year} {1966})\BibitemShut {NoStop}%
\bibitem [{\citenamefont {Guo}\ \emph {et~al.}(2007)\citenamefont {Guo},
  \citenamefont {Sanz~Cillero},\ and\ \citenamefont {Zheng}}]{Guo:2007ff}%
  \BibitemOpen
  \bibfield  {author} {\bibinfo {author} {\bibfnamefont {Z.~H.}\ \bibnamefont
  {Guo}}, \bibinfo {author} {\bibfnamefont {J.~J.}\ \bibnamefont
  {Sanz~Cillero}},\ and\ \bibinfo {author} {\bibfnamefont {H.~Q.}\ \bibnamefont
  {Zheng}},\ }\bibfield  {title} {\bibinfo {title} {{Partial waves and large
  N(C) resonance sum rules}},\ }\href
  {https://doi.org/10.1088/1126-6708/2007/06/030} {\bibfield  {journal}
  {\bibinfo  {journal} {JHEP}\ }\textbf {\bibinfo {volume} {06}},\ \bibinfo
  {pages} {030}},\ \Eprint {https://arxiv.org/abs/hep-ph/0701232}
  {arXiv:hep-ph/0701232} \BibitemShut {NoStop}%
\bibitem [{\citenamefont {Guo}\ \emph {et~al.}(2008)\citenamefont {Guo},
  \citenamefont {Sanz-Cillero},\ and\ \citenamefont {Zheng}}]{Guo:2007hm}%
  \BibitemOpen
  \bibfield  {author} {\bibinfo {author} {\bibfnamefont {Z.~H.}\ \bibnamefont
  {Guo}}, \bibinfo {author} {\bibfnamefont {J.~J.}\ \bibnamefont
  {Sanz-Cillero}},\ and\ \bibinfo {author} {\bibfnamefont {H.~Q.}\ \bibnamefont
  {Zheng}},\ }\bibfield  {title} {\bibinfo {title} {{O(p**6) extension of the
  large - N(C) partial wave dispersion relations}},\ }\href
  {https://doi.org/10.1016/j.physletb.2008.01.073} {\bibfield  {journal}
  {\bibinfo  {journal} {Phys. Lett. B}\ }\textbf {\bibinfo {volume} {661}},\
  \bibinfo {pages} {342} (\bibinfo {year} {2008})},\ \Eprint
  {https://arxiv.org/abs/0710.2163} {arXiv:0710.2163 [hep-ph]} \BibitemShut
  {NoStop}%
\bibitem [{\citenamefont {Briceno}\ \emph
  {et~al.}(2018{\natexlab{a}})\citenamefont {Briceno}, \citenamefont {Dudek},\
  and\ \citenamefont {Young}}]{Briceno:2017max}%
  \BibitemOpen
  \bibfield  {author} {\bibinfo {author} {\bibfnamefont {R.~A.}\ \bibnamefont
  {Briceno}}, \bibinfo {author} {\bibfnamefont {J.~J.}\ \bibnamefont {Dudek}},\
  and\ \bibinfo {author} {\bibfnamefont {R.~D.}\ \bibnamefont {Young}},\
  }\bibfield  {title} {\bibinfo {title} {{Scattering processes and resonances
  from lattice QCD}},\ }\href {https://doi.org/10.1103/RevModPhys.90.025001}
  {\bibfield  {journal} {\bibinfo  {journal} {Rev. Mod. Phys.}\ }\textbf
  {\bibinfo {volume} {90}},\ \bibinfo {pages} {025001} (\bibinfo {year}
  {2018}{\natexlab{a}})},\ \Eprint {https://arxiv.org/abs/1706.06223}
  {arXiv:1706.06223 [hep-lat]} \BibitemShut {NoStop}%
\bibitem [{\citenamefont {Dudek}\ \emph {et~al.}(2013)\citenamefont {Dudek},
  \citenamefont {Edwards},\ and\ \citenamefont {Thomas}}]{Dudek:2012xn}%
  \BibitemOpen
  \bibfield  {author} {\bibinfo {author} {\bibfnamefont {J.~J.}\ \bibnamefont
  {Dudek}}, \bibinfo {author} {\bibfnamefont {R.~G.}\ \bibnamefont {Edwards}},\
  and\ \bibinfo {author} {\bibfnamefont {C.~E.}\ \bibnamefont {Thomas}}
  (\bibinfo {collaboration} {Hadron Spectrum}),\ }\bibfield  {title} {\bibinfo
  {title} {{Energy dependence of the $\rho$ resonance in $\pi\pi$ elastic
  scattering from lattice QCD}},\ }\href
  {https://doi.org/10.1103/PhysRevD.87.034505} {\bibfield  {journal} {\bibinfo
  {journal} {Phys. Rev. D}\ }\textbf {\bibinfo {volume} {87}},\ \bibinfo
  {pages} {034505} (\bibinfo {year} {2013})},\ \bibinfo {note} {[Erratum:
  Phys.Rev.D 90, 099902 (2014)]},\ \Eprint {https://arxiv.org/abs/1212.0830}
  {arXiv:1212.0830 [hep-ph]} \BibitemShut {NoStop}%
\bibitem [{\citenamefont {Briceno}\ \emph {et~al.}(2017)\citenamefont
  {Briceno}, \citenamefont {Dudek}, \citenamefont {Edwards},\ and\
  \citenamefont {Wilson}}]{Briceno:2016mjc}%
  \BibitemOpen
  \bibfield  {author} {\bibinfo {author} {\bibfnamefont {R.~A.}\ \bibnamefont
  {Briceno}}, \bibinfo {author} {\bibfnamefont {J.~J.}\ \bibnamefont {Dudek}},
  \bibinfo {author} {\bibfnamefont {R.~G.}\ \bibnamefont {Edwards}},\ and\
  \bibinfo {author} {\bibfnamefont {D.~J.}\ \bibnamefont {Wilson}},\ }\bibfield
   {title} {\bibinfo {title} {{Isoscalar $\pi\pi$ scattering and the $\sigma$
  meson resonance from QCD}},\ }\href
  {https://doi.org/10.1103/PhysRevLett.118.022002} {\bibfield  {journal}
  {\bibinfo  {journal} {Phys. Rev. Lett.}\ }\textbf {\bibinfo {volume} {118}},\
  \bibinfo {pages} {022002} (\bibinfo {year} {2017})},\ \Eprint
  {https://arxiv.org/abs/1607.05900} {arXiv:1607.05900 [hep-ph]} \BibitemShut
  {NoStop}%
\bibitem [{\citenamefont {Briceno}\ \emph
  {et~al.}(2018{\natexlab{b}})\citenamefont {Briceno}, \citenamefont {Dudek},
  \citenamefont {Edwards},\ and\ \citenamefont {Wilson}}]{Briceno:2017qmb}%
  \BibitemOpen
  \bibfield  {author} {\bibinfo {author} {\bibfnamefont {R.~A.}\ \bibnamefont
  {Briceno}}, \bibinfo {author} {\bibfnamefont {J.~J.}\ \bibnamefont {Dudek}},
  \bibinfo {author} {\bibfnamefont {R.~G.}\ \bibnamefont {Edwards}},\ and\
  \bibinfo {author} {\bibfnamefont {D.~J.}\ \bibnamefont {Wilson}},\ }\bibfield
   {title} {\bibinfo {title} {{Isoscalar $\pi\pi, K\overline{K}, \eta\eta$
  scattering and the $\sigma, f_0, f_2$ mesons from QCD}},\ }\href
  {https://doi.org/10.1103/PhysRevD.97.054513} {\bibfield  {journal} {\bibinfo
  {journal} {Phys. Rev. D}\ }\textbf {\bibinfo {volume} {97}},\ \bibinfo
  {pages} {054513} (\bibinfo {year} {2018}{\natexlab{b}})},\ \Eprint
  {https://arxiv.org/abs/1708.06667} {arXiv:1708.06667 [hep-lat]} \BibitemShut
  {NoStop}%
\bibitem [{\citenamefont {Dudek}\ \emph {et~al.}(2011)\citenamefont {Dudek},
  \citenamefont {Edwards}, \citenamefont {Peardon}, \citenamefont {Richards},\
  and\ \citenamefont {Thomas}}]{Dudek:2010ew}%
  \BibitemOpen
  \bibfield  {author} {\bibinfo {author} {\bibfnamefont {J.~J.}\ \bibnamefont
  {Dudek}}, \bibinfo {author} {\bibfnamefont {R.~G.}\ \bibnamefont {Edwards}},
  \bibinfo {author} {\bibfnamefont {M.~J.}\ \bibnamefont {Peardon}}, \bibinfo
  {author} {\bibfnamefont {D.~G.}\ \bibnamefont {Richards}},\ and\ \bibinfo
  {author} {\bibfnamefont {C.~E.}\ \bibnamefont {Thomas}},\ }\bibfield  {title}
  {\bibinfo {title} {{Phase shift of isospin-2 $\pi\pi$ scattering from lattice
  QCD}},\ }\href {https://doi.org/10.1103/PhysRevD.83.071504} {\bibfield
  {journal} {\bibinfo  {journal} {Phys. Rev. D}\ }\textbf {\bibinfo {volume}
  {83}},\ \bibinfo {pages} {071504} (\bibinfo {year} {2011})},\ \Eprint
  {https://arxiv.org/abs/1011.6352} {arXiv:1011.6352 [hep-ph]} \BibitemShut
  {NoStop}%
\bibitem [{\citenamefont {Dudek}\ \emph {et~al.}(2012)\citenamefont {Dudek},
  \citenamefont {Edwards},\ and\ \citenamefont {Thomas}}]{Dudek:2012gj}%
  \BibitemOpen
  \bibfield  {author} {\bibinfo {author} {\bibfnamefont {J.~J.}\ \bibnamefont
  {Dudek}}, \bibinfo {author} {\bibfnamefont {R.~G.}\ \bibnamefont {Edwards}},\
  and\ \bibinfo {author} {\bibfnamefont {C.~E.}\ \bibnamefont {Thomas}},\
  }\bibfield  {title} {\bibinfo {title} {{S and D-wave phase shifts in
  isospin-2 pi pi scattering from lattice QCD}},\ }\href
  {https://doi.org/10.1103/PhysRevD.86.034031} {\bibfield  {journal} {\bibinfo
  {journal} {Phys. Rev. D}\ }\textbf {\bibinfo {volume} {86}},\ \bibinfo
  {pages} {034031} (\bibinfo {year} {2012})},\ \Eprint
  {https://arxiv.org/abs/1203.6041} {arXiv:1203.6041 [hep-ph]} \BibitemShut
  {NoStop}%
\bibitem [{\citenamefont {Pelaez}\ and\ \citenamefont
  {Rios}(2010)}]{Pelaez:2010fj}%
  \BibitemOpen
  \bibfield  {author} {\bibinfo {author} {\bibfnamefont {J.~R.}\ \bibnamefont
  {Pelaez}}\ and\ \bibinfo {author} {\bibfnamefont {G.}~\bibnamefont {Rios}},\
  }\bibfield  {title} {\bibinfo {title} {{Chiral extrapolation of light
  resonances from one and two-loop unitarized Chiral Perturbation Theory versus
  lattice results}},\ }\href {https://doi.org/10.1103/PhysRevD.82.114002}
  {\bibfield  {journal} {\bibinfo  {journal} {Phys. Rev. D}\ }\textbf {\bibinfo
  {volume} {82}},\ \bibinfo {pages} {114002} (\bibinfo {year} {2010})},\
  \Eprint {https://arxiv.org/abs/1010.6008} {arXiv:1010.6008 [hep-ph]}
  \BibitemShut {NoStop}%
\bibitem [{\citenamefont {Albaladejo}\ and\ \citenamefont
  {Oller}(2012)}]{Albaladejo:2012te}%
  \BibitemOpen
  \bibfield  {author} {\bibinfo {author} {\bibfnamefont {M.}~\bibnamefont
  {Albaladejo}}\ and\ \bibinfo {author} {\bibfnamefont {J.~A.}\ \bibnamefont
  {Oller}},\ }\bibfield  {title} {\bibinfo {title} {{On the size of the sigma
  meson and its nature}},\ }\href {https://doi.org/10.1103/PhysRevD.86.034003}
  {\bibfield  {journal} {\bibinfo  {journal} {Phys. Rev. D}\ }\textbf {\bibinfo
  {volume} {86}},\ \bibinfo {pages} {034003} (\bibinfo {year} {2012})},\
  \Eprint {https://arxiv.org/abs/1205.6606} {arXiv:1205.6606 [hep-ph]}
  \BibitemShut {NoStop}%
\bibitem [{\citenamefont {D\"oring}\ \emph {et~al.}(2018)\citenamefont
  {D\"oring}, \citenamefont {Hu},\ and\ \citenamefont {Mai}}]{Doring:2016bdr}%
  \BibitemOpen
  \bibfield  {author} {\bibinfo {author} {\bibfnamefont {M.}~\bibnamefont
  {D\"oring}}, \bibinfo {author} {\bibfnamefont {B.}~\bibnamefont {Hu}},\ and\
  \bibinfo {author} {\bibfnamefont {M.}~\bibnamefont {Mai}},\ }\bibfield
  {title} {\bibinfo {title} {{Chiral Extrapolation of the Sigma Resonance}},\
  }\href {https://doi.org/10.1016/j.physletb.2018.05.042} {\bibfield  {journal}
  {\bibinfo  {journal} {Phys. Lett. B}\ }\textbf {\bibinfo {volume} {782}},\
  \bibinfo {pages} {785} (\bibinfo {year} {2018})},\ \Eprint
  {https://arxiv.org/abs/1610.10070} {arXiv:1610.10070 [hep-lat]} \BibitemShut
  {NoStop}%
\bibitem [{\citenamefont {Danilkin}\ \emph {et~al.}(2021)\citenamefont
  {Danilkin}, \citenamefont {Deineka},\ and\ \citenamefont
  {Vanderhaeghen}}]{Danilkin:2020pak}%
  \BibitemOpen
  \bibfield  {author} {\bibinfo {author} {\bibfnamefont {I.}~\bibnamefont
  {Danilkin}}, \bibinfo {author} {\bibfnamefont {O.}~\bibnamefont {Deineka}},\
  and\ \bibinfo {author} {\bibfnamefont {M.}~\bibnamefont {Vanderhaeghen}},\
  }\bibfield  {title} {\bibinfo {title} {{Data-driven dispersive analysis of
  the \ensuremath{\pi}\ensuremath{\pi} and \ensuremath{\pi}K scattering}},\
  }\href {https://doi.org/10.1103/PhysRevD.103.114023} {\bibfield  {journal}
  {\bibinfo  {journal} {Phys. Rev. D}\ }\textbf {\bibinfo {volume} {103}},\
  \bibinfo {pages} {114023} (\bibinfo {year} {2021})},\ \Eprint
  {https://arxiv.org/abs/2012.11636} {arXiv:2012.11636 [hep-ph]} \BibitemShut
  {NoStop}%
\bibitem [{\citenamefont {Hanhart}\ \emph {et~al.}(2008)\citenamefont
  {Hanhart}, \citenamefont {Pelaez},\ and\ \citenamefont
  {Rios}}]{Hanhart:2008mx}%
  \BibitemOpen
  \bibfield  {author} {\bibinfo {author} {\bibfnamefont {C.}~\bibnamefont
  {Hanhart}}, \bibinfo {author} {\bibfnamefont {J.~R.}\ \bibnamefont
  {Pelaez}},\ and\ \bibinfo {author} {\bibfnamefont {G.}~\bibnamefont {Rios}},\
  }\bibfield  {title} {\bibinfo {title} {{Quark mass dependence of the rho and
  sigma from dispersion relations and Chiral Perturbation Theory}},\ }\href
  {https://doi.org/10.1103/PhysRevLett.100.152001} {\bibfield  {journal}
  {\bibinfo  {journal} {Phys. Rev. Lett.}\ }\textbf {\bibinfo {volume} {100}},\
  \bibinfo {pages} {152001} (\bibinfo {year} {2008})},\ \Eprint
  {https://arxiv.org/abs/0801.2871} {arXiv:0801.2871 [hep-ph]} \BibitemShut
  {NoStop}%
\bibitem [{\citenamefont {Gao}\ \emph {et~al.}(2022)\citenamefont {Gao},
  \citenamefont {Guo}, \citenamefont {Xiao},\ and\ \citenamefont
  {Zhou}}]{Gao:2022dln}%
  \BibitemOpen
  \bibfield  {author} {\bibinfo {author} {\bibfnamefont {X.-L.}\ \bibnamefont
  {Gao}}, \bibinfo {author} {\bibfnamefont {Z.-H.}\ \bibnamefont {Guo}},
  \bibinfo {author} {\bibfnamefont {Z.}~\bibnamefont {Xiao}},\ and\ \bibinfo
  {author} {\bibfnamefont {Z.-Y.}\ \bibnamefont {Zhou}},\ }\bibfield  {title}
  {\bibinfo {title} {{Scrutinizing \ensuremath{\pi}\ensuremath{\pi} scattering
  in light of recent lattice phase shifts}},\ }\href
  {https://doi.org/10.1103/PhysRevD.105.094002} {\bibfield  {journal} {\bibinfo
   {journal} {Phys. Rev. D}\ }\textbf {\bibinfo {volume} {105}},\ \bibinfo
  {pages} {094002} (\bibinfo {year} {2022})},\ \Eprint
  {https://arxiv.org/abs/2202.03124} {arXiv:2202.03124 [hep-ph]} \BibitemShut
  {NoStop}%
\bibitem [{\citenamefont {van Beveren}\ and\ \citenamefont
  {Rupp}(2023)}]{vanBeveren:2022zfx}%
  \BibitemOpen
  \bibfield  {author} {\bibinfo {author} {\bibfnamefont {E.}~\bibnamefont {van
  Beveren}}\ and\ \bibinfo {author} {\bibfnamefont {G.}~\bibnamefont {Rupp}},\
  }\bibfield  {title} {\bibinfo {title} {{Comment on
  \textquotedblleft{}Scrutinizing \ensuremath{\pi}\ensuremath{\pi} scattering
  in light of recent lattice phase shifts\textquotedblright{}}},\ }\href
  {https://doi.org/10.1103/PhysRevD.107.058501} {\bibfield  {journal} {\bibinfo
   {journal} {Phys. Rev. D}\ }\textbf {\bibinfo {volume} {107}},\ \bibinfo
  {pages} {058501} (\bibinfo {year} {2023})},\ \Eprint
  {https://arxiv.org/abs/2202.08809} {arXiv:2202.08809 [hep-ph]} \BibitemShut
  {NoStop}%
\bibitem [{\citenamefont {Gao}\ \emph {et~al.}(2023)\citenamefont {Gao},
  \citenamefont {Guo}, \citenamefont {Xiao},\ and\ \citenamefont
  {Zhou}}]{Gao:2022tlh}%
  \BibitemOpen
  \bibfield  {author} {\bibinfo {author} {\bibfnamefont {X.-L.}\ \bibnamefont
  {Gao}}, \bibinfo {author} {\bibfnamefont {Z.-H.}\ \bibnamefont {Guo}},
  \bibinfo {author} {\bibfnamefont {Z.}~\bibnamefont {Xiao}},\ and\ \bibinfo
  {author} {\bibfnamefont {Z.-Y.}\ \bibnamefont {Zhou}},\ }\bibfield  {title}
  {\bibinfo {title} {{Reply to \textquotedblleft{}Comment on
  \textquoteleft{}Scrutinizing \ensuremath{\pi}\ensuremath{\pi} scattering in
  light of recent lattice phase shifts\textquoteright{}\textquotedblright{}}},\
  }\href {https://doi.org/10.1103/PhysRevD.107.058502} {\bibfield  {journal}
  {\bibinfo  {journal} {Phys. Rev. D}\ }\textbf {\bibinfo {volume} {107}},\
  \bibinfo {pages} {058502} (\bibinfo {year} {2023})},\ \Eprint
  {https://arxiv.org/abs/2204.01562} {arXiv:2204.01562 [hep-ph]} \BibitemShut
  {NoStop}%
\bibitem [{\citenamefont {Balachandran}\ and\ \citenamefont
  {Nuyts}(1968)}]{Balachandran:1968zza}%
  \BibitemOpen
  \bibfield  {author} {\bibinfo {author} {\bibfnamefont {A.~P.}\ \bibnamefont
  {Balachandran}}\ and\ \bibinfo {author} {\bibfnamefont {J.}~\bibnamefont
  {Nuyts}},\ }\bibfield  {title} {\bibinfo {title} {{Simultaneous partial-wave
  expansion in the Mandelstamm variables: Crossing symmetry for partial
  waves}},\ }\href {https://doi.org/10.1103/PhysRev.172.1821} {\bibfield
  {journal} {\bibinfo  {journal} {Phys. Rev.}\ }\textbf {\bibinfo {volume}
  {172}},\ \bibinfo {pages} {1821} (\bibinfo {year} {1968})}\BibitemShut
  {NoStop}%
\bibitem [{\citenamefont {Roskies}(1969)}]{Roskies:1969pe}%
  \BibitemOpen
  \bibfield  {author} {\bibinfo {author} {\bibfnamefont {R.}~\bibnamefont
  {Roskies}},\ }\bibfield  {title} {\bibinfo {title} {{Crossing constraints on
  pi pi partial wave amplitudes}},\ }\href
  {https://doi.org/10.1016/0370-2693(69)90283-4} {\bibfield  {journal}
  {\bibinfo  {journal} {Phys. Lett. B}\ }\textbf {\bibinfo {volume} {30}},\
  \bibinfo {pages} {42} (\bibinfo {year} {1969})}\BibitemShut {NoStop}%
\bibitem [{\citenamefont {Roskies}(1970)}]{Roskies:1970uj}%
  \BibitemOpen
  \bibfield  {author} {\bibinfo {author} {\bibfnamefont {R.}~\bibnamefont
  {Roskies}},\ }\bibfield  {title} {\bibinfo {title} {{Crossing restrictions on
  pi pi partial waves}},\ }\href {https://doi.org/10.1007/BF02824912}
  {\bibfield  {journal} {\bibinfo  {journal} {Nuovo Cim. A}\ }\textbf {\bibinfo
  {volume} {65}},\ \bibinfo {pages} {467} (\bibinfo {year} {1970})}\BibitemShut
  {NoStop}%
\bibitem [{\citenamefont {Martin}\ \emph {et~al.}(1976)\citenamefont {Martin},
  \citenamefont {Morgan},\ and\ \citenamefont {Shaw}}]{Martin:1976}%
  \BibitemOpen
  \bibfield  {author} {\bibinfo {author} {\bibfnamefont {B.~R.}\ \bibnamefont
  {Martin}}, \bibinfo {author} {\bibfnamefont {D.}~\bibnamefont {Morgan}},\
  and\ \bibinfo {author} {\bibfnamefont {G.~L.}\ \bibnamefont {Shaw}},\
  }\href@noop {} {\emph {\bibinfo {title} {{Pion-pion Interactions in Particle
  Physics}}}}\ (\bibinfo  {publisher} {Academic Press, London},\ \bibinfo
  {year} {1976})\BibitemShut {NoStop}%
\bibitem [{\citenamefont {Pelaez}\ \emph {et~al.}(2019)\citenamefont {Pelaez},
  \citenamefont {Rodas},\ and\ \citenamefont {Ruiz
  De~Elvira}}]{Pelaez:2019eqa}%
  \BibitemOpen
  \bibfield  {author} {\bibinfo {author} {\bibfnamefont {J.~R.}\ \bibnamefont
  {Pelaez}}, \bibinfo {author} {\bibfnamefont {A.}~\bibnamefont {Rodas}},\ and\
  \bibinfo {author} {\bibfnamefont {J.}~\bibnamefont {Ruiz De~Elvira}},\
  }\bibfield  {title} {\bibinfo {title} {{Global parameterization of $\pi \pi $
  scattering up to 2 ${\mathrm {\,GeV}}$}},\ }\href
  {https://doi.org/10.1140/epjc/s10052-019-7509-6} {\bibfield  {journal}
  {\bibinfo  {journal} {Eur. Phys. J. C}\ }\textbf {\bibinfo {volume} {79}},\
  \bibinfo {pages} {1008} (\bibinfo {year} {2019})},\ \Eprint
  {https://arxiv.org/abs/1907.13162} {arXiv:1907.13162 [hep-ph]} \BibitemShut
  {NoStop}%
\bibitem [{\citenamefont {Moussallam}(2000)}]{Moussallam:1999aq}%
  \BibitemOpen
  \bibfield  {author} {\bibinfo {author} {\bibfnamefont {B.}~\bibnamefont
  {Moussallam}},\ }\bibfield  {title} {\bibinfo {title} {{N(f) dependence of
  the quark condensate from a chiral sum rule}},\ }\href
  {https://doi.org/10.1007/s100520050738} {\bibfield  {journal} {\bibinfo
  {journal} {Eur. Phys. J. C}\ }\textbf {\bibinfo {volume} {14}},\ \bibinfo
  {pages} {111} (\bibinfo {year} {2000})},\ \Eprint
  {https://arxiv.org/abs/hep-ph/9909292} {arXiv:hep-ph/9909292} \BibitemShut
  {NoStop}%
\bibitem [{\citenamefont {Martin}\ and\ \citenamefont
  {Spearman}(1970)}]{Martin:1970Elementary}%
  \BibitemOpen
  \bibfield  {author} {\bibinfo {author} {\bibfnamefont {A.~D.}\ \bibnamefont
  {Martin}}\ and\ \bibinfo {author} {\bibfnamefont {T.~D.}\ \bibnamefont
  {Spearman}},\ }\href@noop {} {\emph {\bibinfo {title} {Elementary Particle
  Theory}}}\ (\bibinfo  {publisher} {Amsterdam: North-Holland},\ \bibinfo
  {year} {1970})\BibitemShut {NoStop}%
\bibitem [{\citenamefont {Collins}(2009)}]{Collins:1977jy}%
  \BibitemOpen
  \bibfield  {author} {\bibinfo {author} {\bibfnamefont {P.~D.~B.}\
  \bibnamefont {Collins}},\ }\href {https://doi.org/10.1017/CBO9780511897603}
  {\emph {\bibinfo {title} {{An Introduction to Regge Theory and High-Energy
  Physics}}}},\ Cambridge Monographs on Mathematical Physics\ (\bibinfo
  {publisher} {Cambridge Univ. Press},\ \bibinfo {address} {Cambridge, UK},\
  \bibinfo {year} {2009})\BibitemShut {NoStop}%
\bibitem [{\citenamefont {Pelaez}\ and\ \citenamefont
  {Yndurain}(2003)}]{Pelaez:2003eh}%
  \BibitemOpen
  \bibfield  {author} {\bibinfo {author} {\bibfnamefont {J.~R.}\ \bibnamefont
  {Pelaez}}\ and\ \bibinfo {author} {\bibfnamefont {F.~J.}\ \bibnamefont
  {Yndurain}},\ }\bibfield  {title} {\bibinfo {title} {{On the precision of
  chiral dispersive calculations of pi pi scattering}},\ }\href
  {https://doi.org/10.1103/PhysRevD.68.074005} {\bibfield  {journal} {\bibinfo
  {journal} {Phys. Rev. D}\ }\textbf {\bibinfo {volume} {68}},\ \bibinfo
  {pages} {074005} (\bibinfo {year} {2003})},\ \Eprint
  {https://arxiv.org/abs/hep-ph/0304067} {arXiv:hep-ph/0304067} \BibitemShut
  {NoStop}%
\bibitem [{\citenamefont {Veneziano}(1968)}]{Veneziano:1968yb}%
  \BibitemOpen
  \bibfield  {author} {\bibinfo {author} {\bibfnamefont {G.}~\bibnamefont
  {Veneziano}},\ }\bibfield  {title} {\bibinfo {title} {{Construction of a
  crossing - symmetric, Regge behaved amplitude for linearly rising
  trajectories}},\ }\href {https://doi.org/10.1007/BF02824451} {\bibfield
  {journal} {\bibinfo  {journal} {Nuovo Cim. A}\ }\textbf {\bibinfo {volume}
  {57}},\ \bibinfo {pages} {190} (\bibinfo {year} {1968})}\BibitemShut
  {NoStop}%
\bibitem [{\citenamefont {Lovelace}(1968)}]{Lovelace:1968kjy}%
  \BibitemOpen
  \bibfield  {author} {\bibinfo {author} {\bibfnamefont {C.}~\bibnamefont
  {Lovelace}},\ }\bibfield  {title} {\bibinfo {title} {{A novel application of
  regge trajectories}},\ }\href {https://doi.org/10.1016/0370-2693(68)90255-4}
  {\bibfield  {journal} {\bibinfo  {journal} {Phys. Lett. B}\ }\textbf
  {\bibinfo {volume} {28}},\ \bibinfo {pages} {264} (\bibinfo {year}
  {1968})}\BibitemShut {NoStop}%
\bibitem [{\citenamefont {Shapiro}(1969)}]{Shapiro:1969km}%
  \BibitemOpen
  \bibfield  {author} {\bibinfo {author} {\bibfnamefont {J.~A.}\ \bibnamefont
  {Shapiro}},\ }\bibfield  {title} {\bibinfo {title} {{Narrow-resonance model
  with regge behavior for pi pi scattering}},\ }\href
  {https://doi.org/10.1103/PhysRev.179.1345} {\bibfield  {journal} {\bibinfo
  {journal} {Phys. Rev.}\ }\textbf {\bibinfo {volume} {179}},\ \bibinfo {pages}
  {1345} (\bibinfo {year} {1969})}\BibitemShut {NoStop}%
\bibitem [{\citenamefont {Donnachie}\ \emph {et~al.}(2004)\citenamefont
  {Donnachie}, \citenamefont {Dosch}, \citenamefont {Nachtmann},\ and\
  \citenamefont {Landshoff}}]{Donnachie:2002en}%
  \BibitemOpen
  \bibfield  {author} {\bibinfo {author} {\bibfnamefont {S.}~\bibnamefont
  {Donnachie}}, \bibinfo {author} {\bibfnamefont {H.~G.}\ \bibnamefont
  {Dosch}}, \bibinfo {author} {\bibfnamefont {O.}~\bibnamefont {Nachtmann}},\
  and\ \bibinfo {author} {\bibfnamefont {P.}~\bibnamefont {Landshoff}},\
  }\href@noop {} {\emph {\bibinfo {title} {{Pomeron physics and QCD}}}},\
  Vol.~\bibinfo {volume} {19}\ (\bibinfo  {publisher} {Cambridge University
  Press},\ \bibinfo {year} {2004})\BibitemShut {NoStop}%
\bibitem [{\citenamefont {Gasser}\ and\ \citenamefont
  {Wanders}(1999)}]{Gasser:1999hz}%
  \BibitemOpen
  \bibfield  {author} {\bibinfo {author} {\bibfnamefont {J.}~\bibnamefont
  {Gasser}}\ and\ \bibinfo {author} {\bibfnamefont {G.}~\bibnamefont
  {Wanders}},\ }\bibfield  {title} {\bibinfo {title} {{One channel Roy
  equations revisited}},\ }\href {https://doi.org/10.1007/s100529900086}
  {\bibfield  {journal} {\bibinfo  {journal} {Eur. Phys. J. C}\ }\textbf
  {\bibinfo {volume} {10}},\ \bibinfo {pages} {159} (\bibinfo {year}
  {1999})}\BibitemShut {NoStop}%
\bibitem [{\citenamefont {Wanders}(2000)}]{Wanders:2000mn}%
  \BibitemOpen
  \bibfield  {author} {\bibinfo {author} {\bibfnamefont {G.}~\bibnamefont
  {Wanders}},\ }\bibfield  {title} {\bibinfo {title} {{The Role of the input in
  Roy's equations for pi pi scattering}},\ }\href
  {https://doi.org/10.1007/s100520000459} {\bibfield  {journal} {\bibinfo
  {journal} {Eur. Phys. J. C}\ }\textbf {\bibinfo {volume} {17}},\ \bibinfo
  {pages} {323} (\bibinfo {year} {2000})}\BibitemShut {NoStop}%
\bibitem [{\citenamefont {Schenk}(1991)}]{Schenk:1991xe}%
  \BibitemOpen
  \bibfield  {author} {\bibinfo {author} {\bibfnamefont {A.}~\bibnamefont
  {Schenk}},\ }\bibfield  {title} {\bibinfo {title} {{Absorption and dispersion
  of pions at finite temperature}},\ }\href
  {https://doi.org/10.1016/0550-3213(91)90236-Q} {\bibfield  {journal}
  {\bibinfo  {journal} {Nucl. Phys. B}\ }\textbf {\bibinfo {volume} {363}},\
  \bibinfo {pages} {97} (\bibinfo {year} {1991})}\BibitemShut {NoStop}%
\bibitem [{\citenamefont {Regge}(1958)}]{Regge:1958ft}%
  \BibitemOpen
  \bibfield  {author} {\bibinfo {author} {\bibfnamefont {T.}~\bibnamefont
  {Regge}},\ }\bibfield  {title} {\bibinfo {title} {{Analytic properties of the
  scattering matrix}},\ }\href {https://doi.org/10.1007/BF02815247} {\bibfield
  {journal} {\bibinfo  {journal} {Nuovo Cim.}\ }\textbf {\bibinfo {volume}
  {8}},\ \bibinfo {pages} {671} (\bibinfo {year} {1958})}\BibitemShut {NoStop}%
\bibitem [{\citenamefont {Truong}(1988)}]{Truong:1988zp}%
  \BibitemOpen
  \bibfield  {author} {\bibinfo {author} {\bibfnamefont {T.~N.}\ \bibnamefont
  {Truong}},\ }\bibfield  {title} {\bibinfo {title} {{Chiral Perturbation
  Theory and Final State Theorem}},\ }\href
  {https://doi.org/10.1103/PhysRevLett.61.2526} {\bibfield  {journal} {\bibinfo
   {journal} {Phys. Rev. Lett.}\ }\textbf {\bibinfo {volume} {61}},\ \bibinfo
  {pages} {2526} (\bibinfo {year} {1988})}\BibitemShut {NoStop}%
\bibitem [{\citenamefont {Dobado}\ \emph {et~al.}(1990)\citenamefont {Dobado},
  \citenamefont {Herrero},\ and\ \citenamefont {Truong}}]{Dobado:1989qm}%
  \BibitemOpen
  \bibfield  {author} {\bibinfo {author} {\bibfnamefont {A.}~\bibnamefont
  {Dobado}}, \bibinfo {author} {\bibfnamefont {M.~J.}\ \bibnamefont
  {Herrero}},\ and\ \bibinfo {author} {\bibfnamefont {T.~N.}\ \bibnamefont
  {Truong}},\ }\bibfield  {title} {\bibinfo {title} {{Unitarized Chiral
  Perturbation Theory for Elastic Pion-Pion Scattering}},\ }\href
  {https://doi.org/10.1016/0370-2693(90)90109-J} {\bibfield  {journal}
  {\bibinfo  {journal} {Phys. Lett. B}\ }\textbf {\bibinfo {volume} {235}},\
  \bibinfo {pages} {134} (\bibinfo {year} {1990})}\BibitemShut {NoStop}%
\bibitem [{\citenamefont {Dobado}\ and\ \citenamefont
  {Pelaez}(1997)}]{Dobado:1996ps}%
  \BibitemOpen
  \bibfield  {author} {\bibinfo {author} {\bibfnamefont {A.}~\bibnamefont
  {Dobado}}\ and\ \bibinfo {author} {\bibfnamefont {J.~R.}\ \bibnamefont
  {Pelaez}},\ }\bibfield  {title} {\bibinfo {title} {{The Inverse amplitude
  method in chiral perturbation theory}},\ }\href
  {https://doi.org/10.1103/PhysRevD.56.3057} {\bibfield  {journal} {\bibinfo
  {journal} {Phys. Rev. D}\ }\textbf {\bibinfo {volume} {56}},\ \bibinfo
  {pages} {3057} (\bibinfo {year} {1997})},\ \Eprint
  {https://arxiv.org/abs/hep-ph/9604416} {arXiv:hep-ph/9604416} \BibitemShut
  {NoStop}%
\bibitem [{\citenamefont {Boglione}\ and\ \citenamefont
  {Pennington}(1997)}]{Boglione:1996uz}%
  \BibitemOpen
  \bibfield  {author} {\bibinfo {author} {\bibfnamefont {M.}~\bibnamefont
  {Boglione}}\ and\ \bibinfo {author} {\bibfnamefont {M.~R.}\ \bibnamefont
  {Pennington}},\ }\bibfield  {title} {\bibinfo {title} {{Chiral poles and
  zeros and the role of the left hand cut}},\ }\href
  {https://doi.org/10.1007/s002880050452} {\bibfield  {journal} {\bibinfo
  {journal} {Z. Phys. C}\ }\textbf {\bibinfo {volume} {75}},\ \bibinfo {pages}
  {113} (\bibinfo {year} {1997})},\ \Eprint
  {https://arxiv.org/abs/hep-ph/9607266} {arXiv:hep-ph/9607266} \BibitemShut
  {NoStop}%
\bibitem [{\citenamefont {Nieves}\ \emph {et~al.}(2002)\citenamefont {Nieves},
  \citenamefont {Pavon~Valderrama},\ and\ \citenamefont
  {Ruiz~Arriola}}]{Nieves:2001de}%
  \BibitemOpen
  \bibfield  {author} {\bibinfo {author} {\bibfnamefont {J.}~\bibnamefont
  {Nieves}}, \bibinfo {author} {\bibfnamefont {M.}~\bibnamefont
  {Pavon~Valderrama}},\ and\ \bibinfo {author} {\bibfnamefont {E.}~\bibnamefont
  {Ruiz~Arriola}},\ }\bibfield  {title} {\bibinfo {title} {{The Inverse
  amplitude method in pi pi scattering in chiral perturbation theory to two
  loops}},\ }\href {https://doi.org/10.1103/PhysRevD.65.036002} {\bibfield
  {journal} {\bibinfo  {journal} {Phys. Rev. D}\ }\textbf {\bibinfo {volume}
  {65}},\ \bibinfo {pages} {036002} (\bibinfo {year} {2002})},\ \Eprint
  {https://arxiv.org/abs/hep-ph/0109077} {arXiv:hep-ph/0109077} \BibitemShut
  {NoStop}%
\bibitem [{\citenamefont {Cavalcante}\ and\ \citenamefont
  {Sa~Borges}(2002)}]{Cavalcante:2001yw}%
  \BibitemOpen
  \bibfield  {author} {\bibinfo {author} {\bibfnamefont {I.~P.}\ \bibnamefont
  {Cavalcante}}\ and\ \bibinfo {author} {\bibfnamefont {J.}~\bibnamefont
  {Sa~Borges}},\ }\bibfield  {title} {\bibinfo {title} {{Crossing symmetry
  violation of unitarized pion pion amplitude in the resonance region}},\
  }\href {https://doi.org/10.1088/0954-3899/28/6/315} {\bibfield  {journal}
  {\bibinfo  {journal} {J. Phys. G}\ }\textbf {\bibinfo {volume} {28}},\
  \bibinfo {pages} {1351} (\bibinfo {year} {2002})},\ \Eprint
  {https://arxiv.org/abs/hep-ph/0110392} {arXiv:hep-ph/0110392} \BibitemShut
  {NoStop}%
\bibitem [{\citenamefont {Qin}\ \emph {et~al.}(2002)\citenamefont {Qin},
  \citenamefont {Deng}, \citenamefont {Xiao},\ and\ \citenamefont
  {Zheng}}]{Qin:2002hk}%
  \BibitemOpen
  \bibfield  {author} {\bibinfo {author} {\bibfnamefont {G.-Y.}\ \bibnamefont
  {Qin}}, \bibinfo {author} {\bibfnamefont {W.~Z.}\ \bibnamefont {Deng}},
  \bibinfo {author} {\bibfnamefont {Z.}~\bibnamefont {Xiao}},\ and\ \bibinfo
  {author} {\bibfnamefont {H.~Q.}\ \bibnamefont {Zheng}},\ }\bibfield  {title}
  {\bibinfo {title} {{The [1,2] Pade amplitudes for pi pi scatterings in chiral
  perturbation theory}},\ }\href
  {https://doi.org/10.1016/S0370-2693(02)02312-2} {\bibfield  {journal}
  {\bibinfo  {journal} {Phys. Lett. B}\ }\textbf {\bibinfo {volume} {542}},\
  \bibinfo {pages} {89} (\bibinfo {year} {2002})},\ \Eprint
  {https://arxiv.org/abs/hep-ph/0205214} {arXiv:hep-ph/0205214} \BibitemShut
  {NoStop}%
\bibitem [{\citenamefont {Salas-Bern\'ardez}\ \emph {et~al.}(2021)\citenamefont
  {Salas-Bern\'ardez}, \citenamefont {Llanes-Estrada}, \citenamefont
  {Escudero-Pedrosa},\ and\ \citenamefont {Oller}}]{Salas-Bernardez:2020hua}%
  \BibitemOpen
  \bibfield  {author} {\bibinfo {author} {\bibfnamefont {A.}~\bibnamefont
  {Salas-Bern\'ardez}}, \bibinfo {author} {\bibfnamefont {F.~J.}\ \bibnamefont
  {Llanes-Estrada}}, \bibinfo {author} {\bibfnamefont {J.}~\bibnamefont
  {Escudero-Pedrosa}},\ and\ \bibinfo {author} {\bibfnamefont {J.~A.}\
  \bibnamefont {Oller}},\ }\bibfield  {title} {\bibinfo {title} {{Systematizing
  and addressing theory uncertainties of unitarization with the Inverse
  Amplitude Method}},\ }\href {https://doi.org/10.21468/SciPostPhys.11.2.020}
  {\bibfield  {journal} {\bibinfo  {journal} {SciPost Phys.}\ }\textbf
  {\bibinfo {volume} {11}},\ \bibinfo {pages} {020} (\bibinfo {year} {2021})},\
  \Eprint {https://arxiv.org/abs/2010.13709} {arXiv:2010.13709 [hep-ph]}
  \BibitemShut {NoStop}%
\bibitem [{\citenamefont {Workman}\ \emph {et~al.}(2022)\citenamefont {Workman}
  \emph {et~al.}}]{ParticleDataGroup:2022pth}%
  \BibitemOpen
  \bibfield  {author} {\bibinfo {author} {\bibfnamefont {R.~L.}\ \bibnamefont
  {Workman}} \emph {et~al.} (\bibinfo {collaboration} {Particle Data Group}),\
  }\bibfield  {title} {\bibinfo {title} {{Review of Particle Physics}},\ }\href
  {https://doi.org/10.1093/ptep/ptac097} {\bibfield  {journal} {\bibinfo
  {journal} {PTEP}\ }\textbf {\bibinfo {volume} {2022}},\ \bibinfo {pages}
  {083C01} (\bibinfo {year} {2022})}\BibitemShut {NoStop}%
\bibitem [{\citenamefont {Pelaez}\ \emph {et~al.}(2023)\citenamefont {Pelaez},
  \citenamefont {Rodas},\ and\ \citenamefont {de~Elvira}}]{Pelaez:2022qby}%
  \BibitemOpen
  \bibfield  {author} {\bibinfo {author} {\bibfnamefont {J.~R.}\ \bibnamefont
  {Pelaez}}, \bibinfo {author} {\bibfnamefont {A.}~\bibnamefont {Rodas}},\ and\
  \bibinfo {author} {\bibfnamefont {J.~R.}\ \bibnamefont {de~Elvira}},\
  }\bibfield  {title} {\bibinfo {title} {{f0(1370) Controversy from Dispersive
  Meson-Meson Scattering Data Analyses}},\ }\href
  {https://doi.org/10.1103/PhysRevLett.130.051902} {\bibfield  {journal}
  {\bibinfo  {journal} {Phys. Rev. Lett.}\ }\textbf {\bibinfo {volume} {130}},\
  \bibinfo {pages} {051902} (\bibinfo {year} {2023})},\ \Eprint
  {https://arxiv.org/abs/2206.14822} {arXiv:2206.14822 [hep-ph]} \BibitemShut
  {NoStop}%
\bibitem [{\citenamefont {Bando}\ \emph {et~al.}(1985)\citenamefont {Bando},
  \citenamefont {Kugo}, \citenamefont {Uehara}, \citenamefont {Yamawaki},\ and\
  \citenamefont {Yanagida}}]{Bando:1984ej}%
  \BibitemOpen
  \bibfield  {author} {\bibinfo {author} {\bibfnamefont {M.}~\bibnamefont
  {Bando}}, \bibinfo {author} {\bibfnamefont {T.}~\bibnamefont {Kugo}},
  \bibinfo {author} {\bibfnamefont {S.}~\bibnamefont {Uehara}}, \bibinfo
  {author} {\bibfnamefont {K.}~\bibnamefont {Yamawaki}},\ and\ \bibinfo
  {author} {\bibfnamefont {T.}~\bibnamefont {Yanagida}},\ }\bibfield  {title}
  {\bibinfo {title} {{Is rho Meson a Dynamical Gauge Boson of Hidden Local
  Symmetry?}},\ }\href {https://doi.org/10.1103/PhysRevLett.54.1215} {\bibfield
   {journal} {\bibinfo  {journal} {Phys. Rev. Lett.}\ }\textbf {\bibinfo
  {volume} {54}},\ \bibinfo {pages} {1215} (\bibinfo {year}
  {1985})}\BibitemShut {NoStop}%
\bibitem [{\citenamefont {Locher}\ \emph {et~al.}(1998)\citenamefont {Locher},
  \citenamefont {Markushin},\ and\ \citenamefont {Zheng}}]{Locher:1997gr}%
  \BibitemOpen
  \bibfield  {author} {\bibinfo {author} {\bibfnamefont {M.~P.}\ \bibnamefont
  {Locher}}, \bibinfo {author} {\bibfnamefont {V.~E.}\ \bibnamefont
  {Markushin}},\ and\ \bibinfo {author} {\bibfnamefont {H.~Q.}\ \bibnamefont
  {Zheng}},\ }\bibfield  {title} {\bibinfo {title} {{Structure of f0 (980) from
  a coupled channel analysis of S wave pi pi scattering}},\ }\href
  {https://doi.org/10.1007/s100520050210} {\bibfield  {journal} {\bibinfo
  {journal} {Eur. Phys. J. C}\ }\textbf {\bibinfo {volume} {4}},\ \bibinfo
  {pages} {317} (\bibinfo {year} {1998})},\ \Eprint
  {https://arxiv.org/abs/hep-ph/9705230} {arXiv:hep-ph/9705230} \BibitemShut
  {NoStop}%
\bibitem [{\citenamefont {Baru}\ \emph {et~al.}(2004)\citenamefont {Baru},
  \citenamefont {Haidenbauer}, \citenamefont {Hanhart}, \citenamefont
  {Kalashnikova},\ and\ \citenamefont {Kudryavtsev}}]{Baru:2003qq}%
  \BibitemOpen
  \bibfield  {author} {\bibinfo {author} {\bibfnamefont {V.}~\bibnamefont
  {Baru}}, \bibinfo {author} {\bibfnamefont {J.}~\bibnamefont {Haidenbauer}},
  \bibinfo {author} {\bibfnamefont {C.}~\bibnamefont {Hanhart}}, \bibinfo
  {author} {\bibfnamefont {Y.}~\bibnamefont {Kalashnikova}},\ and\ \bibinfo
  {author} {\bibfnamefont {A.~E.}\ \bibnamefont {Kudryavtsev}},\ }\bibfield
  {title} {\bibinfo {title} {{Evidence that the a(0)(980) and f(0)(980) are not
  elementary particles}},\ }\href
  {https://doi.org/10.1016/j.physletb.2004.01.088} {\bibfield  {journal}
  {\bibinfo  {journal} {Phys. Lett. B}\ }\textbf {\bibinfo {volume} {586}},\
  \bibinfo {pages} {53} (\bibinfo {year} {2004})},\ \Eprint
  {https://arxiv.org/abs/hep-ph/0308129} {arXiv:hep-ph/0308129} \BibitemShut
  {NoStop}%
\bibitem [{\citenamefont {Su}\ \emph {et~al.}(2007)\citenamefont {Su},
  \citenamefont {Xiao},\ and\ \citenamefont {Zheng}}]{Su:2007au}%
  \BibitemOpen
  \bibfield  {author} {\bibinfo {author} {\bibfnamefont {M.-X.}\ \bibnamefont
  {Su}}, \bibinfo {author} {\bibfnamefont {L.~Y.}\ \bibnamefont {Xiao}},\ and\
  \bibinfo {author} {\bibfnamefont {H.~Q.}\ \bibnamefont {Zheng}},\ }\bibfield
  {title} {\bibinfo {title} {{On the scalar nonet in the extended Nambu
  Jona-Lasinio model}},\ }\href
  {https://doi.org/10.1016/j.nuclphysa.2007.06.004} {\bibfield  {journal}
  {\bibinfo  {journal} {Nucl. Phys. A}\ }\textbf {\bibinfo {volume} {792}},\
  \bibinfo {pages} {288} (\bibinfo {year} {2007})},\ \Eprint
  {https://arxiv.org/abs/0706.1823} {arXiv:0706.1823 [hep-ph]} \BibitemShut
  {NoStop}%
\bibitem [{\citenamefont {Ang}\ \emph {et~al.}(2001)\citenamefont {Ang},
  \citenamefont {Xiao}, \citenamefont {Zheng},\ and\ \citenamefont
  {Song}}]{Ang:2001bd}%
  \BibitemOpen
  \bibfield  {author} {\bibinfo {author} {\bibfnamefont {Q.}~\bibnamefont
  {Ang}}, \bibinfo {author} {\bibfnamefont {Z.}~\bibnamefont {Xiao}}, \bibinfo
  {author} {\bibfnamefont {H.~Q.}\ \bibnamefont {Zheng}},\ and\ \bibinfo
  {author} {\bibfnamefont {X.~C.}\ \bibnamefont {Song}},\ }\bibfield  {title}
  {\bibinfo {title} {{A Critical examination to the unitarized pi pi scattering
  chiral amplitudes}},\ }\href {https://doi.org/10.1088/0253-6102/36/5/563}
  {\bibfield  {journal} {\bibinfo  {journal} {Commun. Theor. Phys.}\ }\textbf
  {\bibinfo {volume} {36}},\ \bibinfo {pages} {563} (\bibinfo {year} {2001})},\
  \Eprint {https://arxiv.org/abs/hep-ph/0109012} {arXiv:hep-ph/0109012}
  \BibitemShut {NoStop}%
\bibitem [{\citenamefont {Dai}\ \emph {et~al.}(2019)\citenamefont {Dai},
  \citenamefont {Kang}, \citenamefont {Luo},\ and\ \citenamefont
  {Mei\ss{}ner}}]{Dai:2019zao}%
  \BibitemOpen
  \bibfield  {author} {\bibinfo {author} {\bibfnamefont {L.-Y.}\ \bibnamefont
  {Dai}}, \bibinfo {author} {\bibfnamefont {X.-W.}\ \bibnamefont {Kang}},
  \bibinfo {author} {\bibfnamefont {T.}~\bibnamefont {Luo}},\ and\ \bibinfo
  {author} {\bibfnamefont {U.-G.}\ \bibnamefont {Mei\ss{}ner}},\ }\bibfield
  {title} {\bibinfo {title} {{A study on the correlation between poles and cuts
  in $\pi\pi$ scattering}},\ }\href
  {https://doi.org/10.1088/0253-6102/71/11/1309} {\bibfield  {journal}
  {\bibinfo  {journal} {Commun. Theor. Phys.}\ }\textbf {\bibinfo {volume}
  {71}},\ \bibinfo {pages} {1309} (\bibinfo {year} {2019})},\ \Eprint
  {https://arxiv.org/abs/1903.01685} {arXiv:1903.01685 [hep-ph]} \BibitemShut
  {NoStop}%
\bibitem [{\citenamefont {Wilson}\ \emph {et~al.}(2015)\citenamefont {Wilson},
  \citenamefont {Briceno}, \citenamefont {Dudek}, \citenamefont {Edwards},\
  and\ \citenamefont {Thomas}}]{Wilson:2015dqa}%
  \BibitemOpen
  \bibfield  {author} {\bibinfo {author} {\bibfnamefont {D.~J.}\ \bibnamefont
  {Wilson}}, \bibinfo {author} {\bibfnamefont {R.~A.}\ \bibnamefont {Briceno}},
  \bibinfo {author} {\bibfnamefont {J.~J.}\ \bibnamefont {Dudek}}, \bibinfo
  {author} {\bibfnamefont {R.~G.}\ \bibnamefont {Edwards}},\ and\ \bibinfo
  {author} {\bibfnamefont {C.~E.}\ \bibnamefont {Thomas}},\ }\bibfield  {title}
  {\bibinfo {title} {{Coupled $\pi\pi, K\bar{K}$ scattering in $P$-wave and the
  $\rho$ resonance from lattice QCD}},\ }\href
  {https://doi.org/10.1103/PhysRevD.92.094502} {\bibfield  {journal} {\bibinfo
  {journal} {Phys. Rev. D}\ }\textbf {\bibinfo {volume} {92}},\ \bibinfo
  {pages} {094502} (\bibinfo {year} {2015})},\ \Eprint
  {https://arxiv.org/abs/1507.02599} {arXiv:1507.02599 [hep-ph]} \BibitemShut
  {NoStop}%
\bibitem [{\citenamefont {Danilkin}\ \emph {et~al.}(2022)\citenamefont
  {Danilkin}, \citenamefont {Biloshytskyi}, \citenamefont {Ren},\ and\
  \citenamefont {Vanderhaeghen}}]{Danilkin:2022cnj}%
  \BibitemOpen
  \bibfield  {author} {\bibinfo {author} {\bibfnamefont {I.}~\bibnamefont
  {Danilkin}}, \bibinfo {author} {\bibfnamefont {V.}~\bibnamefont
  {Biloshytskyi}}, \bibinfo {author} {\bibfnamefont {X.-L.}\ \bibnamefont
  {Ren}},\ and\ \bibinfo {author} {\bibfnamefont {M.}~\bibnamefont
  {Vanderhaeghen}},\ }\href@noop {} {\bibinfo {title} {{Analytical dispersive
  parameterization for elastic scattering of spinless particles}}} (\bibinfo
  {year} {2022}),\ \Eprint {https://arxiv.org/abs/2206.15223} {arXiv:2206.15223
  [hep-ph]} \BibitemShut {NoStop}%
\bibitem [{\citenamefont {Bulava}\ \emph {et~al.}(2016)\citenamefont {Bulava},
  \citenamefont {Fahy}, \citenamefont {H\"orz}, \citenamefont {Juge},
  \citenamefont {Morningstar},\ and\ \citenamefont {Wong}}]{Bulava:2016mks}%
  \BibitemOpen
  \bibfield  {author} {\bibinfo {author} {\bibfnamefont {J.}~\bibnamefont
  {Bulava}}, \bibinfo {author} {\bibfnamefont {B.}~\bibnamefont {Fahy}},
  \bibinfo {author} {\bibfnamefont {B.}~\bibnamefont {H\"orz}}, \bibinfo
  {author} {\bibfnamefont {K.~J.}\ \bibnamefont {Juge}}, \bibinfo {author}
  {\bibfnamefont {C.}~\bibnamefont {Morningstar}},\ and\ \bibinfo {author}
  {\bibfnamefont {C.~H.}\ \bibnamefont {Wong}},\ }\bibfield  {title} {\bibinfo
  {title} {{$I=1$ and $I=2$ $\pi-\pi$ scattering phase shifts from
  $N_{\mathrm{f}} = 2+1$ lattice QCD}},\ }\href
  {https://doi.org/10.1016/j.nuclphysb.2016.07.024} {\bibfield  {journal}
  {\bibinfo  {journal} {Nucl. Phys. B}\ }\textbf {\bibinfo {volume} {910}},\
  \bibinfo {pages} {842} (\bibinfo {year} {2016})},\ \Eprint
  {https://arxiv.org/abs/1604.05593} {arXiv:1604.05593 [hep-lat]} \BibitemShut
  {NoStop}%
\bibitem [{\citenamefont {Mai}\ \emph {et~al.}(2019)\citenamefont {Mai},
  \citenamefont {Culver}, \citenamefont {Alexandru}, \citenamefont {D\"oring},\
  and\ \citenamefont {Lee}}]{Mai:2019pqr}%
  \BibitemOpen
  \bibfield  {author} {\bibinfo {author} {\bibfnamefont {M.}~\bibnamefont
  {Mai}}, \bibinfo {author} {\bibfnamefont {C.}~\bibnamefont {Culver}},
  \bibinfo {author} {\bibfnamefont {A.}~\bibnamefont {Alexandru}}, \bibinfo
  {author} {\bibfnamefont {M.}~\bibnamefont {D\"oring}},\ and\ \bibinfo
  {author} {\bibfnamefont {F.~X.}\ \bibnamefont {Lee}},\ }\bibfield  {title}
  {\bibinfo {title} {{Cross-channel study of pion scattering from lattice
  QCD}},\ }\href {https://doi.org/10.1103/PhysRevD.100.114514} {\bibfield
  {journal} {\bibinfo  {journal} {Phys. Rev. D}\ }\textbf {\bibinfo {volume}
  {100}},\ \bibinfo {pages} {114514} (\bibinfo {year} {2019})},\ \Eprint
  {https://arxiv.org/abs/1908.01847} {arXiv:1908.01847 [hep-lat]} \BibitemShut
  {NoStop}%
\bibitem [{\citenamefont {Niehus}\ \emph {et~al.}(2021)\citenamefont {Niehus},
  \citenamefont {Hoferichter}, \citenamefont {Kubis},\ and\ \citenamefont
  {Ruiz~de Elvira}}]{Niehus:2020gmf}%
  \BibitemOpen
  \bibfield  {author} {\bibinfo {author} {\bibfnamefont {M.}~\bibnamefont
  {Niehus}}, \bibinfo {author} {\bibfnamefont {M.}~\bibnamefont {Hoferichter}},
  \bibinfo {author} {\bibfnamefont {B.}~\bibnamefont {Kubis}},\ and\ \bibinfo
  {author} {\bibfnamefont {J.}~\bibnamefont {Ruiz~de Elvira}},\ }\bibfield
  {title} {\bibinfo {title} {{Two-Loop Analysis of the Pion Mass Dependence of
  the $\rho$ Meson}},\ }\href {https://doi.org/10.1103/PhysRevLett.126.102002}
  {\bibfield  {journal} {\bibinfo  {journal} {Phys. Rev. Lett.}\ }\textbf
  {\bibinfo {volume} {126}},\ \bibinfo {pages} {102002} (\bibinfo {year}
  {2021})},\ \Eprint {https://arxiv.org/abs/2009.04479} {arXiv:2009.04479
  [hep-ph]} \BibitemShut {NoStop}%
\bibitem [{\citenamefont {Bijnens}\ \emph {et~al.}(1996)\citenamefont
  {Bijnens}, \citenamefont {Colangelo}, \citenamefont {Ecker}, \citenamefont
  {Gasser},\ and\ \citenamefont {Sainio}}]{Bijnens:1995yn}%
  \BibitemOpen
  \bibfield  {author} {\bibinfo {author} {\bibfnamefont {J.}~\bibnamefont
  {Bijnens}}, \bibinfo {author} {\bibfnamefont {G.}~\bibnamefont {Colangelo}},
  \bibinfo {author} {\bibfnamefont {G.}~\bibnamefont {Ecker}}, \bibinfo
  {author} {\bibfnamefont {J.}~\bibnamefont {Gasser}},\ and\ \bibinfo {author}
  {\bibfnamefont {M.~E.}\ \bibnamefont {Sainio}},\ }\bibfield  {title}
  {\bibinfo {title} {{Elastic pi pi scattering to two loops}},\ }\href
  {https://doi.org/10.1016/0370-2693(96)00165-7} {\bibfield  {journal}
  {\bibinfo  {journal} {Phys. Lett. B}\ }\textbf {\bibinfo {volume} {374}},\
  \bibinfo {pages} {210} (\bibinfo {year} {1996})},\ \Eprint
  {https://arxiv.org/abs/hep-ph/9511397} {arXiv:hep-ph/9511397} \BibitemShut
  {NoStop}%
\bibitem [{\citenamefont {Bijnens}\ \emph {et~al.}(1997)\citenamefont
  {Bijnens}, \citenamefont {Colangelo}, \citenamefont {Ecker}, \citenamefont
  {Gasser},\ and\ \citenamefont {Sainio}}]{Bijnens:1997vq}%
  \BibitemOpen
  \bibfield  {author} {\bibinfo {author} {\bibfnamefont {J.}~\bibnamefont
  {Bijnens}}, \bibinfo {author} {\bibfnamefont {G.}~\bibnamefont {Colangelo}},
  \bibinfo {author} {\bibfnamefont {G.}~\bibnamefont {Ecker}}, \bibinfo
  {author} {\bibfnamefont {J.}~\bibnamefont {Gasser}},\ and\ \bibinfo {author}
  {\bibfnamefont {M.~E.}\ \bibnamefont {Sainio}},\ }\bibfield  {title}
  {\bibinfo {title} {{Pion-pion scattering at low energy}},\ }\href
  {https://doi.org/10.1016/S0550-3213(97)00621-4} {\bibfield  {journal}
  {\bibinfo  {journal} {Nucl. Phys. B}\ }\textbf {\bibinfo {volume} {508}},\
  \bibinfo {pages} {263} (\bibinfo {year} {1997})},\ \bibinfo {note} {[Erratum:
  Nucl.Phys.B 517, 639--639 (1998)]},\ \Eprint
  {https://arxiv.org/abs/hep-ph/9707291} {arXiv:hep-ph/9707291} \BibitemShut
  {NoStop}%
\bibitem [{\citenamefont {Bijnens}\ \emph {et~al.}(1998)\citenamefont
  {Bijnens}, \citenamefont {Colangelo},\ and\ \citenamefont
  {Talavera}}]{Bijnens:1998fm}%
  \BibitemOpen
  \bibfield  {author} {\bibinfo {author} {\bibfnamefont {J.}~\bibnamefont
  {Bijnens}}, \bibinfo {author} {\bibfnamefont {G.}~\bibnamefont {Colangelo}},\
  and\ \bibinfo {author} {\bibfnamefont {P.}~\bibnamefont {Talavera}},\
  }\bibfield  {title} {\bibinfo {title} {{The Vector and scalar form-factors of
  the pion to two loops}},\ }\href
  {https://doi.org/10.1088/1126-6708/1998/05/014} {\bibfield  {journal}
  {\bibinfo  {journal} {JHEP}\ }\textbf {\bibinfo {volume} {05}},\ \bibinfo
  {pages} {014}},\ \Eprint {https://arxiv.org/abs/hep-ph/9805389}
  {arXiv:hep-ph/9805389} \BibitemShut {NoStop}%
\bibitem [{\citenamefont {Bijnens}\ \emph {et~al.}(2000)\citenamefont
  {Bijnens}, \citenamefont {Colangelo},\ and\ \citenamefont
  {Ecker}}]{Bijnens:1999hw}%
  \BibitemOpen
  \bibfield  {author} {\bibinfo {author} {\bibfnamefont {J.}~\bibnamefont
  {Bijnens}}, \bibinfo {author} {\bibfnamefont {G.}~\bibnamefont {Colangelo}},\
  and\ \bibinfo {author} {\bibfnamefont {G.}~\bibnamefont {Ecker}},\ }\bibfield
   {title} {\bibinfo {title} {{Renormalization of chiral perturbation theory to
  order p**6}},\ }\href {https://doi.org/10.1006/aphy.1999.5982} {\bibfield
  {journal} {\bibinfo  {journal} {Annals Phys.}\ }\textbf {\bibinfo {volume}
  {280}},\ \bibinfo {pages} {100} (\bibinfo {year} {2000})},\ \Eprint
  {https://arxiv.org/abs/hep-ph/9907333} {arXiv:hep-ph/9907333} \BibitemShut
  {NoStop}%
\bibitem [{\citenamefont {Bijnens}\ and\ \citenamefont
  {Ecker}(2014)}]{Bijnens:2014lea}%
  \BibitemOpen
  \bibfield  {author} {\bibinfo {author} {\bibfnamefont {J.}~\bibnamefont
  {Bijnens}}\ and\ \bibinfo {author} {\bibfnamefont {G.}~\bibnamefont
  {Ecker}},\ }\bibfield  {title} {\bibinfo {title} {{Mesonic low-energy
  constants}},\ }\href {https://doi.org/10.1146/annurev-nucl-102313-025528}
  {\bibfield  {journal} {\bibinfo  {journal} {Ann. Rev. Nucl. Part. Sci.}\
  }\textbf {\bibinfo {volume} {64}},\ \bibinfo {pages} {149} (\bibinfo {year}
  {2014})},\ \Eprint {https://arxiv.org/abs/1405.6488} {arXiv:1405.6488
  [hep-ph]} \BibitemShut {NoStop}%
\bibitem [{\citenamefont {Rodas}\ \emph {et~al.}(2023)\citenamefont {Rodas},
  \citenamefont {Dudek},\ and\ \citenamefont {Edwards}}]{Rodas:2023twk}%
  \BibitemOpen
  \bibfield  {author} {\bibinfo {author} {\bibfnamefont {A.}~\bibnamefont
  {Rodas}}, \bibinfo {author} {\bibfnamefont {J.~J.}\ \bibnamefont {Dudek}},\
  and\ \bibinfo {author} {\bibfnamefont {R.~G.}\ \bibnamefont {Edwards}},\
  }\href@noop {} {\bibinfo {title} {{Constraining the quark mass dependence of
  the lightest resonance in QCD}}} (\bibinfo {year} {2023}),\ \Eprint
  {https://arxiv.org/abs/2304.03762} {arXiv:2304.03762 [hep-lat]} \BibitemShut
  {NoStop}%
\bibitem [{\citenamefont {Blankenbecler}\ \emph {et~al.}(1961)\citenamefont
  {Blankenbecler}, \citenamefont {Goldberger}, \citenamefont {MacDowell},\ and\
  \citenamefont {Treiman}}]{PhysRev.123.692}%
  \BibitemOpen
  \bibfield  {author} {\bibinfo {author} {\bibfnamefont {R.}~\bibnamefont
  {Blankenbecler}}, \bibinfo {author} {\bibfnamefont {M.~L.}\ \bibnamefont
  {Goldberger}}, \bibinfo {author} {\bibfnamefont {S.~W.}\ \bibnamefont
  {MacDowell}},\ and\ \bibinfo {author} {\bibfnamefont {S.~B.}\ \bibnamefont
  {Treiman}},\ }\bibfield  {title} {\bibinfo {title} {Singularities of
  scattering amplitudes on unphysical sheets and their interpretation},\ }\href
  {https://doi.org/10.1103/PhysRev.123.692} {\bibfield  {journal} {\bibinfo
  {journal} {Phys. Rev.}\ }\textbf {\bibinfo {volume} {123}},\ \bibinfo {pages}
  {692} (\bibinfo {year} {1961})}\BibitemShut {NoStop}%
\bibitem [{\citenamefont {Zhou}\ \emph {et~al.}(2005)\citenamefont {Zhou},
  \citenamefont {Qin}, \citenamefont {Zhang}, \citenamefont {Xiao},
  \citenamefont {Zheng},\ and\ \citenamefont {Wu}}]{Zhou:2004ms}%
  \BibitemOpen
  \bibfield  {author} {\bibinfo {author} {\bibfnamefont {Z.~Y.}\ \bibnamefont
  {Zhou}}, \bibinfo {author} {\bibfnamefont {G.~Y.}\ \bibnamefont {Qin}},
  \bibinfo {author} {\bibfnamefont {P.}~\bibnamefont {Zhang}}, \bibinfo
  {author} {\bibfnamefont {Z.}~\bibnamefont {Xiao}}, \bibinfo {author}
  {\bibfnamefont {H.~Q.}\ \bibnamefont {Zheng}},\ and\ \bibinfo {author}
  {\bibfnamefont {N.}~\bibnamefont {Wu}},\ }\bibfield  {title} {\bibinfo
  {title} {{The Pole structure of the unitary, crossing symmetric low energy pi
  pi scattering amplitudes}},\ }\href
  {https://doi.org/10.1088/1126-6708/2005/02/043} {\bibfield  {journal}
  {\bibinfo  {journal} {JHEP}\ }\textbf {\bibinfo {volume} {02}},\ \bibinfo
  {pages} {043}}\BibitemShut {NoStop}%
\bibitem [{\citenamefont {Li}\ and\ \citenamefont {Zheng}(2022)}]{Li:2021oou}%
  \BibitemOpen
  \bibfield  {author} {\bibinfo {author} {\bibfnamefont {Q.-Z.}\ \bibnamefont
  {Li}}\ and\ \bibinfo {author} {\bibfnamefont {H.-Q.}\ \bibnamefont {Zheng}},\
  }\bibfield  {title} {\bibinfo {title} {{Singularities and accumulation of
  singularities of \ensuremath{\pi}N scattering amplitudes}},\ }\href
  {https://doi.org/10.1088/1572-9494/ac8869} {\bibfield  {journal} {\bibinfo
  {journal} {Commun. Theor. Phys.}\ }\textbf {\bibinfo {volume} {74}},\
  \bibinfo {pages} {115203} (\bibinfo {year} {2022})},\ \Eprint
  {https://arxiv.org/abs/2108.03734} {arXiv:2108.03734 [nucl-th]} \BibitemShut
  {NoStop}%
\bibitem [{\citenamefont {Nebreda}\ \emph {et~al.}(2011)\citenamefont
  {Nebreda}, \citenamefont {Pelaez},\ and\ \citenamefont
  {Rios}}]{Nebreda:2011di}%
  \BibitemOpen
  \bibfield  {author} {\bibinfo {author} {\bibfnamefont {J.}~\bibnamefont
  {Nebreda}}, \bibinfo {author} {\bibfnamefont {J.~R.}\ \bibnamefont
  {Pelaez}},\ and\ \bibinfo {author} {\bibfnamefont {G.}~\bibnamefont {Rios}},\
  }\bibfield  {title} {\bibinfo {title} {{Chiral extrapolation of pion-pion
  scattering phase shifts within standard and unitarized Chiral Perturbation
  Theory}},\ }\href {https://doi.org/10.1103/PhysRevD.83.094011} {\bibfield
  {journal} {\bibinfo  {journal} {Phys. Rev. D}\ }\textbf {\bibinfo {volume}
  {83}},\ \bibinfo {pages} {094011} (\bibinfo {year} {2011})},\ \Eprint
  {https://arxiv.org/abs/1101.2171} {arXiv:1101.2171 [hep-ph]} \BibitemShut
  {NoStop}%
\bibitem [{\citenamefont {Chen}\ \emph {et~al.}(2023)\citenamefont {Chen},
  \citenamefont {Cheng}, \citenamefont {Yan}, \citenamefont {Duan},\ and\
  \citenamefont {Guo}}]{Chen:2023ybr}%
  \BibitemOpen
  \bibfield  {author} {\bibinfo {author} {\bibfnamefont {C.}~\bibnamefont
  {Chen}}, \bibinfo {author} {\bibfnamefont {N.-Q.}\ \bibnamefont {Cheng}},
  \bibinfo {author} {\bibfnamefont {L.-W.}\ \bibnamefont {Yan}}, \bibinfo
  {author} {\bibfnamefont {C.-G.}\ \bibnamefont {Duan}},\ and\ \bibinfo
  {author} {\bibfnamefont {Z.-H.}\ \bibnamefont {Guo}},\ }\bibfield  {title}
  {\bibinfo {title} {{Revisiting the tensor-meson nonet in resonance chiral
  theory}},\ }\href {https://doi.org/10.1103/PhysRevD.108.014002} {\bibfield
  {journal} {\bibinfo  {journal} {Phys. Rev. D}\ }\textbf {\bibinfo {volume}
  {108}},\ \bibinfo {pages} {014002} (\bibinfo {year} {2023})},\ \Eprint
  {https://arxiv.org/abs/2302.11316} {arXiv:2302.11316 [hep-ph]} \BibitemShut
  {NoStop}%
\end{thebibliography}%

\end{document}